\documentclass[11pt, notitlepage]{article} 
\usepackage{a4}
\usepackage{xcolor}
\usepackage{soul}

\definecolor{TUMblue}{RGB}{0, 101, 189}
\definecolor{TUMlightblue}{RGB}{100,160,200}
\definecolor{TUMgreen}{RGB}{162,173,0}
\definecolor{TUMorange}{RGB}{227,114,034}
\definecolor{TUMivory}{RGB}{218,215,203}
\definecolor{fgreen}{RGB}{34,139,34}

\usepackage{natbib}
\usepackage[margin=0.8in]{geometry}
\usepackage{hyperref}
\hypersetup{
	colorlinks=true,
	linkcolor=TUMblue,
	citecolor=TUMblue,
	filecolor=TUMblue,
	urlcolor=TUMblue
}

\usepackage{etoolbox}
\makeatletter

\pretocmd{\NAT@citex}{%
	\let\NAT@hyper@\NAT@hyper@citex
	\def\NAT@postnote{#2}%
	\setcounter{NAT@total@cites}{0}%
	\setcounter{NAT@count@cites}{0}%
	\forcsvlist{\stepcounter{NAT@total@cites}\@gobble}{#3}}{}{}
\newcounter{NAT@total@cites}
\newcounter{NAT@count@cites}
\def\NAT@postnote{}

\def\NAT@hyper@citex#1{%
	\stepcounter{NAT@count@cites}%
	\hyper@natlinkstart{\@citeb\@extra@b@citeb}#1%
	\ifnumequal{\value{NAT@count@cites}}{\value{NAT@total@cites}}
	{\ifNAT@swa\else\if*\NAT@postnote*\else%
		\NAT@cmt\NAT@postnote\global\def\NAT@postnote{}\fi\fi}{}%
	\ifNAT@swa\else\if\relax\NAT@date\relax
	\else\NAT@@close\global\let\NAT@nm\@empty\fi\fi
	\hyper@natlinkend}
\renewcommand\hyper@natlinkbreak[2]{#1}

\patchcmd{\NAT@citex}
{\ifNAT@swa\else\if*#2*\else\NAT@cmt#2\fi
	\if\relax\NAT@date\relax\else\NAT@@close\fi\fi}{}{}{}
\patchcmd{\NAT@citex}
{\if\relax\NAT@date\relax\NAT@def@citea\else\NAT@def@citea@close\fi}
{\if\relax\NAT@date\relax\NAT@def@citea\else\NAT@def@citea@space\fi}{}{}

\makeatother

\usepackage{courier}
\usepackage{amssymb, graphicx}
\usepackage{amsmath}
\usepackage{amsthm}
\usepackage{verbatim}
\usepackage{dsfont}
\usepackage{geometry}
\usepackage{pdflscape}
\usepackage{multirow}
\usepackage{aliascnt}
\usepackage{verbatim}
\usepackage{graphicx}
\usepackage{float}
\usepackage{tikz}
\usetikzlibrary{calc,shapes,arrows,decorations.pathmorphing,graphs,positioning,backgrounds,arrows.meta}
\usepackage{latexsym}
\usepackage{mathtools}
\usepackage{mathrsfs}
\usepackage{bm}
\usepackage{bbm}
\usepackage{shadethm}
\usepackage{enumerate}
\usepackage{colortbl}
\usepackage{framed}
\colorlet{shadecolor}{gray!25}
\usepackage{booktabs}
\usepackage{longtable}
\usepackage{multirow}
\usepackage{rotating}
\usepackage{chngpage}
\usepackage{appendix}
\usepackage{pdfpages}
\usepackage{adjustbox}
\usepackage{subfig}
\usepackage{comment}

\usepackage[ruled,vlined]{algorithm2e}
\usepackage[noend]{algpseudocode}
\algnewcommand{\IIf}[1]{\State\algorithmicif\ #1\ \algorithmicthen}
\algnewcommand{\EndIIf}{\unskip\ \algorithmicend\ \algorithmicif}
\makeatletter
\def\BState{\State\hskip-\ALG@thistlm}
\makeatother

\newcommand{\mynewtheorem}[2]{
	\newaliascnt{#1}{dummy}
	\newtheorem{#1}[#1]{#2}
	\aliascntresetthe{#1}
	\expandafter\def\csname #1autorefname\endcsname{#2}
}

\bibpunct[\textcolor{TUMblue}{, }]{\textcolor{TUMblue}{(}}{\textcolor{TUMblue}{)}}{\textcolor{TUMblue}{;}}{\textcolor{TUMblue}{a}}{\textcolor{TUMblue}{}}{\textcolor{TUMblue}{,}}
\makeatletter
\renewcommand\eqref[1]{%
	\textup{\color{TUMblue}\tagform@{\ref{#1}}}%
}

\newtheorem{Theorem}{Theorem}
\newtheorem{Corollary}{Corollary}

\theoremstyle{definition}
\mynewtheorem{thm}{Theorem}
\mynewtheorem{definition}{Definition}
\mynewtheorem{lem}{Lemma}
\mynewtheorem{cor}{Corollary}
\mynewtheorem{proposition}{Proposition}
\mynewtheorem{example}{Example}
\mynewtheorem{alg}{Algorithm}
\mynewtheorem{remark}{Remark}

\DeclareMathOperator*{\argmax}{arg\,max}

\definecolor{mygray}{gray}{0.85}
\def\bm#1{\mbox{\boldmath $#1$}}

\usetikzlibrary{shapes,arrows,decorations.pathmorphing,graphs,positioning,backgrounds}

\tikzstyle{VineNode} = [ellipse, fill = white, draw = black, text = black, align = center, minimum height = 1cm, minimum width = 1cm]
\tikzstyle{DummyNode}  = [draw = none, fill = none, text = black]
\tikzstyle{TreeLabels} = [draw = none, fill = none, text = black] 

\usepackage{etoolbox}
\makeatletter
\patchcmd{\hyper@makecurrent}{%
	\ifx\Hy@param\Hy@chapterstring
	\let\Hy@param\Hy@chapapp
	\fi
}{%
	\iftoggle{inappendix}{
		\@checkappendixparam{chapter}%
		\@checkappendixparam{section}%
		\@checkappendixparam{subsection}%
		\@checkappendixparam{subsubsection}%
		\@checkappendixparam{paragraph}%
		\@checkappendixparam{subparagraph}%
	}{}%
}{}{\errmessage{failed to patch}}

\newcommand*{\@checkappendixparam}[1]{%
	\def\@checkappendixparamtmp{#1}%
	\ifx\Hy@param\@checkappendixparamtmp
	\let\Hy@param\Hy@appendixstring
	\fi
}
\makeatletter

\newtoggle{inappendix}
\togglefalse{inappendix}

\apptocmd{\appendix}{\toggletrue{inappendix}}{}{\errmessage{failed to patch}}
\apptocmd{\subappendices}{\toggletrue{inappendix}}{}{\errmessage{failed to patch}}

\usepackage{sectsty}
\allsectionsfont{\sffamily}

\usepackage{titlesec}
\titleformat{\chapter}[display]
{\normalfont\sffamily\LARGE\bfseries\centering}
{\chaptertitlename\ \thechapter}{20pt}{\LARGE}

\usepackage{caption}
\usepackage{footnote}

\begin{document}
	{	\renewcommand*{\thefootnote}{\fnsymbol{footnote}}
		\title{\textbf{\sffamily Bivariate vine copula based regression, bivariate level and quantile curves}}
		
		\date{\small \today}
		\newcounter{savecntr1}
		\newcounter{restorecntr1}
		\newcounter{savecntr2}
		\newcounter{restorecntr2}

		\author{Marija Tepegjozova\setcounter{savecntr1}{\value{footnote}}\thanks{Department of Mathematics, Technische Universit{\"a}t M{\"u}nchen, Boltzmannstra{\ss}e 3, 85748 Garching, Germany (email: \href{mailto:m.tepegjozova@tum.de}{m.tepegjozova@tum.de} (corresponding author))} $\;$ and
			Claudia Czado\setcounter{savecntr2}{\value{footnote}}\thanks{Department of Mathematics and Munich Data Science Institute, Technische Universit{\"a}t M{\"u}nchen, Boltzmannstra{\ss}e 3, 85748, Garching, Germany (email: \href{mailto:cczado@ma.tum.de}{cczado@ma.tum.de})}
		}
		\maketitle 
	}
	\begin{abstract}
		The statistical analysis of univariate quantiles is a well developed  research topic. However, there is a need for research in multivariate quantiles. We construct bivariate (conditional) quantiles using the level curves of vine copula based bivariate regression model.
		Vine copulas are graph theoretical models identified by  a sequence of linked trees, which allow for separate modelling of marginal distributions and the dependence structure.  We introduce a novel graph structure model (given by a tree sequence) specifically designed for a symmetric treatment of two responses in a predictive regression setting.  We establish  computational tractability of the model and a straight forward way of obtaining different conditional distributions. Using vine copulas the typical shortfalls of regression, as the need for transformations or interactions of predictors, collinearity or quantile crossings are avoided. We illustrate the copula based bivariate level curves for different copula distributions and show how they can be adjusted to form valid quantile curves.  We apply our approach to weather measurements from Seoul, Korea. This  data example emphasizes the benefits of the joint bivariate response  modelling in contrast to two separate univariate regressions or by assuming conditional independence, for bivariate response data set in the presence of conditional dependence.
	\end{abstract}
	\textit{Keyword: multivariate quantiles, bivariate response, bivariate conditional distribution functions  }

	\section{Introduction}\label{introduction1}
	The topic of predicting quantiles of a response variable conditioned on a set of  predictor variables taking on fixed values, continuously attracts interest.
	The statistical analysis of such univariate quantiles is a well developed  research topic \citep{koenker1978regression, Koenker2005quantile}. Since the introduction of the linear quantile regression by \cite{koenker1978regression}  many extensions have been developed for the case of a univariate response variable. A short summary of developments in quantile regression modelling is given in \cite{koenker2017quantile}. 
	Recent approaches for quantile regression are vine copula based quantile regression methods \citep{kraus2017d, tepegjozova2021nonparametric, chang2019prediction, zhu2021simplified}. Copulas allow for separate modelling of the marginal distributions and  the dependence structure in the data. Vine copulas construct  multivariate copulas using bivariate building blocks only, a so-called pair copula construction. This way, a very flexible model, without assuming  homoscedasticity, or a linear relationship between the response and the predictors, is constructed. Thus, vine based quantile regression methods  overcome two  drawbacks of the standard quantile regression methods. First, by construction quantile crossings and collinearity are avoided, and  second, there is no need for  transformations or  interactions of variables \citep{kraus2017d}. 
	There are several different vine copula tree structures that can be considered, the most  general regular (R-)vine copulas, or subsets such as drawable (D-vines whose tree structure is a sequence of paths) and canonical (C-vines whose tree structure is a sequence of stars) vines. \cite{kraus2017d} developed a parametric  D-vine based quantile regression method by optimizing the conditional log-likelihood and adding predictors until there is no improvement, thus introducing an automatic forward variable selection method. This approach was extended in \cite{tepegjozova2021nonparametric} where a nonparametric D- and C-vine copula based quantile regression was introduced. They also follow the approach to maximize the conditional log likelihood, but introduce an additional step to check for future improvement of the conditional log likelihood, a so called two-step ahead approach. \cite{chang2019prediction} introduced  R-vine based quantile regression model by first finding the  optimal R-vine structure among all predictors and then adding the response variable to each tree in the vine structure as a leaf node. Another R-vine based regression was introduced in \cite{zhu2021simplified} by choosing the R-vine structure which gives the largest sum of the absolute value of the partial correlations in each step of the forward extension with predictor variables, while keeping the response as a leaf node. This approach is motivated by the algorithm and results from \cite{zhu2020common}. All these selected structures allow to express the conditional density of the response given the predictors without integration.
	
	Despite the great attention univariate quantiles  have received, the extension to multivariate response quantiles is not trivial nor well-defined. Several theoretical notions of multivariate quantiles  have been introduced, but there is no consensus which one is the corresponding generalization of the univariate quantiles. These include geometric quantiles based on halfspace depth contours with different concepts of statistical depth (e.g. see \cite{tukey1975mathematics}, \cite{chaudhuri1996geometric}, \cite{hallin2010multivariate}, \cite{chernozhukov2017monge}), vector quantiles (see \cite{carlier2016vector} and \cite{carlier2017vec}), spatial quantiles (see \cite{abdous1992note}). 
	Our goal is to define bivariate quantiles  in terms of copula distribution and to introduce a  vine based regression method able to handle bivariate responses. 
	Copula based models are known to be excelling at modeling tail events and asymmetric dependencies. Bivariate quantiles arising from  copula based models are advantageous in assessing the joint risk of failure or occurrence of two events. 
	One example of specific bivariate quantiles being applied is the analysis of floods by \cite{chebana2011multivariate}. The risk of flood is studied through a joint analysis of flood peaks and flood volumes,  using level sets of bivariate copulas. \cite{requena2013bivariate} present a similar analysis of the risk of flood, focusing on hydrologic dam designs. Additionally, the authors use the same approach  for a joint analysis of reservoir volume and spillway crest length as indicators for the risk of dam overtopping. A multivariate risk of failure analysis based on copulas is presented in \cite{salvadori2015practical} for structural failure assessment in engineering. An application in financial mathematics can be found in \cite{di2014estimation}. The authors propose  estimation of a tail event risk measure based on multivariate level sets of copulas. However,  all these approaches model the two responses with a copula, but do not consider any explanatory variables or predictors. It would be even more beneficial to jointly model two response variables, taking into account the influence of a set of possible predictors. Thus, to fill in this gap, we focus on bivariate conditional quantiles and their estimation using a very flexible vine copula model.
	
	The first  heuristic for a vine based quantile regression with multiple responses is given in \cite{zhu2021simplified}, but the question of multivariate response quantiles is not tackled. The suggested heuristic  for the bivariate response case is limited to modelling only, but not predicting the bivariate quantiles. Further,  this approach has an asymmetric  treatment of the response variables. This might lead to different performance of the regression methods when the order of the response variables is exchanged. Thus, there is still a need for: $(1.)$ a valid definition of (unconditional and conditional) multivariate quantiles  linked to the usage of copulas; $(2.)$ a vine based quantile regression method with a symmetric treatment of the responses; $(3.)$ a numerical method for obtaining the multivariate (unconditional and conditional) quantiles  from the estimated vine based model and evaluating predictions from it.  
	
	Our methodology  deals with the case of a bivariate response variables, i.e. bivariate (unconditional and conditional) quantiles defined by sets that can be characterised as curves. 
	For the conditional case we choose the multivariate vine copulas class since they allow modelling of complex dependence patterns including asymmetric tail dependencies.
	For $(1.) $ we extend the definition of bivariate unconditional and conditional quantiles  in terms of a copula distribution function. We first start by exploring (un)conditional level curves of  copula distributions. 
	Further,  we illustrate the bivariate unconditional level curves for known bivariate copula distributions and the bivariate conditional level curves for a 3-dimensional vine copula distribution. Then, we show that the coverage probabilities of the $\alpha$ level curves is not $\alpha$ and propose to adjust them in such a way that the adjusted level, say $\beta(\alpha)$ results in bivariate quantile curves with coverage $\alpha$.
For $(2.)$, we propose a novel tree structure for vines, called Y-vine tree sequence, which is contained in the set of regular vine tree sequences. It is designed to allow for a  symmetric treatment of the responses. Moreover, we show that using the Y-vine tree sequence the associated bivariate conditional density is analytically expressible as a product of all pair copula terms involving one or both of the response variables. In the case of more than one conditioning variable (predictor) we develop a forward selection method. For this we propose an appropriate fit measure for the selection  of predictors to prevent overfitting and remove non-significant predictors. For $(3.)$ we develop a numerical method to evaluate the bivariate unconditional and conditional level curves and corresponding quantile curves.
Based on the estimated quantiles, we can construct bivariate confidence regions, which is a generalization of univariate confidence intervals. They contain the data with a given probability and can locate parts of a distribution with high density values. Such confidence regions can be effective for visualizing trends, patterns and outliers \citep{korpela2014confidence, korpela2017multivariate, guilbaud2008simultaneous}. Further, for applicability of the proposed method we develop a prediction method  of the two responses given an arbitrary set of predictor values and show how to simulate data from a Y-vine with fixed predictor values.   Finally, we give an application involving a data set with minimal and maximal daily temperatures together with other weather variables. For this application we show that the conditional dependence cannot be ignored and that it is non-Gaussian, thus requiring the full class of pair copula families. In addition, we illustrate our confidence regions and compare them to alternative ways of construction of  confidence regions, assuming (conditional) independence. We highlight the advantages of our modeling approach and its usability in data analysis.

The remainder of the paper is organized as follows. First, we introduce the necessary vine copula concepts for our approach in Section~\ref{sc:theory}. In Section~\ref{sc:quantilereg} we introduce the Y-vine copula based  regression model for bivariate responses. For application purposes, in Section~\ref{sc:prediction} we present  the prediction method for Y-vine copulas. Then, in Section~\ref{BQR:bivariatequantiles} we define bivariate level curves and quantile curves in terms of copulas and develop  numerical methods used for their evaluation.  For the demonstration of the usefulness of our method we include a real data example in Section~\ref{sc:dataapp} that contains dependent bivariate responses. We  highlight the advantages of bivariate response modelling over standard univariate models or models that assume conditional independence.  Finally, in Section~\ref{sc:conclussion} we give conclusions and areas of future research.

\section{Theoretical background}\label{sc:theory}	

Consider any continuous d-dimensional random vector $\mathbf{X}=\left(X_1,\ldots,X_d\right)^T$ with observed values  $\mathbf{x}=\left(x_1,\ldots,x_d\right)^T$. We use capital letters for random
variables and lowercase letters for their observed values, i.e., we write $X_i = x_i$ for $i=1,\ldots, d$.  Let $\mathbf{X}$ have joint distribution function  $F$, joint density $f$ and marginal distributions $F_{X_i},\;i=1,\ldots, d.$ 
Following Sklar's theorem \citep{sklar1959fonctions}, we can express the multivariate distribution function $F$ in terms of the marginal distributions, $F_{X_i}$, and the $d$-dimensional copula $C$ as 
\begin{equation}\label{sklar_dist}
	F\left(x_1,\ldots,x_d\right) = C\left(   F_{X_1}(x_1),\ldots, F_{X_d}(x_d)\right).
\end{equation}

\noindent Since we assume a continuous joint distribution $F$, the copula $C:\left[0,1\right]^d\mapsto\left[0,1\right]$  corresponds to the distribution of the random vector $\mathbf{U} = \left(U_1,\ldots ,U_d\right)^T,$ with the components of $\mathbf{U}$ being the probability integral transforms (PITs or u-scale) of the components of $\mathbf{X}$ (x-scale),  $U_i = F_{X_i}\left(X_i\right)$ for $i=1,\ldots ,d$. Each $U_i$ is   uniformly distributed and their joint distribution function $C$   is  the  copula  associated  with $\mathbf{X}$.
By Sklar's Theorem it is implied that $C$ is unique, in the absolutely continuous case, which we assume. Also, if derivatives of the marginal distributions $F_{X_i}$ exist, the density $f$ can be derived as 
\begin{equation}\label{sklar_den}
	f\left(x_1,\ldots ,x_d\right) = c\left( F_{X_1}(x_1),\ldots, F_{X_d}(x_d)\right) \cdot \prod_{i=1}^d f_{X_i}\left(x_i\right),
\end{equation} 
where $c$ is the $d$-dimensional density corresponding to the copula $C$ and $f_{X_1},\ldots ,f_{X_d}$ are the univariate  marginal densities. However, Equations \eqref{sklar_dist} and \eqref{sklar_den} both incorporate a possibly complicated multivariate copula distribution and density.  As shown by \cite{joe1996families}, a $d$-dimensional copula density can be decomposed into $d\left(d-1\right)/2$ bivariate copula densities. This decomposition is not unique, but a large number of possible decompositions exist. The elements of these decompositions, the bivariate copula densities can be chosen completely independent of each other. A graphical model  introduced by \cite{bedford2002vines}, called regular vine copulas (R-vines),  organizes all such decompositions that lead to a valid density.
Thus, the estimation of a $d$-dimensional copula density is subdivided into the estimation of $d (d-1)/2$  two-dimensional  copula  densities. A regular vine copula  on $d$ uniformly distributed random variables $U_1,\ldots ,U_d,$ consists of a regular vine tree sequence, denoted by $\mathcal{V}$, a set of bivariate copula families (also known as pair copulas) $\mathcal{B}\left(\mathcal{V}\right)$, and a set of parameters corresponding to the bivariate copula families $\Theta\left(\mathcal{B}\left(\mathcal{V}\right)\right)$. The vine tree sequence or tree structure $\mathcal{V}$ consists of a sequence of linked trees, $T_k=\left(N_k,E_k\right),\; k=1,\ldots ,d-1$, satisfying the following conditions:
\begin{itemize}
	\item[(i)] $T_1$ is a tree with node set $N_1=\left\{U_1,\ldots ,U_d\right\}$ and edge set $E_1$.
	\item[(ii)] For $k\geq 2$, $T_k$ is a tree with node set $N_k=E_{k-1}$ and edge set $E_k$. 
	\item[(iii)](Proximity condition) For $k\geq 2$, two nodes of the tree $T_k$ can  be connected by an edge if the corresponding edges of $T_{k-1}$ have a common node.

\end{itemize}
If the vine tree sequence consists of paths only, then we call it a drawable vine (D-vine), and if it consists of stars,  it is called a canonical vine (C-vine) \citep{bedford2002vines}.
The tree sequence uniquely  specifies which bivariate (conditional) copula densities occur in the decomposition. Each edge $e \in E_k$ for $k=1,\ldots ,d-1$ is associated with a  bivariate  copula family $c_{U_{j_e},U_{k_e};\mathbf{U}_{D_e}}\in \mathcal{B}\left(\mathcal{V}\right),$ and a corresponding set of parameters $\boldsymbol{\theta}_{j_e,k_e;D_e}\in\Theta\left(\mathcal{B}\left(\mathcal{V}\right)\right)$.  $U_{j_e}$ and $U_{k_e}$ are the conditioned variables and $\mathbf{U}_{D_e}$ represents the conditioning set corresponding to edge $e$, $\mathbf{U}_{D_e} = \left(U_i\right)_{i\in D_e}$. Denote 
the conditional distribution of $U_{j_e}\vert \mathbf{U}_{D_e} = \mathbf{u}_{D_e}$ with $C_{U_{j_e}\vert\mathbf{U}_{ D_e}}$.  We define the so-called pseudo copula data $u_{j_e\vert D_e}$ as $u_{j_e\vert D_e}  \coloneqq C_{U_{j_e}\vert\mathbf{U}_{D_e}}\left(u_{j_e}\vert \mathbf{u}_{D_e}\right)$. Similarly, $u_{k_e\vert D_e}$ is defined.
Then, $c_{U_{j_e},U_{k_e};\mathbf{U}_{D_e}}$   denotes the density of the copula between the  pseudo copula data  $u_{j_e\vert D_e}$ and $u_{k_e\vert D_e}$.  The corresponding  distribution function 
is denoted as $C_{U_{j_e},U_{k_e};\mathbf{U}_{D_e}}$. 

\noindent  \cite{bedford2002vines} have shown that the graphical model of regular vines, leads to a natural decomposition of the joint copula density $c$ using the pair-copulas defined through the tree sequence as
\begin{equation}\label{vine_den}
	c\left(u_1,\ldots ,u_d\right) = \prod_{k=1}^{d-1}\prod_{e\in E_k} c_{U_{j_e},U_{k_e};\mathbf{U}_{D_e}} \left(C_{U_{j_e}\vert \mathbf{U}_{D_e}}\left(u_{j_e}\vert\mathbf{u}_{D_e}\right),C_{U_{k_e}\vert \mathbf{U}_{D_e}}\left(u_{k_e}\vert\mathbf{u}_{D_e}\right)\vert\mathbf{u}_{D_e}\; \right).
\end{equation}
Using Equation~\eqref{vine_den} we can decompose any given regular vine copula density. However, the individual pair copulas, $c_{U_{j_e},U_{k_e};\mathbf{U}_{D_e}}$ in Equation~ \eqref{vine_den} are dependent on  $\mathbf{u}_{D_e}$. This represents the different conditional dependencies between $U_{j_e}$ and $U_{k_e}$  for different conditioning values of $\mathbf{u}_{D_e}$. To improve computational tractability, it is customary to ignore this influence and simplify Equation~\eqref{vine_den} to: 
\begin{equation}\label{s_vine_den}
	c\left(u_1,\ldots ,u_d\right) = \prod_{k=1}^{d-1}\prod_{e\in E_k} c_{U_{j_e},U_{k_e};\mathbf{U}_{D_e}} \left(C_{U_{j_e}\vert \mathbf{U}_{D_e}}\left(u_{j_e}\vert\mathbf{u}_{D_e}\right),C_{U_{k_e}\vert \mathbf{U}_{D_e}}\left(u_{k_e}\vert\mathbf{u}_{D_e}\right)\;  \right).
\end{equation}
This simplification is known as the simplifying assumption (more in \cite{haff2010simplified} and \cite{stoeber2013simplified}). It is made due to tractability in higher dimensions, and can be further tested for validity (see for example \cite{kurz2022testing} and \cite{derumigny2017tests}). In this case, we talk about pair copula constructions (PCC) of multivariate densities.\\
To derive the  conditional distributions in Equation~\eqref{s_vine_den}, we  use  the recursion formula from \cite{joe1996families}. It defines a recursion for  conditional distributions of a regular vine over its tree sequence.
Let $l\in D_e$ and $D_{-l}\coloneqq D_e\setminus \left\lbrace l\right\rbrace $. Further, let $h_{U_{j_e}\vert U_l;\mathbf{U}_{D_{-l}}}\left(\cdot \vert \cdot \right)$ denote the so-called h-function associated with the pair copula $c_{U_{j_e},U_l;\mathbf{U}_{D_{-l}}}$, defined as $h_{U_{j_e}\vert U_l;\mathbf{U}_{D_{-l}}}(u_{j_e}|u_{l}) \coloneqq  \frac{\partial}{\partial u_{l}} C_{U_{j_e},U_{l};\bm{U}_{D_{-l}}}(u_{j_e},u_{l}).$   Then the following  recursion is valid 
\begin{equation}\label{cond_eq1}
	C_{U_{j_e}\vert \mathbf{U}_{D_e}}\left(u_{j_e}\vert\mathbf{u}_{D_e}\right) = h_{U_{j_e}\vert U_l;\mathbf{U}_{D_{-l}}}\left( C_{U_{j_e}\vert\mathbf{U}_{D_{-l}}}\left(u_{j_e}\vert \mathbf{u}_{D_{-l}}\right) \vert C_{U_l\vert\mathbf{u}_{D_{-l}}}\left(u_l\vert \mathbf{U}_{D_{-l}}\right)\right).
\end{equation}

\section{Vine copula based bivariate  regression}\label{sc:quantilereg}
\subsection{General framework}

Consider the variables $(Y_1,Y_2)^T$ as the 2-dimensional response vector and $\mathbf{X} = \left(X_1,\ldots ,\right.$  $\left. X_p\right)^T$ as the p-dimensional predictor vector. The main interest of the bivariate regression is to model the joint conditional distribution function of the response variables $\mathbf{Y} = \left( Y_1,Y_2\right)^T$ given the outcome of some predictor variables $\mathbf{X}= \mathbf{x}$, denoted as $F_{Y_1,Y_2| \mathbf{X}} \left(y_1,y_2|\mathbf{x}  \right) $. This can be achieved by  joint modelling of $\left(\mathbf{Y}, \mathbf{X}\right)^T$ and subsequently deriving the conditional distribution of the bivariate response vector $\mathbf{Y}$ given $\mathbf{X}= \mathbf{x}$. The same can be achieved by joint modelling of the PIT values of the responses $\mathbf{V} = \left( V_1,V_2\right)^T$,  the predictors $\mathbf{U} = \left(U_1,\ldots ,U_p\right)^T,$ and the corresponding conditional distribution function of $\mathbf{V}$ given $\mathbf{U}= \mathbf{u}$, denoted as $C_{V_1,V_2|\mathbf{U}} \left(v_1,v_2|\mathbf{u}  \right)$. The connection between these two approaches for the joint conditional distribution, on the x- and u-scale, is  derived in Proposition~\ref{prop:cond}.  

\begin{proposition} \label{prop:cond}
	The conditional distribution of  $\mathbf{Y} = \left( Y_1,Y_2\right)^T$ given  $\mathbf{X} = \left(X_1,\ldots ,X_p\right)^T,$ with corresponding PITs $V_j \coloneqq F_{Y_j}\left(Y_j\right),\;j=1,2$ and $U_i \coloneqq F_{X_i}\left(X_i\right),\; i=1, \ldots,p$ can be expressed in terms of a conditional distribution function associated with a copula as
	
	\begin{equation*}
		F_{Y_1,Y_2| \mathbf{X}} \left(y_1,y_2|\mathbf{x}  \right) =
		C_{V_1,V_2|\mathbf{U}} \left(F_{Y_1}(y_1), F_{Y_2}(y_2)| F_{X_1}(x_1),\ldots,  F_{X_p}(x_p) \right).
	\end{equation*}	
\end{proposition}

Proof of Proposition~\ref{prop:cond} is given in Appendix~\ref{app:proposition3}. 
Here $C_{V_1,V_2|\mathbf{U}}$ is the bivariate conditional distribution associated with the $p+2$ dimensional copula $C_{V_1,V_2,\mathbf{U}}$ and does not need to have uniform margins. In general, $C_{V_1,V_2|\mathbf{U}}$ is different than $C_{V_1,V_2;\mathbf{U}}$, as $C_{V_1,V_2;\mathbf{U}}$ is a bivariate copula with uniform marginal distributions and corresponds to the copula associated with the bivariate conditional distribution of $(Y_1,Y_2)$ given $\mathbf{X}=\mathbf{x}$. From Proposition \ref{prop:cond}, in order to model the bivariate conditional distribution function, we need to estimate  the marginal distributions $F_{Y_j}, F_{X_i}$ for $ j =1,2,\; i=1,\ldots,p,$ and the bivariate conditional  distribution $C_{V_1,V_2\vert \mathbf{U}}$. To obtain the later, we need to estimate the $p+2$ dimensional copula $C_{V_1,V_2,\mathbf{U}} $ describing the joint distribution of $(V_1,V_2,\mathbf{U})$.
Following \citet{kraus2017d, noh2013copula}, we estimate the marginal distributions nonparameterically to reduce the bias caused by model misspecification.  Examples of
nonparametric univariate estimators are the continuous kernel smoothing estimator \citep{parzen1962estimation} and
the transformed local likelihood estimator \citep{geenens2014probit}.
A more complex task is  estimating the $p+2$ dimensional copula $C_{V_1,V_2,\mathbf{U}}$ and subsequently, deriving the bivariate conditional distribution from this copula. We propose to model the copula $C_{V_1,V_2,\mathbf{U}}$ using regular vine copulas. However, we also have to take care that deriving the bivariate conditional  distribution $C_{V_1,V_2\vert \mathbf{U}}$ remains numerically  tractable. Thus, to obtain the joint conditional distribution of the response variables using  only pair copulas estimated in the vine copula model,  additional constraints are required.
The constraint for a univariate vine regression  is that the node containing the response in the conditioned set is a leaf node in each tree of the tree sequence, as shown by \cite{kraus2017d} for D-vines and by \cite{tepegjozova2021nonparametric} for C-vines. Following these results, the constraint for the bivariate  vine regression model is that the two response variables are exactly the conditioned set of the edge of the last tree in the vine tree sequence, as also used by \cite{zhu2021simplified}. However, in their approach there is no symmetric treatment of the two responses, which is a drawback. Therefore, we propose a new vine tree structure specifically designed for  bivariate  regression modelling with a symmetric treatment of the responses.  
\subsection{Y-vine copula model}
Let $\mathbf{X}_{-i} $ be a $(p-1)$-dimensional vector defined as 
$\mathbf{X}_{-i} \coloneqq (X_1,\ldots, X_{i-1},X_{i+1},\ldots X_{p} )^T  $ 
and  let $\mathbf{X}_{i : i+k}$  be a $(k+1) $-dimensional vector defined as  $\mathbf{X}_{i : i+k} \coloneqq (X_i,\ldots, X_{i+k} ) ^T.  $  Similar definitions hold for the vectors $\mathbf{x}_{-i}, \mathbf{U}_{-i} $, $\mathbf{u}_{-i}  $, and for $ \mathbf{x}_{i:i+k}, \mathbf{U}_{i:i+k}$  $\mathbf{u}_{i:i+k}$, respectively.  
\begin{definition}\label{vinetree}
	Given the marginal  PIT transformed response variables $V_1,V_2$ and   predictor variables $U_1,\ldots, U_p$, we define the $p+1$ trees of the Y-vine tree sequence for bivariate  regression as the following:
	\begin{itemize}
		\item[$\mathbf{T_1}$] with  $N_1 = \left\lbrace V_1,V_2,U_1,\ldots, U_p\right\rbrace$ and $E_1 = \left\lbrace\left(V_1, U_1\right) , \; \left(V_2, U_1\right) \right\rbrace \; \bigcup_{i=1}^{p-1} \left(U_i,U_{i+1}\right).$
		\item[$\mathbf{T_2}$]  with $	N_2 = \left\lbrace V_1U_1, V_2U_1, U_1U_2, \ldots ,U_{p-1}U_p \right\rbrace$ and \\[0.2cm] $E_2 = \left\lbrace\left(V_1U_1,\; U_1U_2\right) , \; \left(V_2U_1,\; U_1U_2\right) \right\rbrace  \;\bigcup_{i=1}^{p-2} \left(U_iU_{i+1}, \; U_{i+1}U_{i+2}\right).$
		\item[$\mathbf{T_k}$] for $3 \leq k \leq p$ with $	N_k = \bigcup_{j=1,2} \left\lbrace V_jU_{k-1}; \mathbf{U}_{1:k-2}\right\rbrace \bigcup_{i=1}^{p-k+1} \left\lbrace U_iU_{i+k-1}; \mathbf{U}_{i+1:i+k-2}   \right\rbrace $\\[0.2cm]
		and $E_k =\bigcup_{j=1,2}\left\lbrace\left(V_jU_{k-1}; \mathbf{U}_{1:k-2} , \; \; U_1U_k; \mathbf{U}_{2:k-1} \right)\right\rbrace $\\
		\hspace*{1cm}	$\;	\; \quad \bigcup_{i=1}^{p-k} \left\lbrace\left( U_iU_{i+k-1}; \mathbf{U}_{i+1:i+k-2} ,\; \; U_{i+1}U_{i+k}; \mathbf{U}_{i+2:i+k-1}  \right)\right\rbrace .$
		\item[$ \mathbf{T_{p+1}} $] with $	N_{p+1} = \bigcup_{j=1,2} \left\lbrace V_jU_p  ;    \mathbf{U}_{1:p-1}  \right\rbrace $ and $	E_{p+1} = \left\lbrace\left(V_1U_p;  \mathbf{U}_{1:p-1} ,\; \; V_2U_p;  \mathbf{U}_{1:p-1}  \right)\right\rbrace .$
	\end{itemize}
\end{definition}  
\begin{figure}\label{y-vine}
	\centering
	\includegraphics[width= 16.5cm, height=20cm]{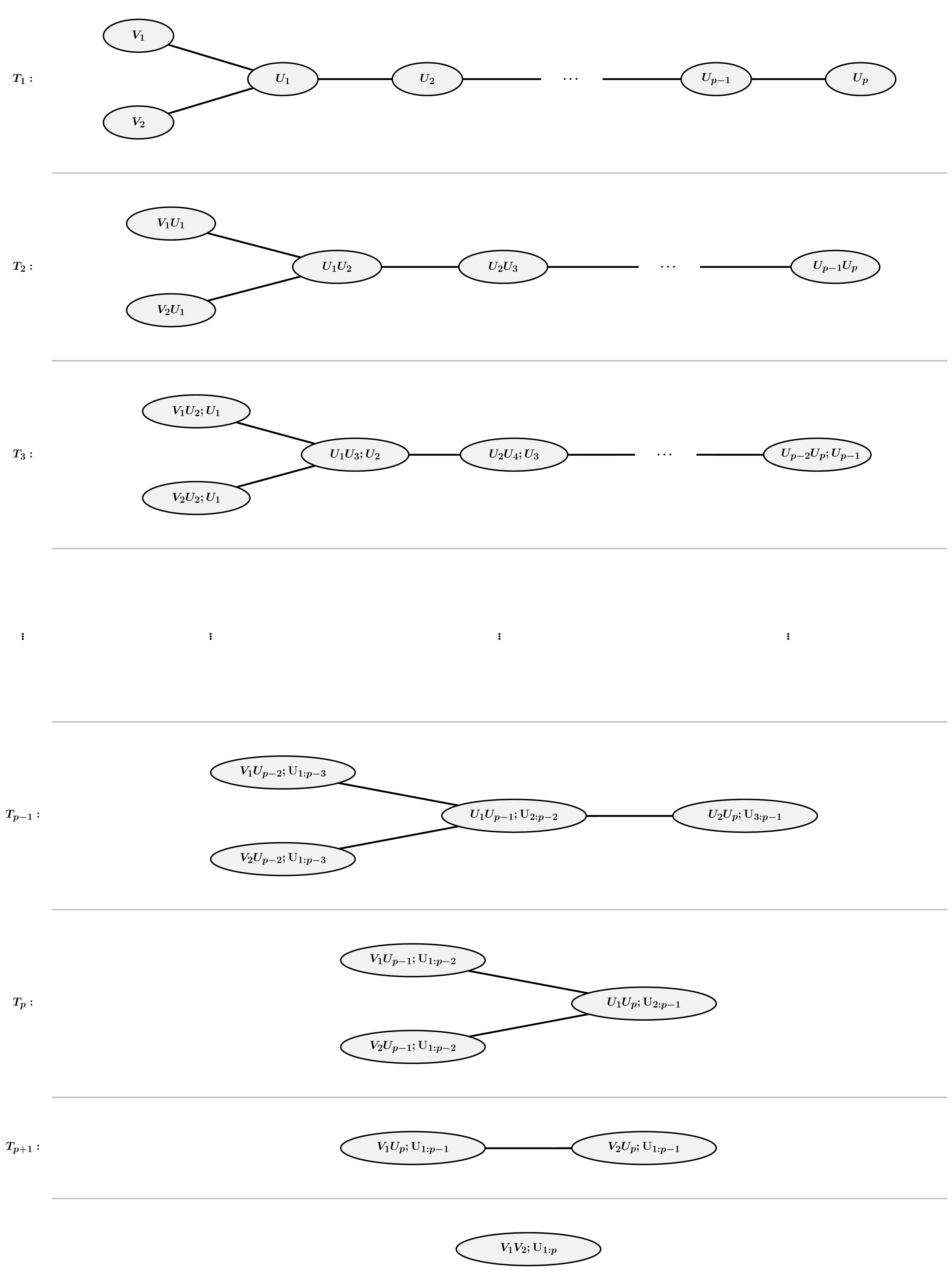}
	\caption{Y-vine tree sequence on the u-scale.}
	\label{figure:vine}
\end{figure}
\noindent The newly proposed  Y-vine tree sequence is illustrated in Figure \ref{figure:vine}. In each tree of the vine tree sequence the nodes containing  the predictor variables in the conditioned set are arranged in a path, while the nodes containing  the response variables in the conditioned set are added as leafs of the path on one end. The subset of the sequence that contains a single response and all predictors   forms a D-vine tree sequence.
This tree structure allows for symmetric treatment of the response variables,  especially important since an asymmetric treatment might lead to different performances of the regression models  depending on the order. 
The proposed  Y-vine tree sequence satisfies the regular vine tree sequence conditions $(i)$-$(iii)$ from Section \ref{sc:theory} and thus, represents a valid regular vine tree sequence. The proof is provided in the supplementary material. 

\noindent A regular vine copula associated with a $Y$-vine tree sequence together with   a set of bivariate copulas $\mathcal{B}\left(\mathcal{V}\right)$ and the corresponding pair copula parameters $\Theta\left(\mathcal{B}\left(\mathcal{V}\right)\right)$ is called a  $Y$-vine copula and we  denote it by $\mathcal{Y}$. 
The joint density $f_{Y_1,Y_2,\mathbf{X}}$ using a $Y$-vine tree sequence can be expressed by Equation\eqref{s_vine_den} as	
\begin{equation}\label{eq:jointdens}
	\begin{split}
		f_{ Y_1,Y_2,\mathbf{X}}& \left(y_1, y_2, \mathbf{x}\right)= \prod_{k=1} ^{p-1}    \left[  \prod_{i=1}^{p-k} c_{U_{i},U_{i+k};\mathbf{U}_{i+1:i+k-1}} \left(  F_{X_{i}|\mathbf{X}_{i+1:i+k-1}} (x_{i}|\mathbf{x}_{i+1:i+k-1}) , \right. \right. \\
		&  \hspace{6.6cm} \left.  F_{X_{i+k}|\mathbf{X}_{i+1:i+k-1}} (x_{i+k}|\mathbf{x}_{i+1:i+k-1})	
		\right) \Bigg] \\
		& \hspace{2cm} \cdot \prod_{i=1}^p  \left[ \prod_{j=1,2}  c_{V_j,U_i; \mathbf{U}_{1:i-1}}  \left( F_{Y_j|\mathbf{X}_{1:i-1}}  \left( y_j|\mathbf{x}_{1:i-1}  \right), \right. \right. 
		\left. F_{\mathbf{X}_i|X_{1:i-1} } \left( x_i|\mathbf{x}_{1:i-1}  \right)    \right) \Bigg]     \\  
		& \hspace{2cm} \cdot c_{V_1,V_2; \mathbf{U} } \left( F_{Y_1|\mathbf{X} } \left( y_1|\mathbf{x}  \right), F_{Y_2|\mathbf{X} } \left( y_2|\mathbf{x} \right)    \right) 
		\cdot \prod_{i=1}^{p}f_{X_{i}} (x_{i}) \cdot \prod_{j=1,2}f_{Y_{j}} (y_{j}) .
	\end{split}
\end{equation}

\newpage
\begin{Theorem}\label{thm:density}
	The joint conditional  density of  $(Y_1,Y_2)$ given the predictors $\mathbf{X}=(X_1,\ldots ,X_p)^T$ denoted by $f_{ Y_1,Y_2|\mathbf{X} }$ in a Y-vine copula  is given as
	\begin{equation}\label{eq:densitycond}
		\begin{aligned}
			f_{Y_1,Y_2\vert \mathbf{X}}\left(y_1, y_2| \mathbf{x}\right) & = 
			\prod_{i=1}^{p} \left[ \prod_{j=1,2}
			c_{V_j,U_i; \mathbf{U}_{1:i-1} } \left( F_{Y_j|\mathbf{X}_{1:i-1} } \left( y_j|\mathbf{x}_{1:i-1}  \right),  F_{X_i|\mathbf{X}_{1:i-1} } \left( x_i|\mathbf{x}_{1:i-1}  \right)    \right) \right.  \\
			& \hspace{1.4cm}   
			\cdot c_{V_1,V_2; \mathbf{U}}  \left( F_{Y_1|\mathbf{X} } \left( y_1|\mathbf{x}  \right), F_{Y_2|\mathbf{X} } \left( y_2|\mathbf{x} \right) \Bigg.   \right) \Bigg] \cdot \prod_{j=1,2} f_{Y_{j}} (y_{j}).
		\end{aligned}
	\end{equation}	
\end{Theorem}
\noindent Proof of Theorem \ref{thm:density} is given in Appendix \ref{app:proofcond}.\\
\noindent In order to determine the joint and the bivariate conditional  density,  $c_{V_1,V_2, \mathbf{U}}$ and  $c_{V_1,V_2\vert \mathbf{U}}$, we only need to set  the marginals to uniform  densities, i.e.  $f_{Y_j}(y_j)=1 \;, j = 1,2$ and $f_{X_i}(x_j) =1 \;, i = 1,\ldots ,p$  in Equation~\eqref{eq:jointdens} and Equation~\eqref{eq:densitycond} respectively. Thus, with the proposed Y-vine copula  we can express the conditional bivariate density as a product of pair copula densities occurring in the Y-vine tree sequence that contain a response in the conditioned set, and the marginal densities of the responses. No  integration is needed.
\noindent In addition to the analytic form of the joint conditional density  
$f_{Y_1,Y_2\vert \mathbf{X}} $, from the Y-vine we can also derive other conditional densities in an analytic  form. 
\begin{Corollary} \label{corollary1}
	From the Y-vine copula  associated with the Y-vine tree sequence of Definition~\ref{vinetree}, we can derive the following conditional densities:\\
		
		\noindent    a. for $j=1,2$ it holds
		\begin{equation}\label{eq:densitycond5}
			f_{Y_j \vert \mathbf{X}}\left(y_j| \mathbf{x}\right)  = f_{Y_{j}} (y_{j})
			\prod_{i=1}^{p} 
			c_{V_j,U_i; \mathbf{U}_{1:i-1} } \left( F_{Y_j|\mathbf{X}_{1:i-1} } \left( y_j|\mathbf{x}_{1:i-1}  \right),  F_{X_i|\mathbf{X}_{1:i-1} } \left( x_i|\mathbf{x}_{1:i-1}  \right)    \right) ;   	        
		\end{equation} 
		
		\noindent b. for $j,k \in \left\lbrace 1,2 \right\rbrace$ with $j\neq k$, it holds
		
		\begin{equation}\label{eq:densitycond4}
			\begin{aligned}
				f_{Y_k \vert \mathbf{X},Y_j}\left( y_k| \mathbf{x},y_j\right) = &  
				\prod_{i=1}^{p} \Big[
				c_{V_k,U_i; \mathbf{U}_{1:i-1} } \left( F_{Y_k|\mathbf{X}_{1:i-1} } \left( y_k|\mathbf{x}_{1:i-1}  \right),  F_{X_i|\mathbf{X}_{1:i-1} } \left( x_i|\mathbf{x}_{1:i-1}  \right)    \right)  \Big] \\
				& 
				\cdot c_{V_1,V_2; \mathbf{U}_{1:p} } \left( F_{Y_1|\mathbf{X}_{1 :p} } \left( y_1|\mathbf{x}  \right), F_{Y_2|\mathbf{X} } \left( y_2|\mathbf{x} \right)    \right) \cdot 	f_{Y_{k}} (y_{k}) .
			\end{aligned}
		\end{equation}	
\end{Corollary}
\noindent Proof of Corollary \ref{corollary1} is given in Appendix \ref{app:proofcorr}.  For the  associated  univariate conditional  densities $c_{V_1 \vert \mathbf{U}}\left( v_1| \mathbf{u}\right),$  $c_{V_2 \vert \mathbf{U}}\left( v_2| \mathbf{u}\right),$ and 
$c_{V_1 \vert \mathbf{U},V_2}\left( v_1| \mathbf{u},v_2\right),$ $c_{V_2 \vert \mathbf{U},V_1}\left( v_2| \mathbf{u},v_1\right),$ 
we set $f_{Y_j}(y_j)= 1,\; j = 1,2$, 
in Equation~\eqref{eq:densitycond5} and Equation~\eqref{eq:densitycond4} respectively.
The univariate conditional distribution functions $C_{V_1 \vert \mathbf{U}},$ $C_{V_2 \vert \mathbf{U}},$ and 
$C_{V_1 \vert \mathbf{U},V_2},$  $C_{V_2 \vert \mathbf{U},V_1}$ can be obtained through integration of these associated conditional densities. 
The bivariate conditional distribution function $C_{V_1,V_2\vert \mathbf{U}_{1:p}}$ is  as:
\begin{equation}\label{cond_int}
	\begin{aligned}
		C_{V_1,V_2\vert \mathbf{U}_{1:p}}\left(v_1,v_2\vert \mathbf{u}_{1:p}\right) & = \int_{0}^{v_1} \int_{0}^{v_2} c_{V_1,V_2\vert \mathbf{U}_{1:p}}\left(v_1',v_2'\vert \mathbf{u}_{1:p}\right) dv_2' dv_1' \\
		& = \int_{0}^{v_1} \int_{0}^{v_2} c_{V_2\vert \mathbf{U}_{1:p}}\left(v_2'\vert \mathbf{u}_{1:p}\right) \cdot c_{V_1\vert V_2, \mathbf{U}_{1:p}}\left(v_1'\vert v_2', \mathbf{u}_{1:p}\right) dv_2' dv_1' \\
		& = \int_{0}^{v_2} c_{V_2\vert \mathbf{U}_{1:p}}\left(v_2'\vert \mathbf{u}_{1:p}\right) \cdot \left[ \int_{0}^{v_1} c_{V_1\vert V_2, \mathbf{U}_{1:p}}\left(v_1'\vert v_2', \mathbf{u}_{1:p}\right) dv_1' \right] dv_2' \\ 
		& = \int_{0}^{v_2} c_{V_2, \mathbf{U}_{1:p}}\left(v_2', \mathbf{u}_{1:p}\right) \cdot C_{V_1\vert V_2, \mathbf{U}_{1:p}}\left(v_1\vert v_2', \mathbf{u}_{1:p}\right) dv_2'.
	\end{aligned}
\end{equation}
One can also condition on $V_1$ instead of $V_2$ in Equation~\eqref{cond_int}.

\subsection{Sequential forward selection of predictors }

Until now, we ordered the predictors as $X_1$ to $X_p$, however other permutations are possible.
Let's denote the associated permutation of the  Y-vine $\mathcal{Y}$ from Figure \ref{figure:vine} by $\mathcal{O} (\mathcal{Y}) \coloneqq (1, 2, \ldots, p-1, p)$. It is the order  in which the predictors  appear in  $T_1$ of the tree sequence.
One can choose the order of the predictors randomly, but the predictive power of the fit greatly depends on the chosen order. Different orders will produce different  Y-vine fits, as the influence over the two responses varies with the predictors. There are $p$ possible permutations of this order,  computing and comparing each of them is not feasible and the optimal permutation is in general unknown. Thus, we propose an algorithm that automatically constructs a Y-vine by sequentially ordering predictors. In addition, we apply a stopping criteria to prevent overfitting, meaning that the least influential predictors will not be considered in the model. This way we obtain an automatic forward selection of predictors for the  bivariate regression model. Similar ordering approaches are introduced in  \cite{kraus2017d} for univariate D-vine  regression and in \cite{tepegjozova2021nonparametric} for C-vine and D-vine copulas with an additional step to check for possible future improvement. In addition,  our framework does not depend on the selection method of the order of  predictors. 
Other approaches can be used, for example,  the D-vine selection with an additional step to check for improvements from \cite{tepegjozova2021nonparametric}, approaches based on the feature ordering by conditional  independence testing by \cite{azadkia2021simple}  or   background knowledge specifying a predefined order, or different fit measures and selection criteria. 
\subsubsection{Joint conditional log-likelihood}
The goal is to find the order of the predictors that has the greatest explanatory power. To compare and quantify the explanatory power of different bivariate regression models we propose a log-likelihood approach. Inspired by the one dimensional vine based  regression \citep{kraus2017d}, we would  like to associate the fit measure with the target function of the bivariate vine based  regression. A suitable choice is the  log-likelihood of $c_{V_1,V_2\vert \mathbf{U}_{1:p}},$ since $c_{V_1,V_2\vert \mathbf{U}_{1:p}}$ is the corresponding density of the target function. However, before deciding on the fit measure we take a more precise look at the proposed log-likelihood.
Following  \cite{killiches2018model}, the conditional copula density $c_{V_j\vert \mathbf{U}_{1:p}}$ can be rewritten as a product of all pair-copulas that contain the response $V_j$ in  a D-vine copula.
In the bivariate response case using Y-vines, we can express $c_{V_1,V_2\vert \mathbf{U}_{1:p}}$ as a product of all pair-copulas that contain the responses $V_1$ and $V_2$, as shown in  Equation~\eqref{eq:densitycond} by setting the marginals to uniform densities.
Thus, the  log-likelihood of $c_{V_1,V_2\vert \mathbf{U}_{1:p}}$ associated with a Y-vine,
can be written  as
\begin{equation*}
	\begin{split}
		\ell\left( c_{V_1,V_2\vert \mathbf{U}_{1:p}}\right) = & \; \; \ell \left(c_{V_1,V_2; \mathbf{U}_{1:p}}\right) + \ell\left(c_{V_1\vert  \mathbf{U}_{1:p}}\right) + \ell\left(c_{V_2\vert  \mathbf{U}_{1:p}}\right) \\
		= &\;  \ell\left(c_{V_1,V_2; \mathbf{U}_{1:p}}\right) + \sum_{j=1,2}\left[\ell\left(c_{V_j,U_1}\right)  + \sum_{k=2}^{p} \ell\left(c_{V_j,U_k;\mathbf{U}_{1:k-1}}\right) \right],
	\end{split}
\end{equation*}
where $\ell(f)$ denotes the log-likelihood associated to a statistical model with density  $f$ and a given independent and identically distributed sample. Here we used the predictor order as given in Figure~\ref{figure:vine}. 
The pair-copula density $c_{V_j,U_k;\mathbf{U}_{1:k-1}}$ represents the behaviour between $U_k$ and $V_j$ given that the effects of the conditioning values $U_1,\ldots,  U_{k-1}$ are adjusted. Therefore, a large value of the log-likelihood  $\ell\left(c_{V_j,U_k;\mathbf{U}_{1:k-1}}\right)$
indicates an influence of $U_k$ on the response $V_j$. This implies that  the log-likelihoods  associated with the  pair copulas $c_{V_j,U_k;\mathbf{U}_{1:k-1}}$ are suitable for a fit measure since we can interpret an increase in the fit measure as an increase in influence from a certain predictor.  But what importance does the  copula between the responses given the predictors $c_{V_1,V_2;\mathbf{U}_{1:k-1}}$ have on the predictive power of the model is a valid question for $k=2,\ldots, p$.
The term  $c_{V_1,V_2; \mathbf{U}_{1:k}}$ represents the behaviour between $V_1$ and $V_2$ given that the effects of $U_1,\ldots,  U_{k}$ are adjusted. This implies that neither an increase nor a decrease in the log-likelihood  $\ell \left(c_{V_1,V_2; \mathbf{U}_{1:k}} \right)$ can be interpreted as an increase in influence for a single predictor.
Thus, $c_{V_1,V_2; \mathbf{U}_{1:k}}$ for $k=2, \ldots, p$ fails to quantify the marginal effect of any predictor on the responses and we exclude it from our proposed fit measure. Finally, we formally introduce the  \textit{adjusted conditional log-likelihood} as our fit measure.

\begin{definition}\label{cll}
	The  adjusted conditional log-likelihood  of a bivariate Y-vine based  regression model, denoted by $ac\ell\ell$, with PIT transformed response and predictor variables $V_1,V_2,U_1,\ldots ,U_p$, 
	is defined as 
	\begin{equation}\label{eq:ccloglik}
		\begin{split}
			acll\left(\mathcal{Y} \right) & \coloneqq	\ell \left( c_{V_1,V_2\vert \mathbf{U}_{1:p}}\right) -  \ell\left(c_{V_1,V_2; \mathbf{U}_{1:p} }\right)  		\\
			& =  \sum_{j=1,2} \left[ \ell \left( c_{V_j,U_1} \right) + \sum_{k=2}^{p} \ell \left( c_{V_j,U_k;\mathbf{U}_{1:k-1}} \right) \right] . 
		\end{split}
	\end{equation}
\end{definition}
Since we are interested in forward selection of  predictors, we need to  easily  compare nested models with one predictor difference. Let $\mathcal{Y}_{p-1}$ and $\mathcal{Y}_p$ be two nested Y-vine based  regression models with response variables $V_1,V_2$, where $\mathcal{Y}_{p-1}$ includes the predictors $U_1,\ldots, U_{p-1}$ in that order and $\mathcal{Y}_{p}$ includes the predictors $U_1,\ldots, U_{p-1}, U_p$. Then the connection between the  adjusted conditional log-likelihoods of those nested models is given as 
\begin{equation}\label{cllnested}
	acll\left(\mathcal{Y}_p\right) = \ell\left(c_{V_1,U_p;\mathbf{U}_{1:p-1}}\right) + \ell\left(c_{V_2,U_p;\mathbf{U}_{1:p-1}}\right) + acll\left(\mathcal{Y}_{p-1}\right),
\end{equation}
and we  use this result for forward selection of predictors.

\subsubsection{Automatic forward selection algorithm}
\noindent Assume we start with the PIT transformed response and predictors  $V_1,V_2,U_1,\ldots ,U_p$, and their observations $\mathbf{v}_n=\left(v_1^n,v_2^n\right)^T ,\; \mathbf{u}_n=\left(u_1^n,\ldots ,u_p^n\right)^T$, for $n=1,\ldots ,N$. We would like to fit a Y-vine copula model to the data, given that $V_1,V_2$ are the responses. First, we build a Y-vine copula model with one predictor only. To see which predictor needs to be on the first place in the order, we fit all possible one-predictor Y-vines. We derive their adjusted conditional log-likelihoods using Equation~\eqref{eq:ccloglik}, and the predictor that maximizes it, say $U_{r_{1}}$ becomes the first predictor in the order of the Y-vine model. Let's denote the fitted Y-vine model with one predictor as $\hat{\mathcal{Y}}_1$ with order $\mathcal{O} (\hat{\mathcal{Y}}_1)= (r_{1})$. In the next step, we need to choose the second predictor to be added to the model. To do so, we  fit the additional pair-copulas that need to be estimated for  the adjusted conditional log-likelihood. Following Equation~\eqref{cllnested}, we need to estimate two more copulas for each of the remaining predictors, derive the adjusted conditional log-likelihoods and the predictor that maximizes it, say $r_{2}$ becomes the second predictor in the order. Thus, at the end of the second step we have a fitted  Y-vine model with two predictors denoted as $\hat{\mathcal{Y}}_2$ with order $\mathcal{O} (\hat{\mathcal{Y}}_2)= ( r_{1},r_{2}$). We continue this forward selection algorithm until we order all predictors or if none of
the remaining predictors is able to increase the   conditional log-likelihood of the model, similar as in \citep{kraus2017d}.
The full estimation procedure and the pseudo code for the algorithm is given in Appendix \ref{app:code}.

\section{Prediction for bivariate  regression}\label{sc:prediction}

Assume we have fitted a bivariate Y-vine regression model $\hat{\mathcal{Y}}$ on a bivariate response vector $(V_1,V_2)^T$ with order of predictors  $\mathcal{O} (\hat{{\mathcal{Y}}})= (1,\ldots ,p)$. The fitted vine has a tree sequence and pair-copula family sets denoted by $\hat{\mathcal{V}}$ and $\hat{\mathcal{B}}(\hat{\mathcal{V}})$, respectively.  
Given a new realization  $\mathbf{u}^{new}=(u^{new}_1,\ldots ,u^{new}_p)^T,$ our target is to obtain the set of points
$	Q_{\alpha}^V\left(\mathbf{u}^{new}\right) = \left\lbrace \left(v_1,v_2\right)\in \left[0,1\right]^2 \; ;\; C_{V_1,V_2| \mathbf{U}} (v_1, v_2| \mathbf{u}^{new}) = \alpha \right\rbrace .$
To estimate the set $Q_{\alpha}^V\left(\mathbf{u}^{new}\right)$ we employ the same numerical  procedure as explained in Section \ref{pseudoinvrs}. 
In addition, we need to be able to evaluate the function $C_{V_1,V_2\vert  \mathbf{U}}$ at every integration point $\mathbf{v}^{inp} = (v_1^{inp},v_2^{inp})^T \in \left[0,1 \right]^2$ and  determine the integral given in Equation ~\eqref{cond_int}. We apply the chosen adaptive quadrature algorithm for integration (see more in \cite{piessens2012quadpack}), which requires the ability to evaluate the function under the integral at all points of the integration interval.
Therefore, given a point $\mathbf{v}^{inp} = (v_1^{inp},v_2^{inp})^T$ we define the integrand, denoted by $IN\left(z\right)$ for any $0<z<v_2^{inp} $, as
\begin{equation}\label{integrand_prod}
	IN(z) \coloneqq c_{V_2\vert \mathbf{U}}\left(z\vert \mathbf{u}^{new} \right) \cdot C_{V_1 \vert V_2, \mathbf{U}}\left(v_1^{inp}\vert z, \mathbf{u}^{new} \right).
\end{equation}
The integration is carried out over the interval $(0,v_2^{inp})$. While the first term in Equation~\eqref{integrand_prod} is available analytically since it is the conditional density associated with the D-vine $V_2-U_1-\ldots -U_p$, the second term needs further consideration. For this  we define the pseudo copula data for $\mathbf{u}^{new}$ as the following\\ $	u_{i\vert i-1}^{new} = h_{U_i\vert U_{i-1}} \left(u^{new}_i\vert u^{new}_{i-1}\right), \quad u_{i-1\vert i}^{new} = h_{U_{i-1}\vert U_{i}} \left(u^{new}_{i-1}\vert u^{new}_{i}\right) \quad \forall i =2, ,\ldots,  p,$
where the $h$-functions $h_{U_i\vert U_{i-1}}$ and $h_{U_{i-1}\vert U_{i}}$ are obtained from the  pair copula $c_{U_i,U_{i-1}} \in  \hat{\mathcal{B}}(\hat{\mathcal{V}})$. For any $k = 2, \ldots, p-1$ it holds $u_{i\vert i-k:i-1}^{new} = h_{U_i\vert U_{i-k}; \mathbf{U}_{i-k+1:i-1}} \left(u_{i \vert i-k+1 :  i-1}^{new}\vert u_{i-k \vert i-k+1 :  i-1}^{new}\right), $ and  similarly $u_{i-k\vert i-k+1:i}^{new} = h_{U_{i-k}\vert U_{i}; \mathbf{U}_{i-k+1:i-1}} \\ \left(u_{i-k \vert i-k+1 :  i-1}^{new} \vert u_{i \vert i-k+1 : i-1}^{new}\right) \; \forall i = 2, ,\ldots,  p, $
where the $h$-functions $ h_{U_i\vert U_{i-k}; \mathbf{U}_{i-k+1:i-1}} $ and $ h_{U_{i-k}\vert U_{i}; \mathbf{U}_{i-k+1:i-1}}$ are  determined from the  pair copula $c_{U_i,U_{i-k}; \mathbf{U}_{i-k+1:i-1}} \in \hat{\mathcal{B}}(\hat{\mathcal{V}})$. In addition,  based on this pseudo-copula data  estimated from the fitted Y-vine we introduce the following two matrices, $W \in [0,1]^{p \times p}$ and $W' \in [0,1]^{p \times p}$, as
\begin{equation*}
	W \left(\mathbf{u}^{new}; \hat{\mathcal{B}}(\hat{\mathcal{V}})\right) \coloneqq
	\begin{pmatrix}
		u^{new}_1 & u^{new}_2 & u^{new}_3 & \ldots & u^{new}_{p-1} & u^{new}_p \\
		u_{2\vert 1}^{new} & u_{3\vert 2}^{new} & u_{4\vert 3}^{new} & \ldots & u_{p\vert p-1}^{new} & \\
		\vdots & \vdots & \vdots & \ldots & & \\
		u_{p-1\vert 1: p-2}^{new} & u_{p-2\vert 2: p-3}^{new} &  &  & &  \\
		u_{p\vert 1:  p-1}^{new} &  &  &  & &  
	\end{pmatrix} 
\end{equation*}
\begin{equation*}
	W' \left(\mathbf{u}^{new}; \hat{\mathcal{B}}(\hat{\mathcal{V}}) \right) \coloneqq \begin{pmatrix}
		u^{new}_1 & u^{new}_2 & u^{new}_3 & \ldots & u^{new}_{p-1} & u^{new}_p \\
		u_{1\vert 2}^{new} & u_{2\vert 3}^{new} & u_{3\vert 4}^{new} & \ldots & u_{p-1\vert p}^{new} & \\
		\vdots & \vdots & \vdots & \ldots & & \\
		u_{1\vert 2: p-1}^{new} & u_{2\vert 3: p-2}^{new} &  &  & &  \\
		u_{1\vert 2:  p}^{new} &  &  &  & &  
	\end{pmatrix}.
\end{equation*}

\noindent Using matrices $	W$ and $W'$ , we define the following pseudo copula data for $j=1,2,$
$u_{v_j\vert 1}= h_{V_j\vert U_1} \left(w\vert u^{new}_1\right)$ and $u_{1\vert v_j}=h_{U_1\vert V_j} \left( u^{new}_1\vert w\right)$, where $h_{V_j\vert U_1}$ and $h_{U_1\vert V_j}$ are estimated from the pair copula $c_{V_j,U_1} \in \hat{\mathcal{B}} $. Further, for $i=2,\ldots, p$, define $u_{v_j\vert 1: i}=	h_{V_j\vert U_i; \mathbf{U}_{1:i-1}} \left(u_{v_j\vert 1:  i-1}^{new}\vert u_{i\vert1:  i-1 }^{new}\right)$ and $u_{i\vert v_j1: i-1}=h_{U_i\vert V_j;\mathbf{U}_{1:i-1}} \\ \left( u_{i\vert 1:  i-1}^{new}\vert u_{v_j\vert 1:  i-1}^{new}\right)$. These h-function are estimated from the pair copula $c_{V_j,U_i ;\mathbf{U}_{1:i-1}}\in \hat{\mathcal{B}}$.  Then, we also define the matrix  $W^2 \in [0,1]^{(p+1) \times 2}$ with $j\in \left\lbrace 1,2 \right\rbrace $ as
\begin{equation*}
	\begin{split}
		&	W^2 \left(w, j; W, W'\right)   \coloneqq
		\begin{pmatrix}
			w & w \\ 
			u_{v_j\vert 1} & u_{1\vert v_j} \\
			u_{v_j\vert 12} & u_{2\vert v_j1} \\
			u_{v_j\vert 1:3} & u_{3\vert v_i12} \\
			\vdots & \vdots \\
			u_{v_j\vert 1: p} & u_{p\vert v_j1: p-1} \\ 
		\end{pmatrix}.  
	\end{split}
\end{equation*}
For a fixed input $v_1^{inp}$ we can evaluate 
$	C_{V_1 \vert V_2, \mathbf{U}}\left(v_1^{inp}\vert z^{new}, \mathbf{u}^{new} \right) = h_{V_1\vert V_2 ;\mathbf{U}} \left(u_{v_1\vert 1:  p}\vert u_{v_2\vert 1:  p}\right),$ at $z=z^{new}$, such that 
$u_{v_1\vert 1: p}$ is obtained from $W^2 \left(w = v_1^{inp}, j=1; W, W'\right)$ and $u_{v_2\vert 1:  p}$ is obtained from  \\$W^2 \left(w = z^{new}, j=2; W, W'\right)$. The $h$-function $h_{V_1\vert V_2 ;\mathbf{U}}$ is estimated from the pair copula $c_{V_1,V_2;\mathbf{U}} \in \hat{\mathcal{B}}(\hat{\mathcal{V}})$. 
$c_{V_2\vert \mathbf{U}} $ is evaluated as
\begin{equation*}
	c_{V_2\vert \mathbf{U}}\left(z^{new}\vert \mathbf{u}^{new} \right)  = \frac{c_{V_2,\mathbf{U}}}{c_{\mathbf{U}}} 
	= c_{V_2,U_1}\left(z^{new}, u_1^{new}\right)\prod_{i=2}^{p} c_{V_2,U_i;\mathbf{U}_{1:i-1}} \left(u_{v_2\vert 1: i-1}, u_{i:  i-1}^{new}\right),
\end{equation*}
where $c_{V_2,U_1}, c_{V_2,U_i;\mathbf{U}_{1:i-1}} \in \hat{\mathcal{B}}(\hat{\mathcal{V}})$ for $i=1\ldots,p.$
Therefore, the integrand in Equation \eqref{integrand_prod} can be evaluated  with no further calculations  from the Y-vine as 
\begin{equation}\label{final_integrand}
	\begin{aligned}
		IN\left(z^{new}\right) = & c_{V_2,U_1}\left(z^{new}, u_1^{new}\right)  \prod_{i=2}^{p} c_{V_2,U_i;\mathbf{U}_{1:i-1}} \left(u_{v_2\vert 1:  i-1}, u_{i\vert 1: i-1}^{new}\right)  \\
		&\cdot h_{V_1\vert V_2 ;\mathbf{U}} \left(u_{v_1\vert 1:  p}\vert u_{v_2\vert 1:  p}\right).
	\end{aligned} 
\end{equation}
To summarize, given the integration point $\mathbf{v}^{inp} = (v_1^{inp},v_2^{inp})^T$, the integrand $IN\left(z^{new}\right)$ at a point $z^{new} \in (0,v_2^{inp} )$ conditioned on $\mathbf{u}^{new}$, can be computed using  the matrices $W,W',$ $W^2 \left(w = z^{new}, j=2; W, W'\right)$, $W^2 \left(w = v_1^{inp}, j=1; W,W'\right)$ and h-functions obtained from the pair copulas defined by $\hat{\mathcal{B}}(\hat{\mathcal{V}})$. This implies that we can efficiently evaluate the function $C_{V_1,V_2\vert  \mathbf{U}}$ using Equation \eqref{final_integrand}. 

\subsection{ Simulation of bivariate data in a Y-vine copula regression}\label{simulations}
Simulation of multivariate vine copula data is possible for general R-vine copulas (see details in \cite[Ch.6]{czado2019analyzing} and \citet[Chapter 5]{DA-JeffreyDissann} ).  It is based on the  multivariate transformation introduced by \cite{rosenblatt1952remarks}. Here we are interested in simulating  $\left(  v_1(\mathbf{u}), v_2(\mathbf{u}) \right)$ from $C_{V_1, V_2| \mathbf{U}}\left(\cdot, \cdot| \mathbf{u}\right)$ for a fixed value $\mathbf{u}$. For this, start by getting a sample $v_1(\mathbf{u})$, by setting $v_1(\mathbf{u}) = C_{V_1| \mathbf{U}}^{-1} (a_1| \mathbf{u}) $ for a value $a_1$ sampled from a uniform distribution on $[0,1]$. Then, we set $v_2(\mathbf{u}) = C^{-1}_{V_2 \vert V_1, \mathbf{U}}\left(a_2\vert v_1, \mathbf{u}\right) $ for a uniform $[0,1]$ sampled value $a_2$. This allows us to get the desired sample  $\left(  v_1(\mathbf{u}), v_2(\mathbf{u}) \right)$ from $C_{V_1, V_2| \mathbf{U}}\left(\cdot, \cdot| \mathbf{u}\right)$  in a step wise fashion.

\section{Bivariate  level and quantile curves}	\label{BQR:bivariatequantiles}
Our proposed definition of multivariate quantiles is linked to multivariate level curves, so we start by defining and exploring the level curves of bivariate unconditional and conditional distribution functions. 

\subsection{Bivariate unconditional level curves}\label{BQR:bivariateuncondquant}

Let $Y_1$ and $Y_2$ be two continuous random variables with observed values $y_1, y_2$ and a joint distribution function $ F_{Y_1,Y_2} (y_1, y_2) $.
\begin{definition}\label{def:biquant}
	The bivariate  level curve for continuous random variables $Y_1,Y_2$  at level  $\alpha \in \left( 0,1\right)$  is a curve in $\mathbb{R}^2$ defined by the set
	\begin{equation*}
		\begin{aligned}
			Q_\alpha^{Y} \coloneqq &  \{ (y_1, y_2) \in \mathbb{R}^2 \; ; \; F_{Y_1,Y_2} (y_1, y_2)  = \alpha \} \\
			= & \{ (y_1, y_2) \in \mathbb{R}^2 \; ; \; \mathbb{P}_{Y_1,Y_2} (Y_1 \leq y_1, Y_2 \leq y_2)  = \alpha \} . 
		\end{aligned}
	\end{equation*}
\end{definition} 
\noindent We require that the joint distribution function is strictly  monotonically increasing in order to have unique solutions of $ F_{Y_1,Y_2} (y_1, y_2)  = \alpha $. Without this assumption the quantile sets exist, but are not curves, as there might be multiple solutions or plateaus in the distributions. 
\noindent Define  the probability integral transforms of the random variable $Y_j$ as $V_j \coloneqq F_{Y_j}\left(Y_j\right)$, with  corresponding observed values $v_j \coloneqq F_{Y_j}\left(y_j\right)$ for $j=1,2$. Applying Sklar's Theorem (Equation~\eqref{sklar_dist})  to the joint distribution function of $Y_1,Y_2,$ we obtain $  F_{Y_1,Y_2} (y_1, y_2)  = C( F_{Y_1} (y_1), F_{Y_2} ( y_2)  )    = C_{V_1,V_2} (v_1, v_2)$. So we can   rewrite the bivariate level curves from Definition~\ref{def:biquant} in terms of copulas as, 
$Q_\alpha^{Y}  	= \{( F^{-1}_{Y_1}(v_1), F^{-1}_{Y_2}(v_2) ) \in  \mathbb{R}^2  \; ;  \; C_{V_1,V_2} (v_1, v_2)  = \alpha, \; \; v_1, v_2 \in (0,1) \}. 
$
We can also define the bivariate level curves of the probability integral transformed  variables on the unit square $[0,1]^2$. 
The bivariate  level curves at $\alpha \in \left( 0,1\right)$ for the  continuous random variables $Y_1,Y_2$ with random PITs  $V_1, V_2$ is a  curve in $[0,1]^2$ defined by the set
\begin{equation}
	\begin{aligned}
		\label{eq:biunthequant}
		Q_\alpha^{V}  \coloneqq & \{ (v_1, v_2) \in [0,1]^2  \; ; \; C_{V_1,V_2} (v_1, v_2)  = \alpha \} \\
		= &  \{ (v_1, v_2) \in [0,1]^2  \; ; \; \mathbb{P} (V_1\leq v_1,V_2 \leq v_2)  = \alpha \}.
	\end{aligned}
\end{equation}
The difference between $Q_\alpha^{Y}$ and $Q_\alpha^{V}$ is that  $Q_\alpha^{Y}$ is defined on  $\mathbb{R}^2 $, while  $Q_\alpha^{V}$ is defined on  $[0,1]^2$. They are connected by  $Q_\alpha^{Y}  	= \{( F^{-1}_{Y_1}(v_1), F^{-1}_{Y_2}(v_2) ) \in  \mathbb{R}^2  \; ; \; (v_1, v_2) \in  Q_\alpha^{V} \}$. Sklar's Theorem implies that a transformation of the bivariate level curves between the x- and u-scale is obtained using inverses of the univariate marginal distributions $ F^{-1}_{Y_1}, F^{-1}_{Y_2}, $ rather that the bivariate joint distribution $F_{Y_1,Y_2}$.

\subsection{Bivariate conditional level curves}\label{subsec:condquan}

\begin{definition}\label{eq:biquant}
	The bivariate conditional level curves for a continuous bivariate vector $\mathbf{Y} = \left( Y_1,Y_2\right)^T$ given the outcome of  a p-dimensional random vector  $(p\geq 1)$,  $\mathbf{X} = \mathbf{x}$ at level $\alpha \in \left( 0,1\right)$ is a  curve in $\mathbb{R}^2$ defined by the set
	\begin{equation*}
		\begin{aligned}
			Q_\alpha^{Y} \left(   \mathbf{x} \right) 
			\coloneqq & \{ (y_1, y_2) \in \mathbb{R}^2 \; ; \; F_{Y_1,Y_2 |\mathbf{X}}(y_1, y_2| \mathbf{x} )  = \alpha \}  \\
			= & \{ (y_1, y_2) \in \mathbb{R}^2 \; ; \; \mathbb{P}_{Y_1,Y_2 |\mathbf{X}}(Y_1 \leq y_1, Y_2 \leq y_2| \mathbf{X} = \mathbf{x} )  = \alpha \}.
		\end{aligned}
	\end{equation*}
\end{definition}
\noindent In order to derive the level curves in terms of copulas, we need to express the conditional distribution of $Y_1,Y_2|\mathbf{X}$ in terms of a copula distribution function. For this we use the results from  Proposition \ref{prop:cond}. 
Thus, the bivariate level curve $	Q_\alpha^{Y} \left(   \mathbf{x} \right) $ can be rewritten as 
$	Q_\alpha^{Y} \left(   \mathbf{x} \right)  =  \{( F^{-1}_{Y_1}(v_1), F^{-1}_{Y_2}(v_2) ) \in  \mathbb{R}^2  \; ; \; C_{V_1,V_2| \mathbf{U}} (v_1, v_2| \mathbf{u})  = \alpha, \; \; v_1, v_2 \in (0,1) \}, $ where $\mathbf{u} = \left(u_1,\ldots ,u_p\right)^T$ are realizations of the random vector $\mathbf{U} = \left(U_1,\ldots ,U_p\right)^T$.
\noindent Similarly, we define the bivariate conditional  level curves of the probability integral transformed  variables on the unit square $[0,1]^2$.
The bivariate conditional  level curves at $\alpha \in \left( 0,1\right)$ for the  continuous random variables $Y_1,Y_2$ with random PITs $V_1, V_2$  given the outcome of the random vector $\mathbf{X} = \mathbf{x},$ with PITs   $\mathbf{U} =\mathbf{u}$ is a  curve in $[0,1]^2$ defined by the set
\begin{equation}
	\begin{aligned}
		\label{eq:biquant2}
		Q_\alpha^{V}(\mathbf{u})  \coloneqq  & \{ (v_1, v_2) \in [0,1]^2  \; ; \; C_{V_1,V_2|\mathbf{U}} (v_1, v_2| \mathbf{u})  = \alpha \} \\
		= & \{ (v_1, v_2) \in [0,1]^2  \; ; \; \mathbb{P}_{V_1,V_2|\mathbf{U}} (V_1 \leq v_1, V_2\leq  v_2| \mathbf{U} = \mathbf{u})  = \alpha \} . 
	\end{aligned}
\end{equation}

\subsection{Numerical evaluation of bivariate level curves}\label{pseudoinvrs}

\subsubsection*{Algorithms}
\begin{figure}%
	\centering
	\subfloat[\centering]{{\includegraphics[width=0.4\textwidth]{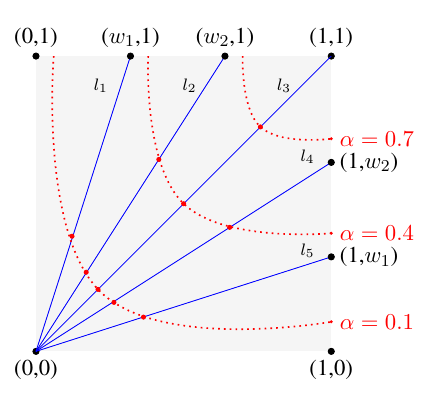} }}%
	\qquad
	\subfloat[\centering]{{\includegraphics[width=0.4\textwidth]{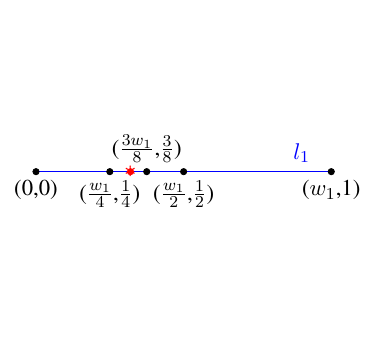} }}%
	\caption{Graphical representation of the numerical estimation procedure.	}%
	\label{fig:grid}%
\end{figure}

Let $C\left(a,b\right)$ be a bivariate (conditional) distribution defined on the unit square $\left[0,1\right]^2$ with no closed form solution for the bivariate level curve. Assume that $C\left(a,b\right)$ can be evaluated at all points $\left(a,b\right)\in \left[0,1\right]^2$ and that the bivariate (conditional) distribution function does not have any plateau, that it is strictly monotonically increasing. The goal is to obtain a numerical estimate of the set defining the (conditional) bivariate level curves, given in Equation~\eqref{eq:biunthequant} (or Equation~\eqref{eq:biquant2} for the conditional case). 
Given a granularity parameter $m\in\mathbb{N}^+$ and $\alpha \in \left(0,1\right)$
we employ the following procedure:
\begin{itemize}
	\item[1.] The set $M=\left\lbrace w_1,\ldots ,w_m\right\rbrace$ is initialized as $m$ equidistant points in the interval $\left[0,1\right]$. 
	\item[2.] We define the set of lines $L$ as follows:
	\begin{equation*}
		L= \left\lbrace \left(\left(0,0\right), \left(w_i,1\right)\right) \vert \forall w_i \in M  \right\rbrace \cup \left\lbrace \left(\left(0,0\right), \left(1,w_i\right)\right) \vert\forall w_i \in M  \right\rbrace.
	\end{equation*}
	\item[3.] Each line $l_s\in L$ is treated as a separate optimization problem and a line search procedure is employed to obtain the point $\left(a_s, b_s\right) \in l_s$ for which
	$		C\left(a_s, b_s\right) = \alpha.$	
	Consider any line $l_s$ and  two points on the line, denoted as  $(a_1, a_2)$ and $(b_1, b_2)$, such that $a_1 \leq b_1$ and $a_2 \leq b_2$. Then,  $P(V_1 \leq a_1, V_2 \leq a_2) \leq P(V_1 \leq b_1, V_2 \leq b_2) $ holds. This follows since $C$ is a bivariate distribution function and is continuous. 
	Also, we assume that the values of $C(\cdot, \cdot),$ along any line $l_s$ starting from  $(0,0)$, are increasing. 
	\item[4.] For each $l_s\in L$ a line search is guaranteed to converge to a solution, if $C\left(w_{s_1}, w_{s_2}\right) \geq \alpha $, where $\left(w_{s_1}, w_{s_2}\right)$ is the endpoint of line $l_s$. In the case $C\left(w_{s_1}, w_{s_2}\right) <  \alpha$ there is no solution on the line $l_s.$  (The same arguments hold for the conditional copula distribution function as well.)
	\item[5.] Finally, the remaining points $\left(a_s,b_s\right)$ for $s=1,\ldots ,2m$ for which a solution exists, are smoothed to obtain a curve representing an estimate of the (conditional) bivariate level curve for a given  $\alpha$. 
\end{itemize}
The  algorithms  used for this numerical evaluation of bivariate level curves  are given in Algorithm~\ref{LineAlg} and \ref{BinAlg} in Appendix \ref{appendix:algs}. The bivariate distribution function $C\left(a,b\right)$ is equivalent to $C_{V_1,V_2} (v_1, v_2)$ (or $C_{V_1,V_2|\mathbf{U}} (v_1, v_2| \mathbf{u}) $) if unconditional (or conditional) bivariate level curves are evaluated.
In Figure~\ref{fig:grid} we show a graphical representation of the numerical procedure for evaluating bivariate level curves. In the left panel, on the unit square $[0,1]^2$  shown are 5 exemplary lines, $l_1= ((0,0), (w_1,1) ),\; l_2=((0,0), (w_2,1) ), \;l_3=((0,0), (1,1) ),\; l_4=((0,0), (1,w_2) ),\; l_5=((0,0), (1,w_1) )  $  on which a line search is employed to find the pair $(a^*,b^*)$ such that $C(a^*,b^*)= \alpha$ holds. The dotted lines represent the solution of the line search, in our case, the bivariate level curves for $\alpha=0.1, 0.4, 0.7$. In the right  panel, we illustrate the binary line search for an exemplary line, say line  $l_1= ((0,0), (w_1,1) )$. First,
the desired function is evaluated at the middle point of the line $l_1$, at $C(\frac{w_1}{2}$,$\frac{1}{2})$. Here it holds $C(\frac{w_1}{2}$,$\frac{1}{2}) > \alpha$, so the middle point of the line $((0,0), (\frac{w_1}{2}$,$\frac{1}{2}) )$ is evaluated next, $C(\frac{w_1}{4}$,$\frac{1}{4})$. Then, it holds $C(\frac{w_1}{4}$,$\frac{1}{4}) < \alpha$, so the middle point of the line $ ((\frac{w_1}{4}$,$\frac{1}{4}), (\frac{w_1}{2}$,$\frac{1}{2}))$ is evaluated next, $C(\frac{3w_1}{8}$,$\frac{3}{8})$. Here $C(\frac{3w_1}{8}$,$\frac{3}{8}) > \alpha$, so we consider the middle point of the line $((\frac{w_1}{4}$,$\frac{1}{4}), (\frac{3w_1}{8}$,$\frac{3}{8}))$ next and iteratively continue until the algorithm converges to a solution. The red dot (star), say $(a^*, b^*)$ is the point at which $C(a^*, b^*) = \alpha$.

\subsubsection*{Illustration of bivariate level curves on the unit square }\label{illustration}
\begin{figure}[]
	\includegraphics[width = 0.9\textwidth]{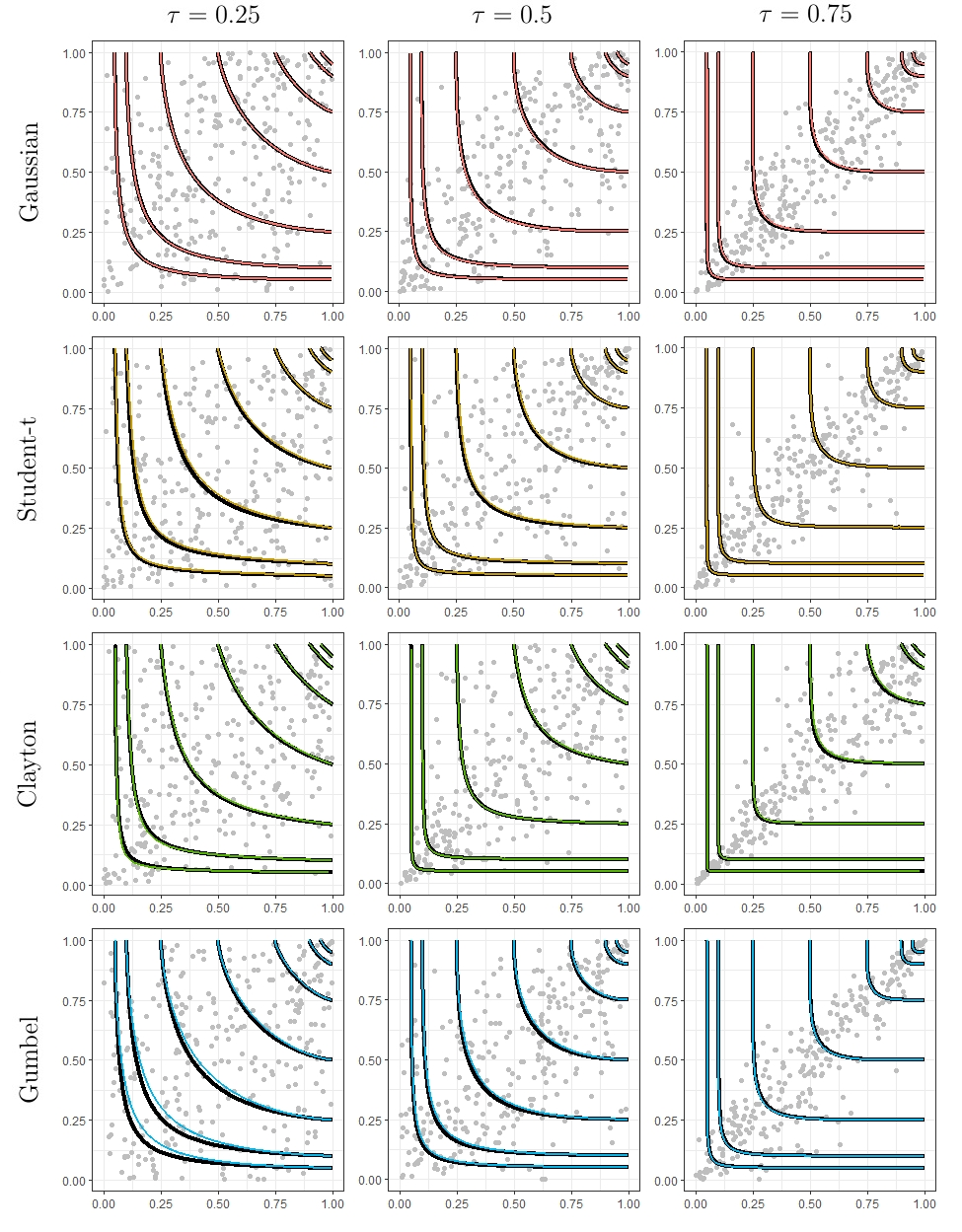}
	\centering
	\caption{x-axis: $V_1$, y-axis: $V_2$. Gray points: simulated data from copula (n=300). Black curves: theoretical quantile curves. Colored curves: estimated quantile curves. Depicted are quantile curves for $\alpha = 0.05,0.1,0.25,0.5,0.75,0.90,0.95$ (left bottom to right top in each panel) for Gaussian, Student-t ($df=5$), Clayton and Gumbel copulas (top to bottom) and $\tau = 0.25,0.5,0.75$ (left to right).}
	\label{fig:uncondquantiles}
\end{figure}

We illustrate  the  bivariate unconditional level curves for known pair copula distributions and the bivariate conditional level curves for a 3-dimensional vine structure. They  correspond to the case of no predictors or 1 predictor in a regression setting, respectively. 
In Figure~\ref{fig:uncondquantiles}  we explore plots of the unconditional level  curves on the unit square for the bivariate Gauss, Student-t, Clayton and Gumbel copulas (rows) with  different strengths of dependency, expressed through  Kendall's $ \tau$ \citep{kendall1938new}, with  $\tau = 0.25,0.5,0.75$ (columns). The level curves  can be obtained in an analogous way for any other copula family.
The theoretical level curves  of a bivariate random vector $\left(V_1,V_2\right)^T$ with bivariate distribution function $C_{V_1,V_2}\left(v_1,v_2;\theta\right)$ and a parameter $\theta$ are derived using Equation~\eqref{eq:biunthequant}
for a given $\alpha$ and are depicted with thick black lines. 
Further, we  estimate the bivariate level curves for the given pair copulas.  For this we simulate  data from the given copula and  based on the simulated data, a pair copula is estimated. The gray points  are $300$ data points simulated from the given copulas.  Subsequently, level curves are evaluated and plotted.  The coloured lines represent the corresponding estimated level curves.
In the supplementary material we give a detailed description on how  the theoretical and the estimated level curves are obtained for each of the four copula families. The panels of Figure \ref{fig:uncondquantiles}  showcase  bivariate level curves at  $\alpha = 0.05,0.1,0.25,0.5,0.75,0.90,0.95$.

Differences can be  spotted between  estimated and theoretical level curves only for the Gumbel level curves, in the case when Kendall's $\tau = 0.25$. In all other cases,  differences between the theoretical and estimated level curves are not visible. When it comes to differences in the level curves for different copula families, the Clayton copula level curve has a significantly smaller area below the $\alpha = 0.05$ level curve caused by its heavy lower tail (expected realizations are closer to the lower diagonal as compared to a lighter lower tail copula) compared to the other copula families at the $\alpha = 0.05$ level curve. On the other hand, the heavy upper tail of the Gumbel copula is causing a bigger area above the $\alpha = 0.95$ level curve compared to the Clayton copula.
In contrast, the Gaussian copula has no tails at all and the Student-t copula has a symmetric tail dependence governed by a single parameter. Their area below the $\alpha = 0.05$ level curve is greater than the corresponding  area in the lower heavy-tailed Clayton copula, and the area above the $\alpha = 0.95$ level curve is smaller than the upper heavy-tailed Gumbel copula.  Considering the  $\alpha = 0.5$ level curve,  the greatest area below it has the Gumbel copula, due to it upper heavy tail, and the smallest area below the $\alpha = 0.5$ level curve has the Clayton copula, again due to the heavy lower tail. This holds for all Kendall's $\tau$ values. Also, as the dependence between the variables increases, the data is more centered around the diagonal, so the curves have sharper curvature around the diagonal.  
Further, Figure \ref{fig:3dplot} in Appendix \ref{app:3dplot}, shows the associated  bivariate distribution functions in a 3-dimensional plot in which the theoretical level curves are shown at given $\alpha$ levels. 

\begin{figure}[H]
	\centering
	\includegraphics[width=12cm, height=3cm]{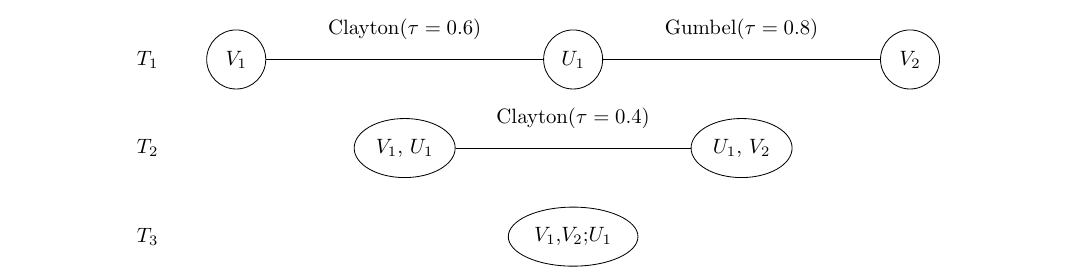}
	\caption{Vine tree sequence of $\mathcal{D}_3$  with the pair copula families and Kendall's $\tau$ corresponding to parameters.}
	\label{dvinesmall}
\end{figure}

Next we consider conditional bivariate level curves arising from a 3-dimensional regular vine distribution $\mathcal{D}_3$. Let $\left(V_1,V_2,U_1\right)^T \sim \mathcal{D}_3$ with vine tree sequence and pair copulas of $\mathcal{D}_3$ given by Figure~\ref{dvinesmall}.
\noindent The corresponding parameters of the copulas are 
$\theta_{V_1,U_1} = 3 \; (\tau= 0.6), \theta_{U_1,V_2} = 5\; (\tau= 0.8$ )and $ \theta_{V_1,V_1;U_1} = 1.33\; (\tau= 0.4)$.
To obtain theoretical level curves from $\mathcal{D}_3$ we employ the following procedure. To evaluate $C_{V_1,V_2\vert U_1}$ at a specific point $\left(\tilde{v}_1, \tilde{v}_2\right)$ conditioned on $U_1 = \tilde{u}_1$ we use Equation \eqref{cond_int}.
The corresponding conditional level curve is evaluated using the numerical evaluation procedure from Section~\ref{pseudoinvrs} and  Equation \eqref{cond_int}. We are also interested in the estimated conditional level curves. To obtain them, 
we simulate a data set $\mathbf{W} \in \left[0,1\right]^{504\times 3}$ from $\mathcal{D}_3$ and  split $\mathbf{W}$ into $\mathbf{W}_{train}\in \mathbb{R}^{500\times 3}$ and $\mathbf{W}_{test}\in \mathbb{R}^{4\times 3}$. On the training set $\mathbf{W}_{train}$ we fit a vine model $\hat{\mathcal{D}}_3$ with the same vine tree structure and order of the variables as the data generator $\mathcal{D}_3$. In 3 dimensions, a C- and a D-vine tree structure coincide, so by  order we mean the order from left to right in which the variables appear in the first tree of the sequence, as defined for a general D-vine copula.  The  estimated pair copulas are $\hat{C}_{V_1,U_1} \sim\; Clayton \left( \hat{\tau} = 0.57, \; \hat{\theta}_{V_1,U_1} = 2.65\right), $ 
$\hat{C}_{U_1,V_2} \sim\; Gumbel \left(\hat{\tau}=0.79,\; \hat{\theta}_{U_1,V_2} = 4.92\right),$ 
$\hat{C}_{V_1,V2;U_1} \sim\; Clayton \left(\hat{\tau} = 0.40,\; \hat{\theta}_{V_1,V_2;U_1} = 1.34 \right) .$

The corresponding conditional level curves of $\hat{\mathcal{D}}_3$ are obtained using the numerical evaluation procedure from Section~\ref{pseudoinvrs}  and evaluating $\hat{C}_{V_1,V2;U_1}$ in a similar manner as in Equation~\eqref{cond_int}, using the estimates of each term.
\begin{figure}
	\includegraphics[width = 0.6\textwidth]{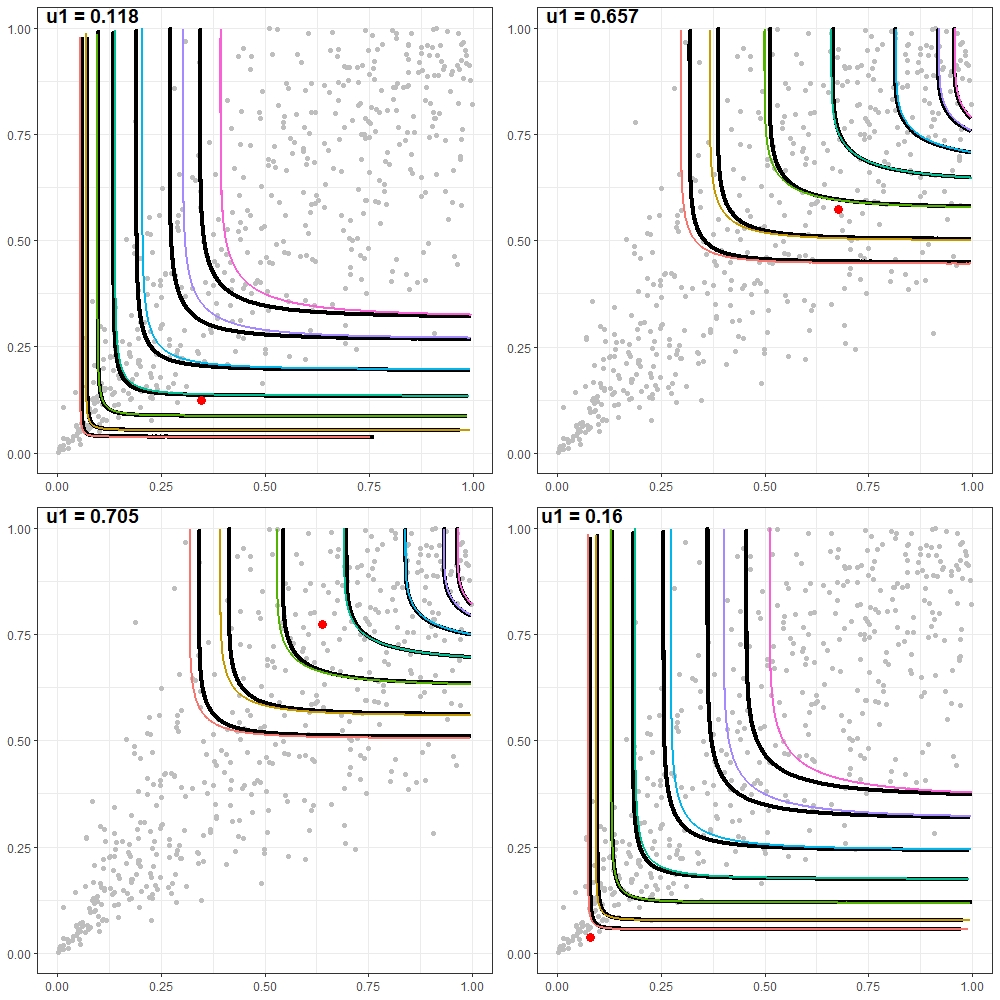}
	\centering
	\caption{x-axis: $V_1$, y-axis: $V_2$. Gray points: simulated data from vine distribution (n=500). Black curves: theoretical conditional level curves. Colored curves: estimated conditional level curves. Depicted are level curves at  $\alpha = 0.05,0.1,0.25,0.5,0.75,0.90,0.95$ (left bottom to right top in each panel). Red dot: associated values of $(v_1, v_2)$ with $u_1$ as conditioning value.}
	\label{fig:conditional}
\end{figure}
\noindent Note that the estimated and the data-generating vine are approximately very close, due to the use of the same tree structure in both data generation and estimation.
But in practice this is not the case, as the underlying tree structure is unknown. However, we can use the Y-vine regression model developed in Section \ref{sc:quantilereg} to model the tree structure and the pair copulas, in a way that the joint conditional distribution is easy to be estimated. Figure \ref{fig:conditional} shows the theoretical  and the  estimated level curves for  4 conditioning values of $u_1$. The values of $u_1$ are chosen from $ \mathbf{W}_{test} $. The level curves depend on the conditioning value. If the value of $u_1$ is low (top-left and bottom-right plot) the level curves are more restricted to the lower left corner. For greater values of $u_1$ (top-right and bottom-left plot) the level curves are more restricted to the top right corner. These occurrences can be explained by the high positive dependence of the  pairs $(V_1,U_1)$ and $(V_2,U_1)$ in the first tree of the vine structure, meaning that low values of $u_1$ correspond to  low values of both $v_1,v_2$. Thus, Figures \ref{fig:uncondquantiles} and \ref{fig:conditional} show that the numerical procedure for obtaining  both unconditional and conditional level curves  properly determines the bivariate level curves. We will employ this method to estimate conditional level curves in the case of more than one conditioning value (corresponding to more than one predictor in a regression setting).  In the case where we have bivariate regression data available we will fit a Y-vine regression model and use the estimated parameters to determine the associated bivariate level curves.

\subsection{Bivariate quantile curves}\label{bqrquantilescurvesnew}

\begin{figure}
	\includegraphics[width=0.3\textwidth]{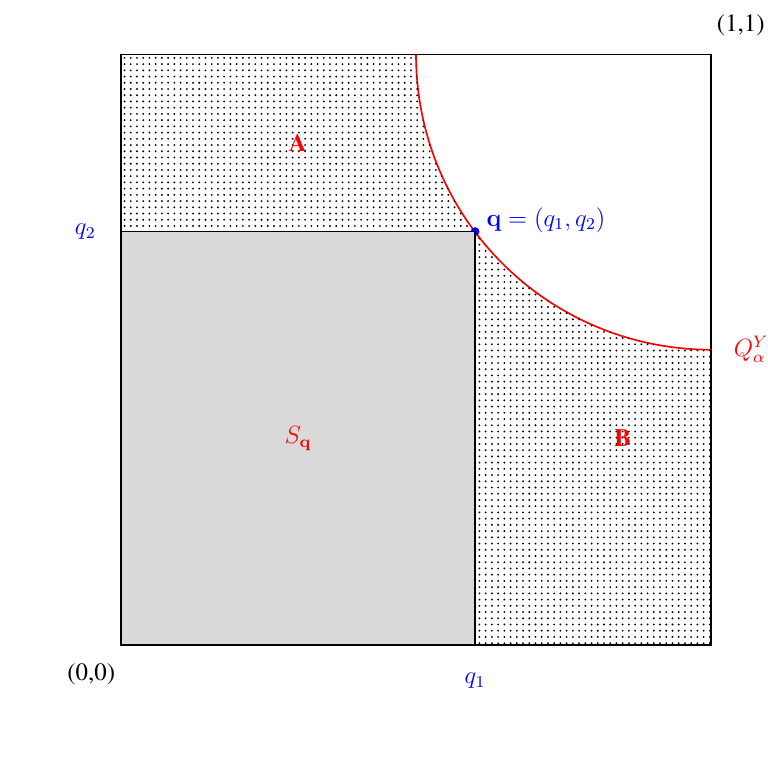}
	\centering
	\caption{A randomly chosen $\mathbf{q}=\left(q_1,q_2\right)$ vector, its corresponding $S_{\mathbf{q}}$ and $S_{\alpha}^{lower},$ where $S_{\alpha}^{lower} = S_{\mathbf{q}} \bigcup A \bigcup B$.}
	\label{fig:bqrexactconfreg}
\end{figure}

The notion of multivariate quantiles is not trivial nor well-defined. In the past, the level sets or curves of a multivariate distribution are considered as multivariate quantile, however in this case the coverage probability is not exact. For example, in
\citet{fernandez2002central}  the bivariate unconditional quantiles are defined as the  level sets of a bivariate  distribution function. The authors state that this definition is a natural generalization of the univariate quantile sets \citep{lewis1981dispersive}, however later in \citet{belzunce2007quantile} it is shown that the level curves 
do not have the property that the $\alpha$-th level curve  separates the lowest $\alpha\times 100$ percent of the observations from the remaining $\left(1 - \alpha\right)\times 100$ percent of the observations. 
Thus, we suggest to define the bivariate quantile curves as  adjusted level curves where the coverage probability is exact. Since we are interested in a regression setting, the following study is done on the conditional case, however, the same methodology can be applied for the unconditional case. Also, we define the bivariate quantiles on the u-scale (copula level), and to transform the bivariate quantiles on the x-scale, we use the same analogy as for the level curves.
Consider any bivariate vector $\mathbf{q} = \left(q_1, q_2\right) \in Q_{\alpha}^V (\mathbf{u})$ that lies on the level curve $Q_{\alpha}^V (\mathbf{u})$ for some $\alpha \in (0,1)$. Further, let $S_{\mathbf{q}}$ be a set of bivariate vectors defined by
$	S_{\mathbf{q}} \coloneqq \{ (v_1, v_2) \in [0,1]^2 \; ; \; v_1\leq q_1, v_2\leq q_2  \}.$
Then for any random vector $\mathbf{W}=(W_1,W_2)^T \sim C_{V_1,V_2|\mathbf{U}} (\cdot, \cdot| \mathbf{u})$ it holds that 
$	P\left(\mathbf{W} \in S_{\mathbf{q}}\right) = \alpha,$
by following  Equation \eqref{eq:biquant2}. In Figure \ref{fig:bqrexactconfreg} we can see an exemplary illustration for the set $	S_{\mathbf{q}} $.
Let the region $S_{\alpha}^{lower}$ be defined as  the set of bivariate vectors below the level curve $Q_{\alpha}^V (\mathbf{u})$,
\begin{equation*}
	S_{\alpha}^{lower} \coloneqq \bigcup_{\forall \left(q_1,q_2\right) \in Q_{\alpha}^V (\mathbf{u})} \left\{(v_1, v_2) \in [0,1]^2 ; v_1 < q_1, v_2 < q_2 \right\}.
\end{equation*}
Then for the random vector $\mathbf{W}=(w_1,w_2)^T $ it holds that 
\begin{equation}\label{bqr:extenprob}
	\mathbb{P}\left(\mathbf{W} \in S_{\alpha}^{lower}\right)  > \alpha,
\end{equation}
since $ S_{\mathbf{q}}\subset  S_{\alpha}^{lower}$ for all $\mathbf{q} = \left(q_1, q_2\right) \in Q_{\alpha}^V (\mathbf{u}),$ as also noted in \citet{fernandez2002central}. (See Figure \ref{fig:bqrexactconfreg} to observe the $S_{\alpha}^{lower}$  region.)
This implies that the level curve  $Q_{\alpha}^V (\mathbf{u})$ divides the $\left[0,1\right]^2$ square into a region for which it holds that $		\mathbb{P}\left(\mathbf{W} \in S_{\alpha}^{lower}\right)  \neq \alpha.$ It also follows that $\mathbb{P}\left(\mathbf{W} \notin S_{\alpha}^{lower}\right)  < 1- \alpha.$ 
Thus, for the definition of a $\alpha$ quantile curve we want to find an adjusted $\beta(\alpha)$ level curve which  will divide the observation space into $\alpha$ and $1- \alpha$ percent, i.e. for which  $\mathbb{P}\left(\mathbf{W} \in S_{\beta(\alpha)}^{lower}\right)  = \alpha$ holds.
\begin{definition}\label{bqr:exactquantilenew}
	The bivariate conditional  quantile  for $\alpha \in \left( 0,1\right)$, a  transformation $\beta :  (0,1) \longmapsto (0,1)$ and   continuous random variables $Y_1,Y_2$ with random PITs $V_1, V_2$  given the outcome of the random vector $\mathbf{X} = \mathbf{x},$ with PITs   $\mathbf{U} =\mathbf{u}$ is a  curve in $[0,1]^2$ defined by the set
	\begin{equation}	\label{eq:biquantadjusted}
		q_{\alpha}^{V}(\mathbf{u})  \coloneqq  \{ (v_1, v_2) \in [0,1]^2  \; ; \; C_{V_1,V_2|\mathbf{U}} (v_1, v_2| \mathbf{u})  = \beta(\alpha) \},
	\end{equation}
	so that the observation space is divided  into $\alpha$ and $1- \alpha$ percent regions, i.e. $\mathbb{P}\left(\mathbf{W} \in S_{\beta(\alpha)}^{lower}\right)  = \alpha$ holds.
\end{definition}
\noindent Following Definition \ref{bqr:exactquantilenew}, we can also define an exact $100\times\left(1-\alpha\right)\%$ confidence region arising from the quantile curves $q_{\alpha/2}^V (\mathbf{u})$ and  $q_{1-\alpha/2}^V (\mathbf{u})$.
\begin{definition}\label{confregion}
	The $100\times\left(1-\alpha\right)\%$ bivariate confidence region for  $\alpha \in\left(0,1 \right)$ and a continuous bivariate vector
	continuous random variables $Y_1,Y_2$ with random PITs $V_1, V_2$  given the outcome of the random vector $\mathbf{X} = \mathbf{x},$ with PITs   $\mathbf{U} =\mathbf{u}$, is set of points in $[0,1]^2$ enclosed by the quantile curves $q^V_{\alpha/2}\left(\mathbf{u}\right)$ and $q^V_{1-\alpha/2}\left(\mathbf{u}\right)$, i.e. 
	\begin{equation*}
		\small
		\begin{aligned}
			CI^{V_1, V_2 | \mathbf{U}}_{\alpha} \coloneqq \bigg\{ \left(w_1^*,w_2^*\right) \in [0,1]^2 \; \big|\; \exists \; \left(v^{1}_1,v^{1}_2\right) \in q^{V}_{\alpha/2}\left(\mathbf{u}\right),&\; \left(v^{2}_1,v^{2}_2\right) \in q^{V}_{1-\alpha/2}\left(\mathbf{u}\right)\; such\; that:  \\  
			&\;\;  v^{1}_1\leq w_1^* \leq v^{2}_1\;\; and\;\; v^{1}_2\leq w_2^* \leq v^{2}_2 
			\bigg\}.
		\end{aligned}
	\end{equation*}
\end{definition}
\noindent In this case, 
$	\mathbb{P}( \mathbf{W} \in CI^{V_1,  V_2 | \mathbf{U}}_{\alpha} ) = \mathbb{P}( \mathbf{W} \in S_{\beta(\alpha/2)}^{lower} )
-  \mathbb{P}( \mathbf{W} \in S_{\beta(1-\alpha/2)}^{lower} ) = \alpha/2 - (1-\alpha/2)= 1-\alpha,$
implying that $CI^{V_1, V_2 | \mathbf{U}}_{\alpha}$ is an exact $100\times\left(1-\alpha\right)\%$ confidence region. 
Returning back to the problem of estimating the transformation $\beta(\alpha)$, for  $\beta :  (0,1) \longmapsto (0,1)$, so that the quantile curves $q_{\alpha}^{V}(\mathbf{u})$ are estimated, we suggest a numerical procedure. 
Basically, we need to change the $\alpha$-level  curve to a new $\beta(\alpha)$- level curve so that $\mathbb{P}(\mathbf{W} \in S_{\beta(\alpha)}^{lower})  = \alpha$ holds true. To achieve this, we define the function 
\begin{equation}
	\label{eq:Gprobfunction}
	\begin{aligned}
		G(\beta) & \coloneq \mathbb{P}\left(\mathbf{W} \in S_{\beta}^{lower}\right) \\
		& = \mathbb{P} \left(C_{V_1,V_2|\mathbf{U}} (\cdot, \cdot| \mathbf{u}) \leq \beta \right) \; \; \forall \; \beta \in (0,1)  .
	\end{aligned}
\end{equation}
From Equation \eqref{bqr:extenprob} we can see that $G(\alpha) > \alpha$. However, we are interested to find the value $\beta(\alpha)$ so that it holds that $G(\beta(\alpha)) = \alpha$, thus $\beta(\alpha) = G^{-1}(\alpha).$ To do so, we suggest a numerical procedure.  As the function $G(\beta)$ is  difficult to evaluate analytically,  we suggest to estimate it using a simulated sample from the Y-vine copula with $\mathbf{U}=\mathbf{u}$ fixed. For $n =1,\cdots, N$ we simulate observations  $\left(  v_1^n(\mathbf{u}), v_2^n(\mathbf{u}) \right) \sim C_{V_1, V_2| \mathbf{U}}\left(\cdot, \cdot| \mathbf{u}\right)$, as described in Section \ref{simulations}. 
Then, we estimate $G(\beta)$ as the proportion of the simulated data below the $\alpha$-quantile over the sample size N, i.e.
\begin{equation*}
	\hat{G}\left(\beta\right) = \frac{1}{N}\sum_{n=1}^{N} \mathbb{I}\left(\left(  v_1^n(\mathbf{u}), v_2^n(\mathbf{u}) \right)\in S_{\beta}^{lower}\right),
\end{equation*}
where $\mathbb{I}$ is an indicator function, being equal to 1 when the condition $\left(  v_1^n(\mathbf{u}), v_2^n(\mathbf{u})  \right)\in S_{\beta}^{lower}$ is satisfied, and equal to 0, otherwise.
To find the desired $\beta(\alpha)$ we use a line search algorithm on the $(0,1)$ interval and  obtain the estimated $\hat{\beta}(\alpha)$ such that  $\hat{G}(\hat{\beta}(\alpha)) = \alpha$. 
This way  the suggested methodology from Section \ref{BQR:bivariatequantiles} can be extended to find the bivariate quantiles   $ q_{\alpha}^{V}(\mathbf{u})$    such that 
$	\hat{G}(\beta(\alpha)) =\mathbb{P}\left(\mathbf{W} \in S_{\beta(\alpha)}^{lower}\right)=  \alpha,$
holds, i.e. the $\beta(\alpha)$-th level set  separates the lowest $\alpha\times 100$ percent of the observations from the remaining $\left(1 - \alpha\right)\times 100$ percent of the observations.

Another concept for the construction of exact confidence regions is been developed in \citet{coblenz2018confidence}. The authors propose,  to construct an exact confidence region for unconditional bivariate copula distribution functions. They use  the Kendall distribution function of a bivariate copula $C$   at a level $\alpha \in (0,1)$, $K(C,\alpha) $ defined as
$	K(C,\alpha) \coloneqq \mathbb{P} \left(  C(U,V) \leq \alpha, \; (U,V) \sim C \right), $
in \citet{genest1993statistical} and \citet{barbe1996kendall}.  
In comparison to our methodology it holds that $ G(\beta(\alpha))= K (C, \alpha) $ in the unconditional case, as shown in \citet{chakak2000bivariate}. For bivariate copula distribution functions computing the Kendall distribution function is possible and certain approaches are available \citep{chakak2000bivariate, ezzerg1999estimacion}, however it is very computationally expensive \citep{brechmann2013hierarchical}. Once $K$ is estimated, $\beta (\alpha)$ can be obtained as the inverse of the Kendall distribution function evaluated at $\alpha$, i.e. $\beta (\alpha) = K^{-1}(C, \alpha)$. 
Estimating the Kendall distribution functions in the conditional case is difficult in general and computationally expensive, however  the same results are expected to follow as for the unconditional case.

\section{Data application}\label{sc:dataapp}
The implementation of the Y-vine quantile regression is done in the statistical software \texttt{R} \citep{software}. As an application to real data we consider the Seoul weather data set, which contains two dependent responses, daily minimum and maximum air temperature. The data originates from the UCI machine learning repository \citep{Dua:2019}, it can be downloaded using \href{https://archive.ics.uci.edu/ml/datasets/Bias+correction+of+numerical+prediction+model+temperature+forecast}{https://archive.ics.uci.edu/ml/ \linebreak datasets/Bias+correction+of+numerical+prediction+model+temperature+forecast} and was first studied by \cite{cho2020comparative}. 
It contains daily data for 25 weather stations in Seoul, South Korea between June 30th and August 30th in the period 2013-2017. \cite{cho2020comparative} use it for   enhancing next-day maximum and  minimum air temperature forecasts based on the Local Data Assimilation and Prediction System (LDAPS) model.
To illustrate the proposed vine based bivariate quantile regression model, we consider the station located in central Seoul (station 25) and and we model the temporal dependence
	in the responses, by considering the present minimum and maximum air temperature
	(including two lagged variables into the regression model) when modeling next day
	values.  Disregarding geographical markers and precipitation measurements, we are left with a data set containing  two response variables and 13 continuous predictors, with  307 data points representing summer days of the years 2013 to 2017. In the supplementary material we provide a description of the considered variables.  We divide the data set into a training and testing set, consisting of 246  data points from 2013-2016, and 61  data points from 2017, respectively. In the supplement,  we also show the empirical normalized contour plots for pairs of variables from the training set, which shows strong non-Gaussian dependence structure in the data, indicated by non-elliptical shapes. This shows that the data is suited for application of the proposed Y-vine copula  class.
In the estimation of our Y-vine quantile regression model we model the marginals distributions using a nonparametric approach, while we model the pair copulas in a  parametric  approach, resulting in a semiparametric model. Modeling the marginals as well as the copulas parametrically might cause the
resulting fully parametric estimator to be biased and inconsistent if one of the parametric models is misspecified \cite{noh2013copula}. Modeling them both using a nonparametric  approach  leads to a fully nonparametric approach that might overfit the data, because penalization is still an open research topic in the nonparametric case, as noted in \cite{tepegjozova2021nonparametric}. Thus, we opt for a semiparametric approach. The marginals are estimated using a univariate nonparametric kernel density estimator implemented in the \texttt{R} package \texttt{kde1d} \citep{kde1d}, and the pair copulas are fitted using a parametric maximum-likelihood approach with the Akaike Information Criterion penalization \citep{akaike1973theory} (AIC) implemented in the \texttt{R} package \texttt{rvinecopulib} \citep{rvinecopulib}. Further, we use as selection criteria for the forward selection of predictors the AIC penalized  $acll,$ (defined in \eqref{eq:ccloglik}) in order to favour a more sparse model.

The automatically chosen order of the predictors in the fitted Y-vine regression model $\hat{{\mathcal{Y}}}$ is given by 
\begin{equation*}
	\begin{split}
		O(\hat{{\mathcal{Y}}})= & \left( \texttt{
			LDAPS\_Tmin\_lapse, LDAPS\_Tmax\_lapse, LDAPS\_CC1,      LDAPS\_WS, } \right. \\ 
		& \left. \texttt{ Present\_Tmin, LDAPS\_RHmax, LDAPS\_CC3,  LDAPS\_LH, Present\_Tmax} \right).
	\end{split}
\end{equation*}

\begin{figure}[]
	\centering
	\subfloat[$x$-scale]{{\includegraphics[width=5.5cm]{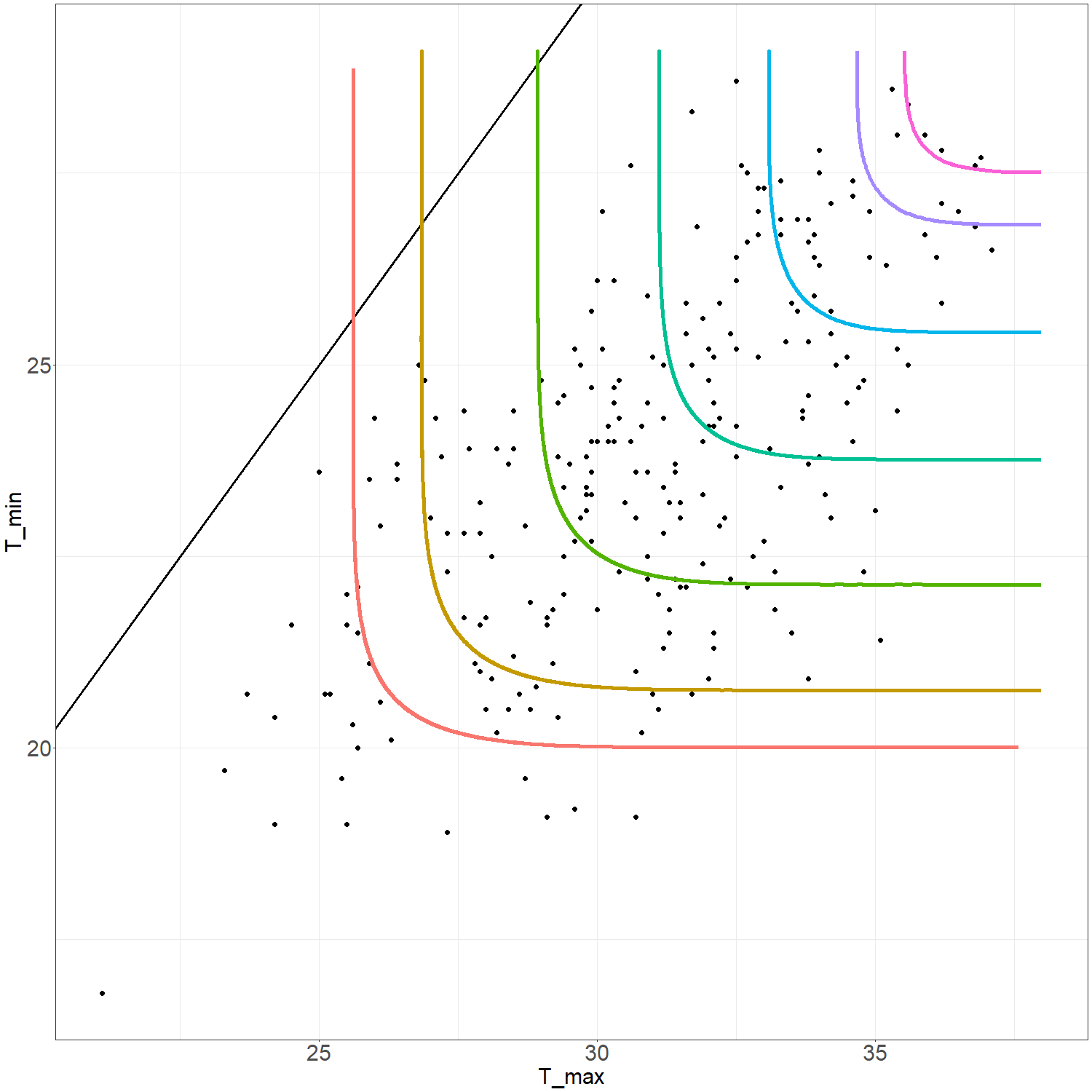} }}
	\qquad
	\subfloat[$u$-scale]{{\includegraphics[width=5.5cm]{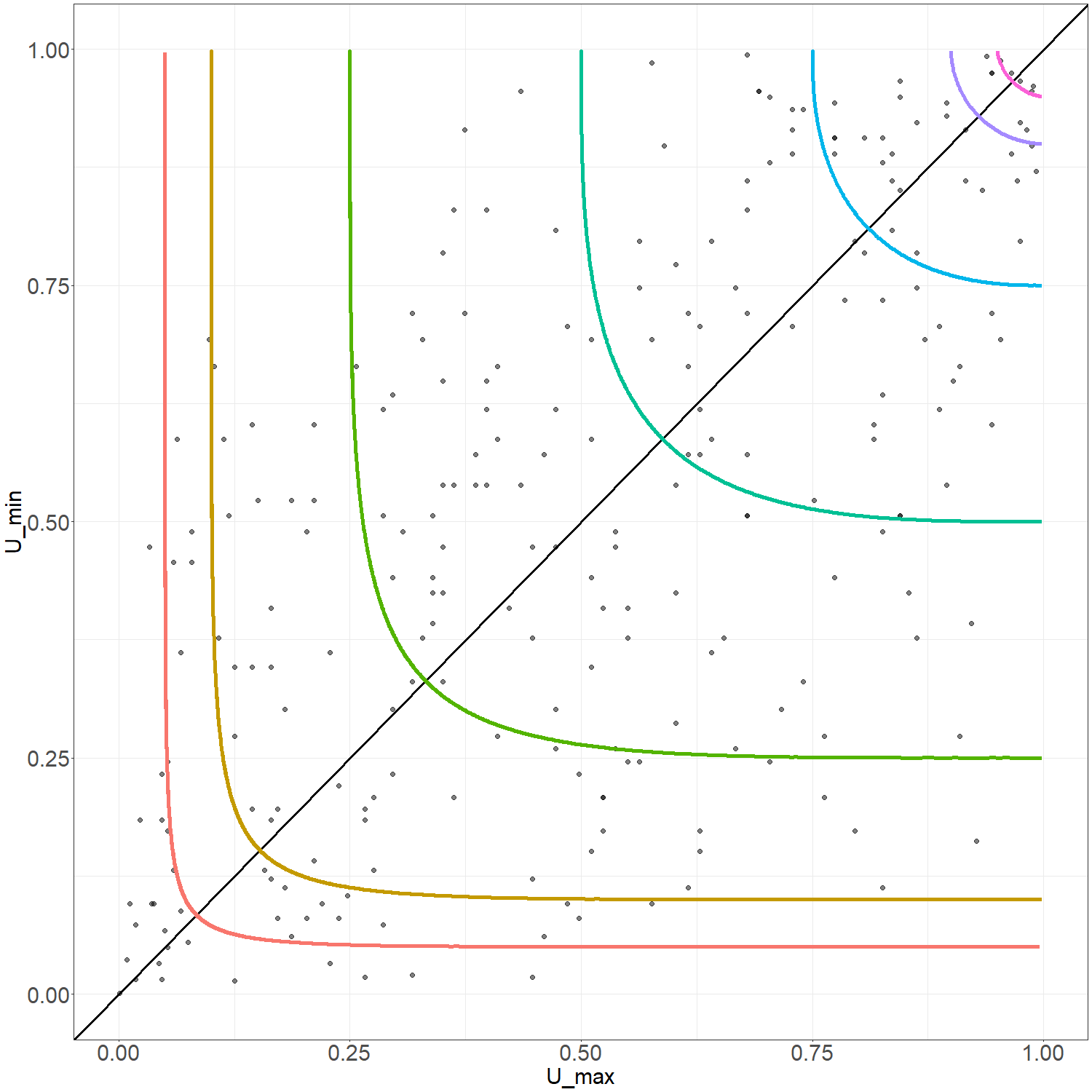} }}
	\caption{ Black points:  data from 2013-2016 (n=246). Colored curves: estimated unconditional level curves at $\alpha = 0.05,0.1,0.25,0.5,0.75,0.90,0.95$ (left bottom to right top).}
	\label{fig:data_example_uncond}
\end{figure}

It orders the predictors by their influence over the two responses. Also, only 9 out of 13 possible predictors are chosen to be in the model. The 4 non-influential predictors, based on the Y-vine model are LDAPS\_CC2, LDAPS\_CC4,  LDAPS\_RHmax and solar radiation. More details on the fitted pair copulas selected by the Y-vine regression model is given in the supplement, in Tables 2 and 3.
The fitted  pair copula between the responses given the 9 chosen predictors, $\hat{c}_{V_1,V_2;\mathbf{U}}$ is a Joe copula with an estimated Kendall's $\tau$ of 0.09. This implies that  after the effect of the predictors is adjusted in the model, there is little dependence between the responses. 
For illustrating  the unconditional level curves of the joint unconditional bivariate distribution of the two responses, U\_max and U\_min we fit a 
pair copula between them.  The estimated pair copula is the Gaussian copula with a parameter of 0.66. The unconditional quantile curves are defined as in Definition \ref{bqr:exactquantilenew} by using the pair copula distribution function between the responses  $C_{V_1,V_2} $, instead of the bivariate conditional distribution $C_{V_1,V_2 | \mathbf{U}}, $ and we denote them as $q_{\alpha}^V$  for $\alpha \in (0,1)$. 
The level curves of this copula, on both the x- and the u-scale are given in Figure  \ref{fig:data_example_uncond}. Note that  the maximum temperature is always  greater than the minimum temperature. However, this ordering constraint does not imply an ordering constraint on the PITs on the u-scale (as the marginal distributions are separately and independently modeled). For illustration see Figure~\ref{fig:data_example_uncond}, where the ordering is visible in panel $(a)$, as all the data is below the diagonal, while this ordering is lost in panel $(b).$
\subsection{Bivariate quantile curves, confidence regions and advantages of  joint modeling of dependent responses}
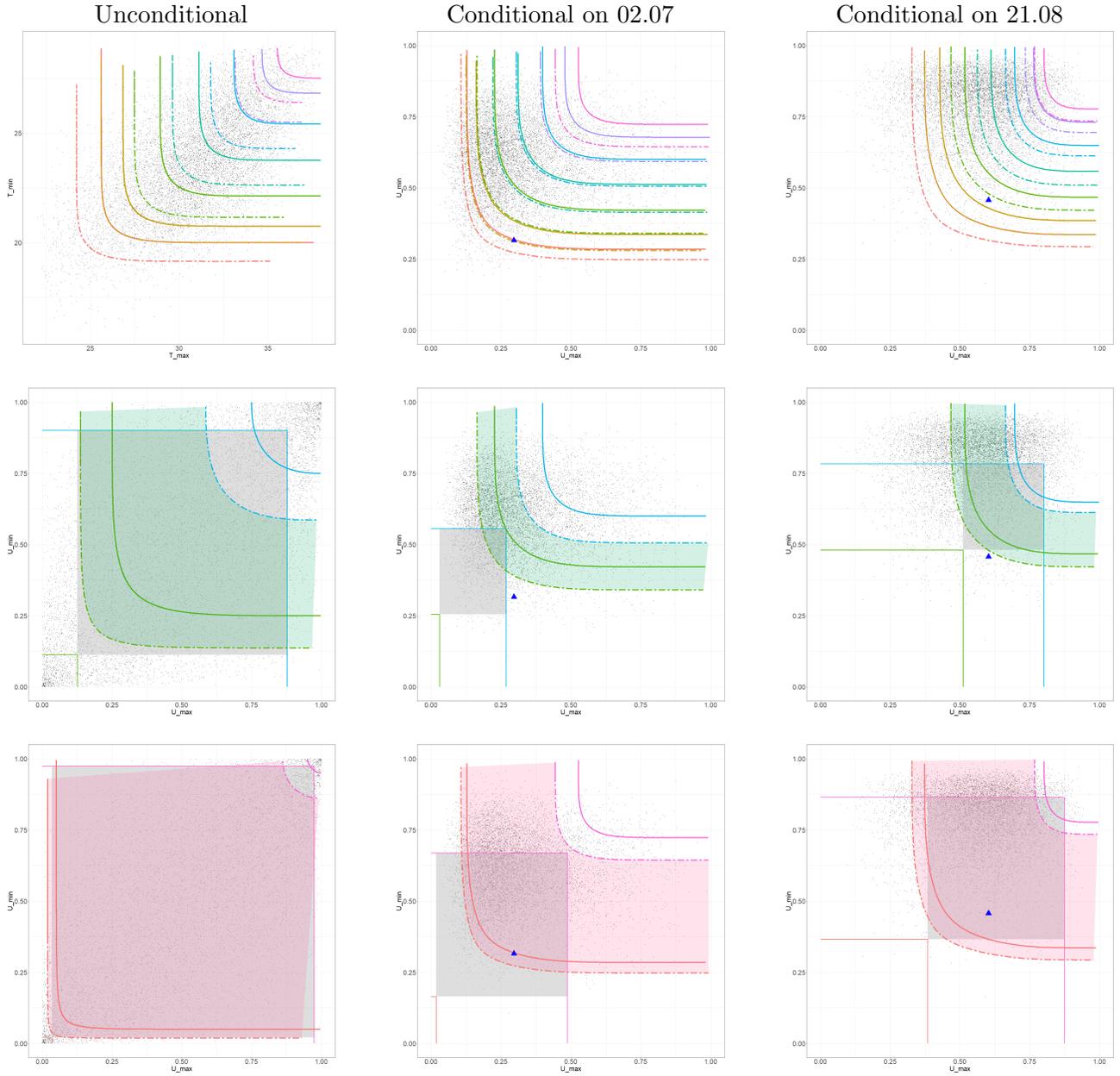
\begin{figure}[]
		\centering
		\begin{tikzpicture}
			
			
			\node[inner sep=0pt] (Col1) at (-20, 10)  {\includegraphics[width=0.29\textwidth]{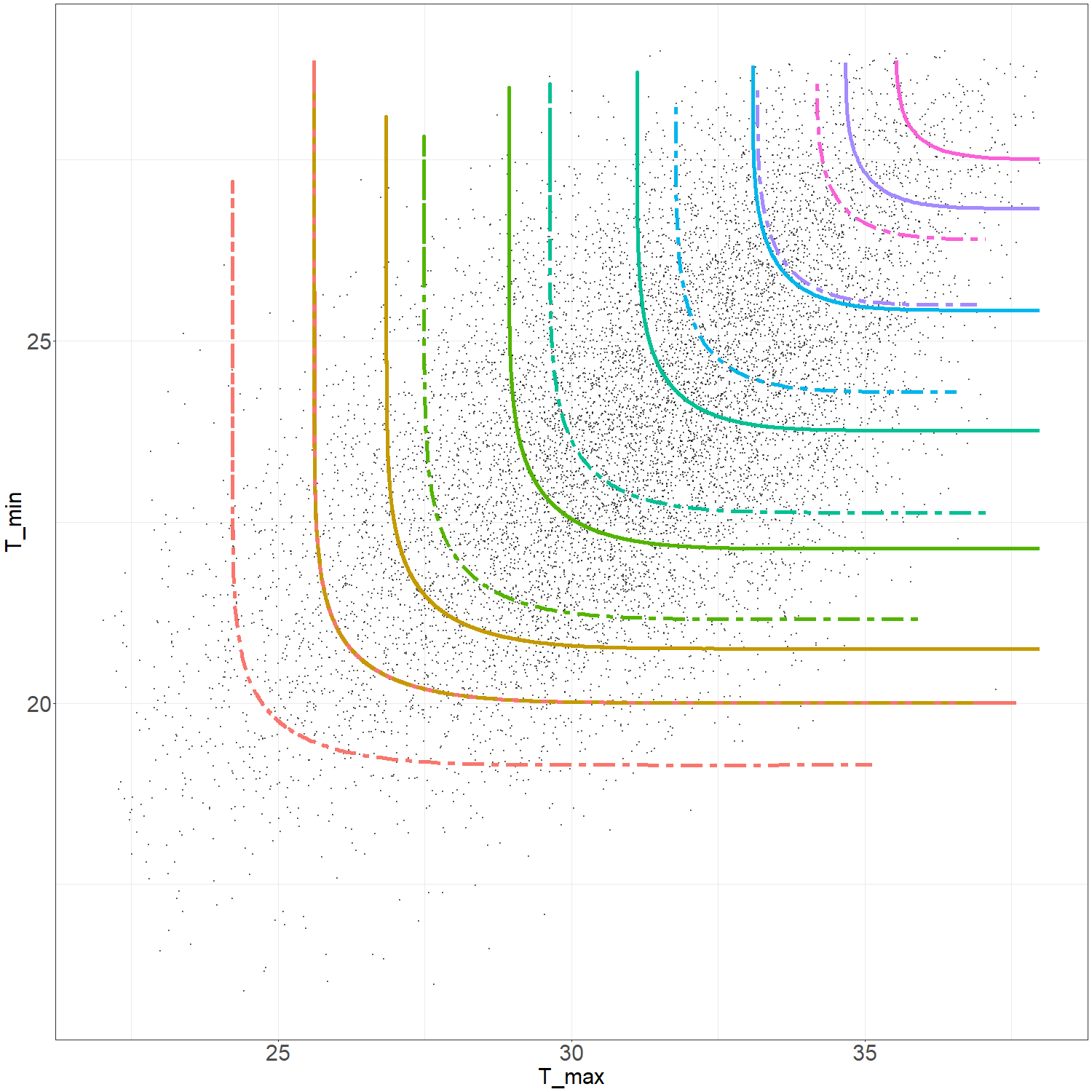}};
			
			\node (Col111) at (-20, 12.8)  {Unconditional};
			
			\node[inner sep=0pt] (Col1) at (-14, 10)  {\includegraphics[width=0.29\textwidth]{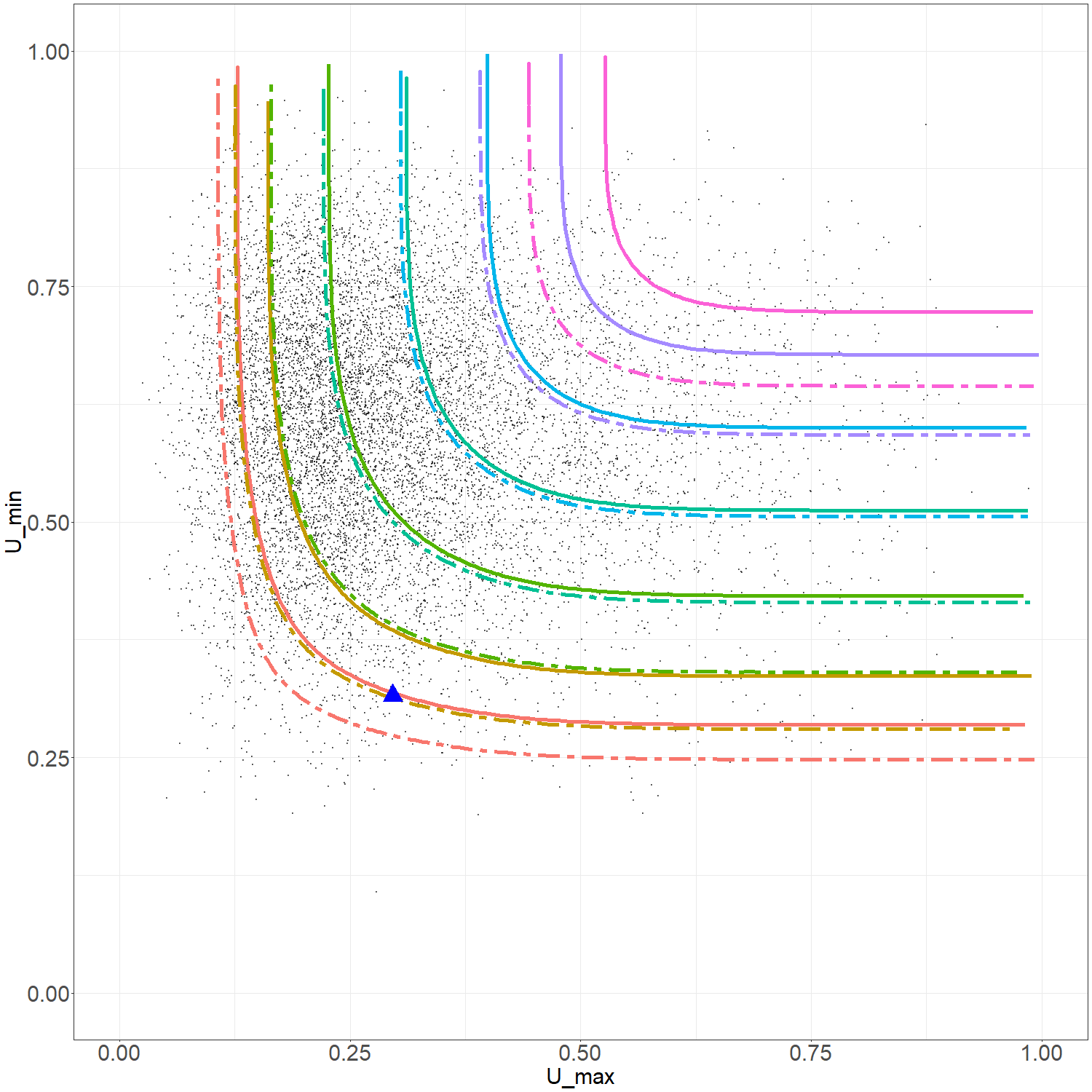}};
			
			\node (Col111) at (-14, 12.8)  {Conditional on 02.07};
			\node[inner sep=0pt] (Col1) at (-8, 10)  {\includegraphics[width=0.29\textwidth]{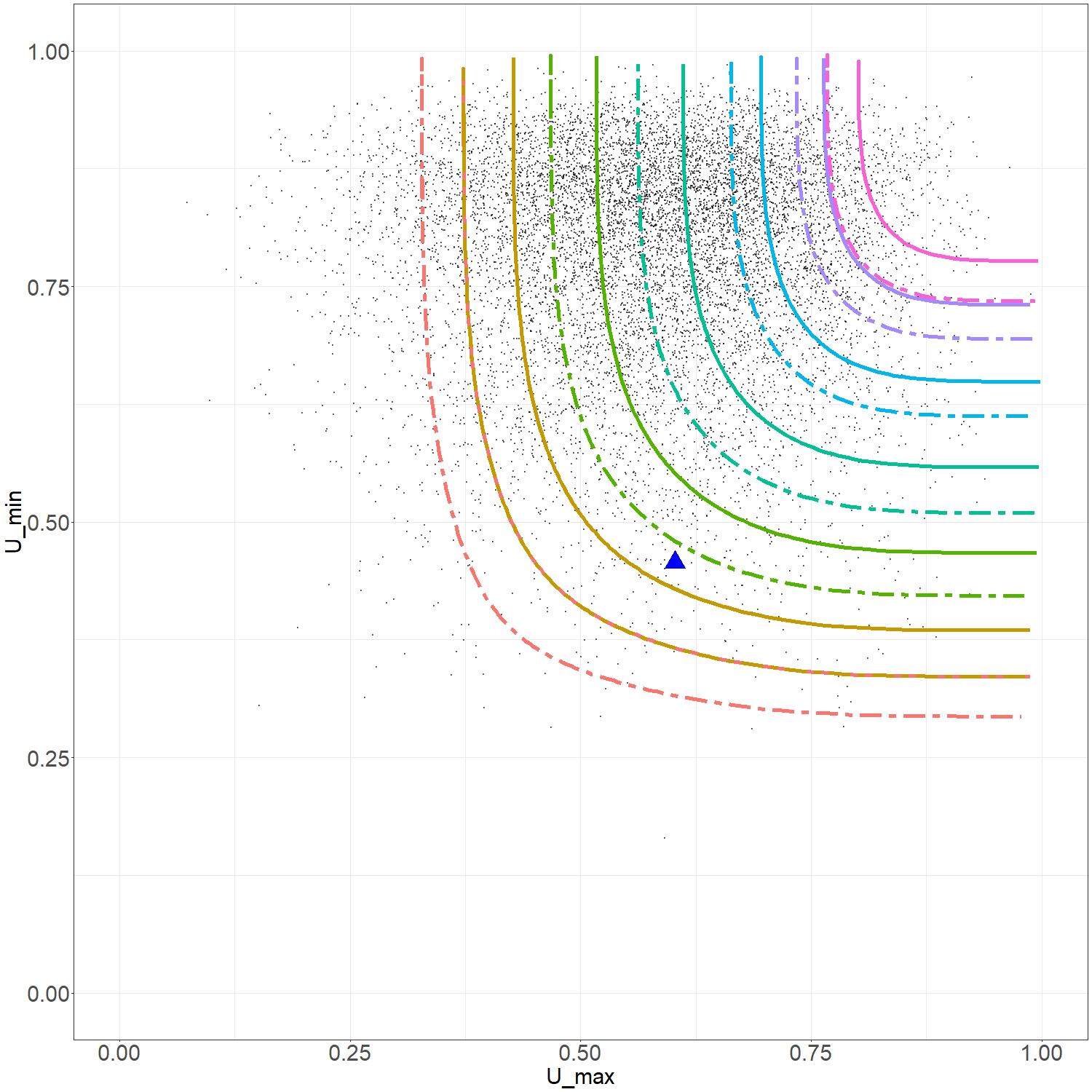}};
			\node (Col111) at (-8, 12.8)  {Conditional on 21.08};
			
			\node[inner sep=0pt] (Col1) at (-20, 4.5)  {\includegraphics[width=0.29\textwidth]{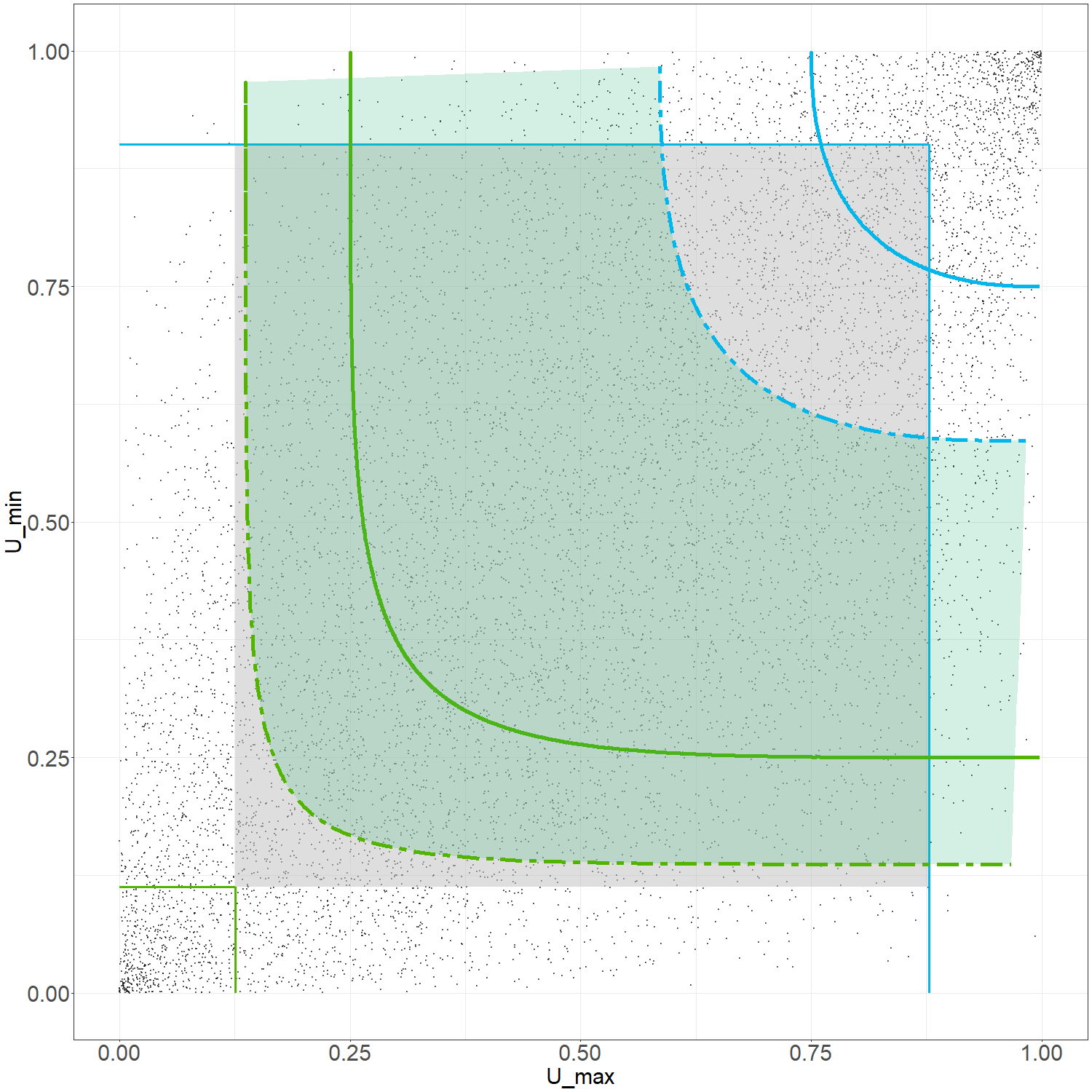}};

			\node[inner sep=0pt] (Col1) at (-14, 4.5)  {\includegraphics[width=0.29\textwidth]{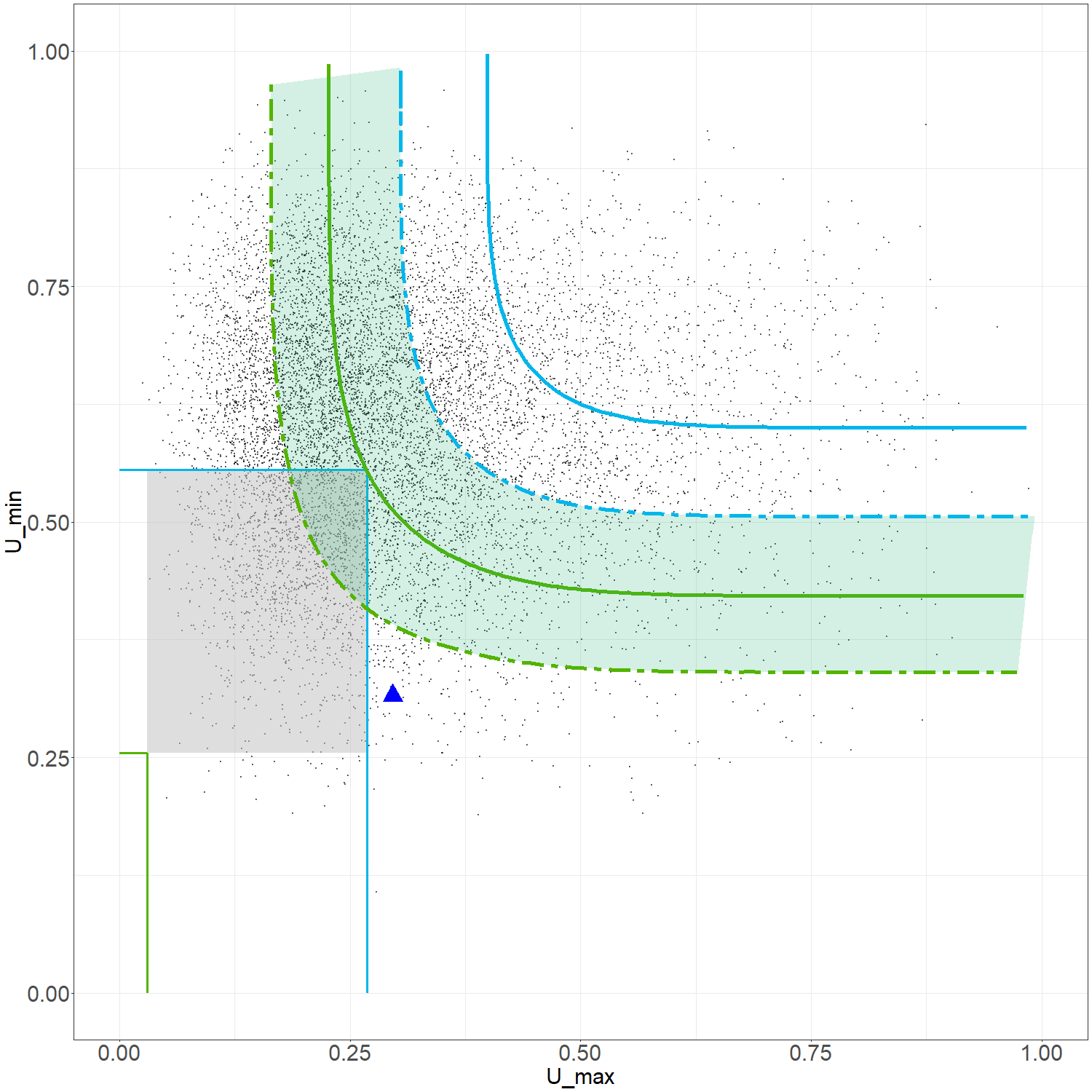}};

			\node[inner sep=0pt] (Col1) at (-8, 4.5)  {\includegraphics[width=0.29\textwidth]{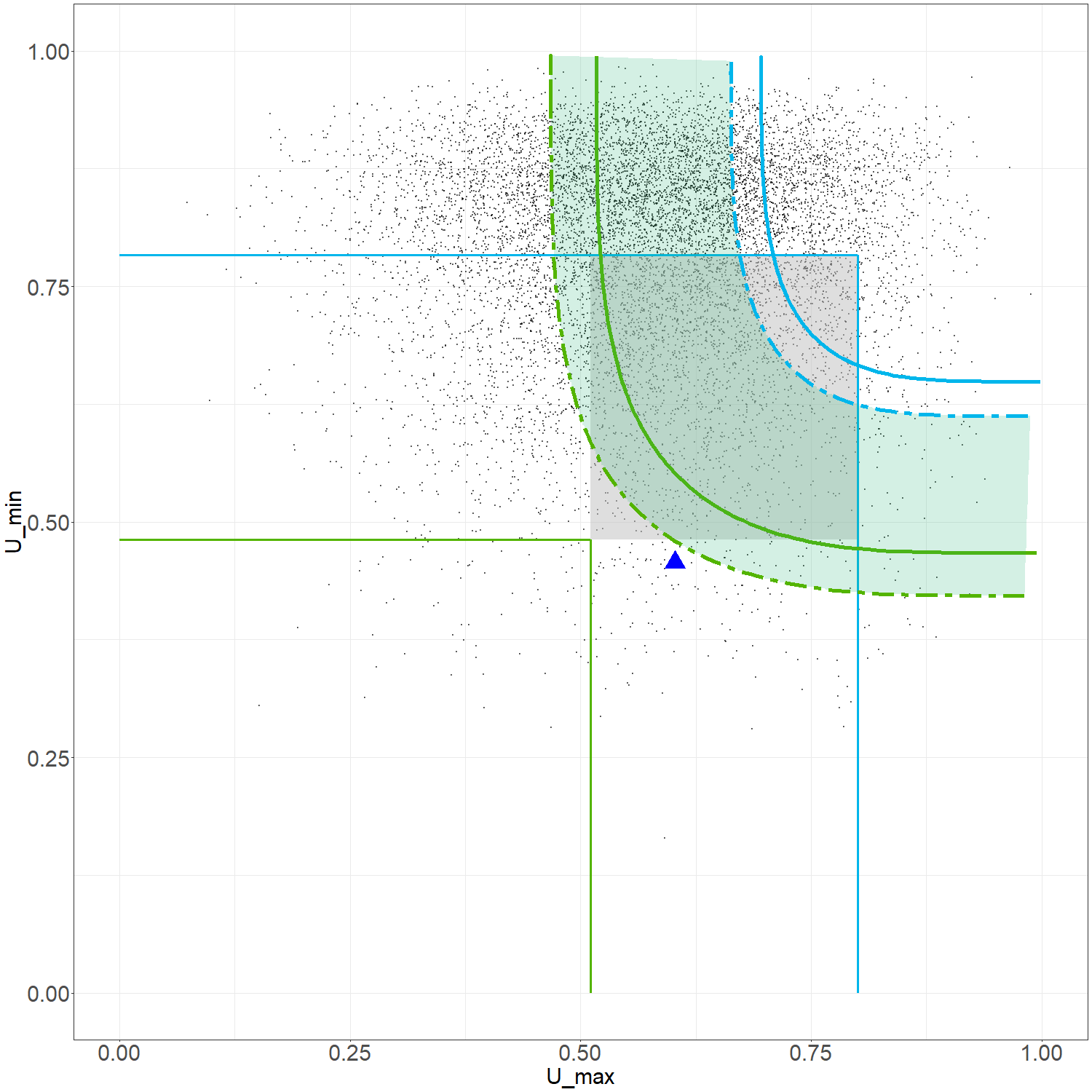}};
			
			\node[inner sep=0pt] (Col1) at (-20, -1)  {\includegraphics[width=0.29\textwidth]{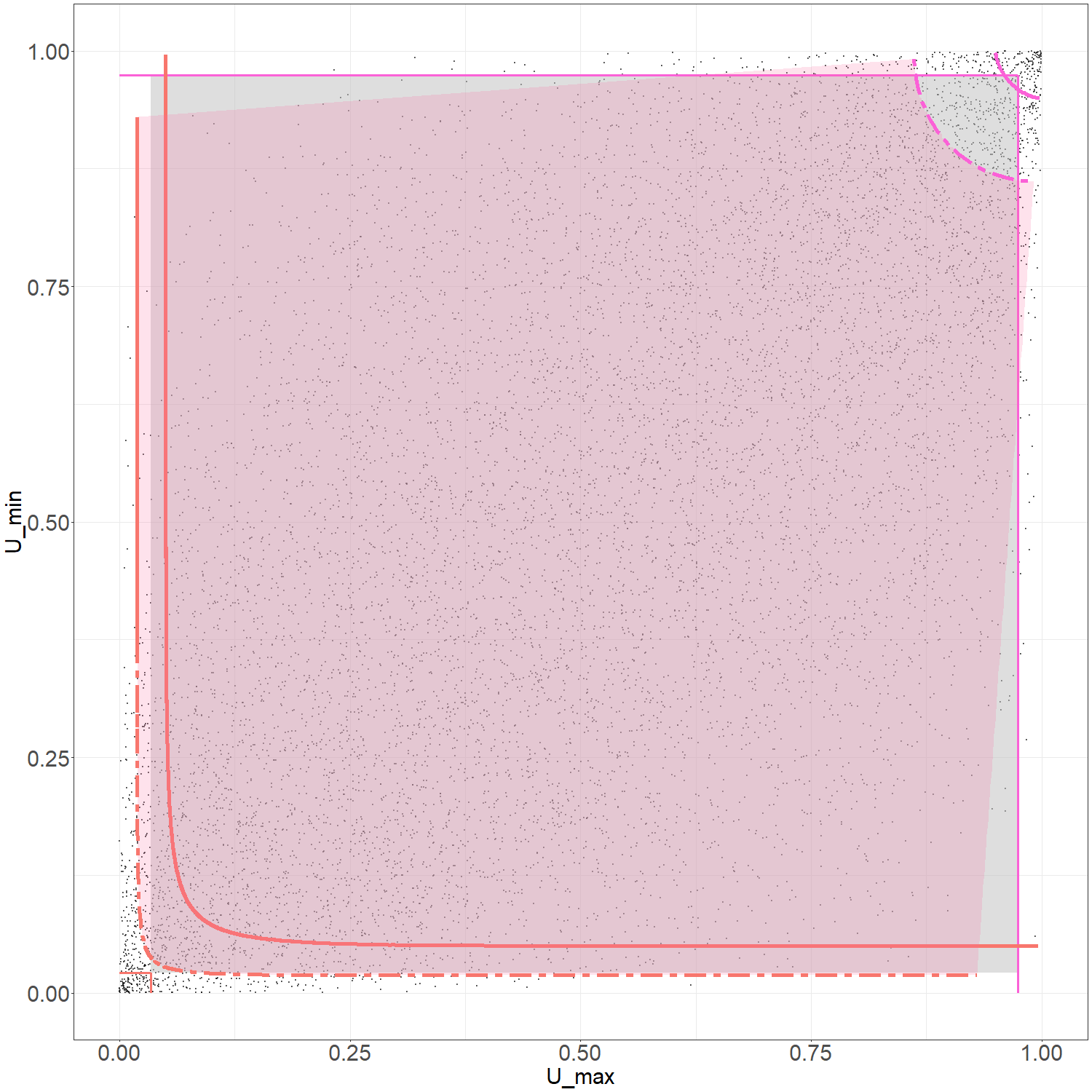}};

			\node[inner sep=0pt] (Col1) at (-14, -1)  {\includegraphics[width=0.29\textwidth]{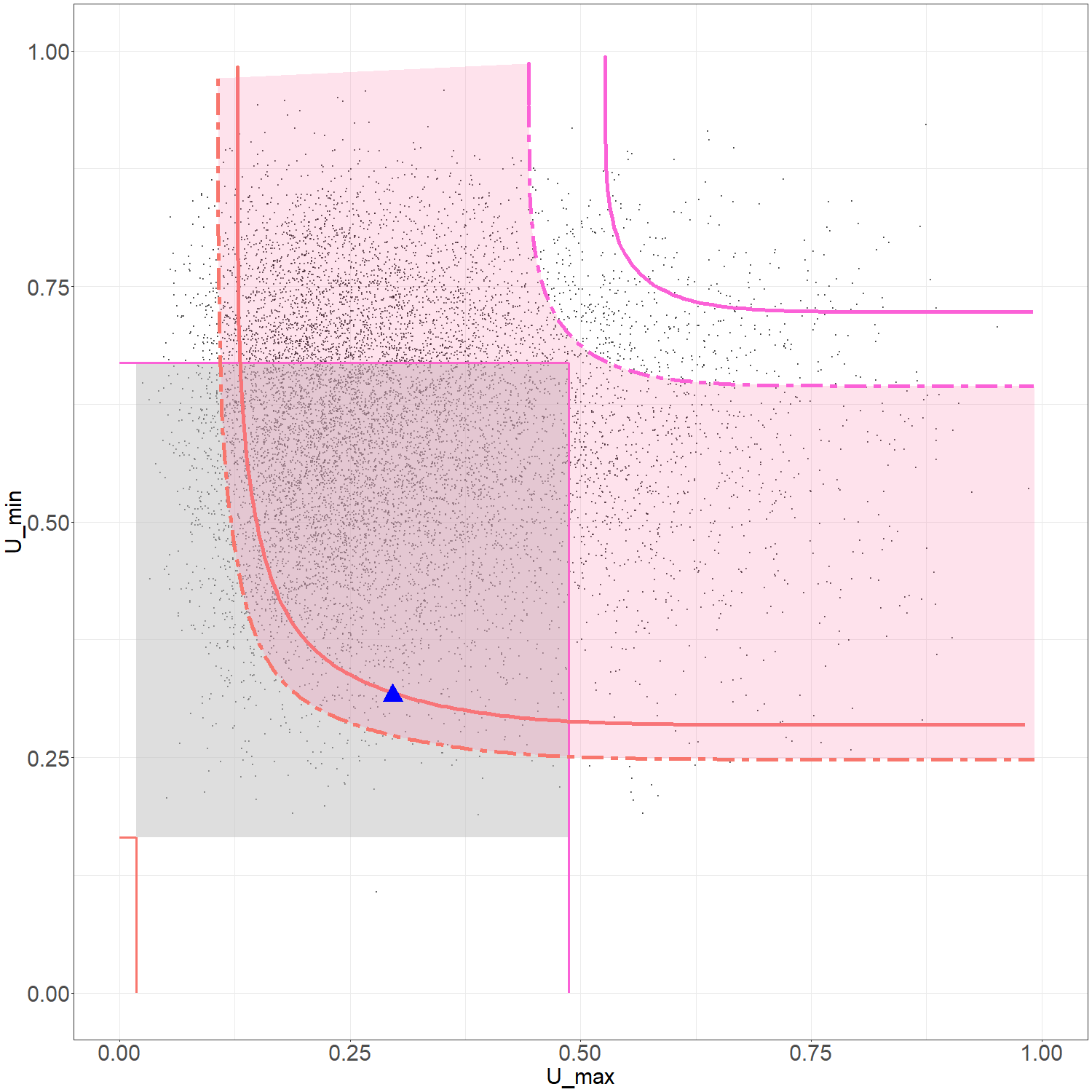}};

			\node[inner sep=0pt] (Col1) at (-8, -1)  {\includegraphics[width=0.29\textwidth]{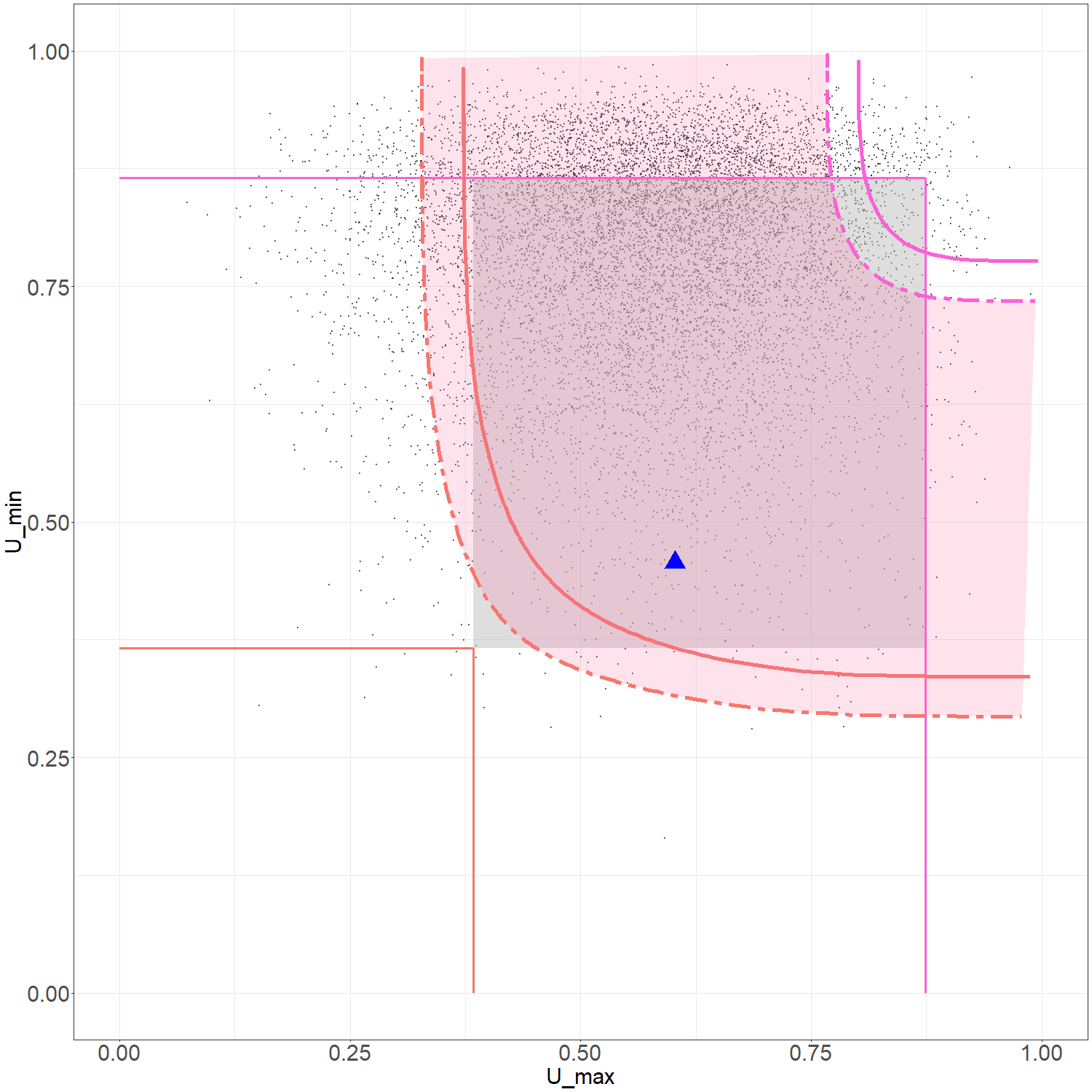}};
			
		\end{tikzpicture}
		\caption{First row: level curves (solid lines) and corresponding quantile curves (dashed lines) at $\alpha = 0.05,0.1,0.25,0.5, 0.75, 0.90,0.95$. Second row: first column $CI^{V_1,  V_2 }_{0.50}$ (green  region, case 1.) ) and $CI^{V_1 \perp V_2 }_{0.50}$ (gray  region, , case 2.)), second and third column  $CI^{V_1,  V_2 | \mathbf{U}}_{0.50}$ (green  region, case 4.)) and $CI^{V_1 \perp V_2 | \mathbf{U} }_{0.50}$ (gray  region, case 3.)). Third row: first column $CI^{V_1,  V_2 }_{0.90}$  (red  region) and $CI^{V_1 \perp V_2 }_{0.90}$ (gray  region), second and third column $CI^{V_1,  V_2 | \mathbf{U} }_{0.90}$ (red  region) and $CI^{V_1 \perp V_2 | \mathbf{U} }_{0.90}$ (gray region). . The blue triangle is the true observed value.}
		\label{figquantilesunconditionall}
	\end{figure}

For comparison purposes we consider 4 different scenarios: 1.) $U_{max}$ and $U_{min}$ are jointly modeled using a bivariate copula,  2.) $U_{max}$ and $U_{min}$ are independent,  3.) $U_{max}$ and $U_{min}$ are conditional independent given the predictors and 4.) $U_{max}$ and $U_{min}$ are jointly modeled with the predictors. The first 2 cases are the unconditional  cases, and the last 2 are the conditional case. 
For 1.) consider the unconditional quantile curves and the corresponding confidence region obtained from fitting a bivariate copula between the responses $U_{max}$ and $U_{min}$, and the confidence region obtained by assuming dependence between the responses. 
The unconditional quantile curves are defined as in Definition \ref{bqr:exactquantilenew} by using the pair copula between the responses  $C_{V_1,V_2} $, instead of the bivariate conditional distribution $C_{V_1,V_2 | \mathbf{U}},$ and are denoted as $q_{\alpha}^V$  for $\alpha \in (0,1)$. Using Definition \ref{confregion},  by substituting the conditional quantile curves with the unconditional ones, we can define the corresponding unconditional confidence region $CI^{V_1, V_2}_{\alpha}$ as
set of points in $[0,1]^2$ enclosed by the quantile curves $q^V_{\alpha/2}$ and $q^V_{1-\alpha/2}$ for some $\alpha \in (0,1)$, i.e. 
\begin{equation*}
	\small
	\begin{aligned}
		CI^{V_1, V_2}_{\alpha} \coloneqq \bigg\{ \left(w_1^*,w_2^*\right) \in [0,1]^2 \; \big|\; \exists \; \left(v^{1}_1,v^{1}_2\right) \in q^{V}_{\alpha/2},&\; \left(v^{2}_1,v^{2}_2\right) \in q^{V}_{1-\alpha/2}\; such\; that:  \\  
		&\;\;  v^{1}_1\leq w_1^* \leq v^{2}_1\;\; and\;\; v^{1}_2\leq w_2^* \leq v^{2}_2 
		\bigg\}.
	\end{aligned}
\end{equation*}
For 2.) we construct a bivariate quantile region from the univariate empirical quantiles, denoted as $q_{\alpha,emp}^{V_1}$ for $\alpha \in (0,1),$ using the Bonferroni correction for multiple testing \citep{bonferroni1936teoria}. We are interested in the bivariate quantile region with coverage probability at  $\alpha \in (0,1)$ meaning that the two univariate empirical quantiles, from which we construct the bivariate quantile region, need to be evaluated at $\frac{\alpha}{4}$ and $1-\frac{\alpha}{4}$, and we denote the corresponding confidence region of the univariate empirical quantiles as $CI^{V_1 \perp V_2 }_{\alpha}$, i.e. 
\begin{equation*}
	CI^{V_1 \perp V_2 }_{\alpha}  \coloneq\left [q_{\frac{\alpha}{4},emp}^{V_1} , q_{1-\frac{\alpha}{4},emp}^{V_1}     \right] \times   \left[q_{\frac{\alpha}{4},emp}^{V_2} , q_{1-\frac{\alpha}{4},emp}^{V_2}    \right].
\end{equation*}
		%
		%
		%
		%
		%
		%
		%
		%
	For 3.)  we treat the response variables as conditionally independent given a set of predictors. Basically, the tasks of predicting maximal and minimal temperatures given the predictors are treated as completely independent problems and univariate conditional quantiles are estimated for both response variables. For this purpose, two univariate $D$-vine regression models with the same predictor order as the Y-vine regression are fitted. The D-vine is  a natural subset model of the Y-vine tree sequence when considering a single response variable. This way we can construct a bivariate quantile region from the univariate quantiles using the Bonferroni correction for multiple testing \citep{bonferroni1936teoria}, similar as before for the unconditional case. We denote these univariate D-vine based quantiles as, $q_{\alpha,Dvine}^V (\mathbf{u})$  for $\alpha \in (0,1).$ We are interested in the bivariate quantile region with coverage probability of  $\alpha$ meaning that the two univariate quantiles, need to be evaluated at $\frac{\alpha}{4}$ and $1-\frac{\alpha}{4}$, and we denote the corresponding confidence region using the univariate conditional quantiles as $CI^{V_1 \perp V_2  |\mathbf{U}}_{\alpha}$, i.e. 
	\begin{equation*}
		CI^{V_1 \perp V_2 |\mathbf{U} }_{\alpha}  \coloneq \left[q_{\frac{\alpha}{4},Dvine}^{V_1} (\mathbf{u}), q_{1-\frac{\alpha}{4},Dvine}^{V_1} (\mathbf{u})    \right] \times   \left[q_{\frac{\alpha}{4},Dvine}^{V_2} (\mathbf{u}), q_{1-\frac{\alpha}{4},Dvine}^{V_2} (\mathbf{u})    \right].
	\end{equation*}
	For case 4.)  we use the fitted Y-vine copula as discussed in Sections \ref{subsec:condquan} and \ref{bqrquantilescurvesnew}. The first row of Figure \ref{figquantilesunconditionall}, shows the bivariate  level  curves (solid lines) and quantile curves (dashed lines), where the first column is the unconditional case (case 1.) and 2.)), while second and third columns are conditional cases for two randomly chosen dates 02.07 and 21.08 (case 3.) and 4.)), respectively.  The adjusted level curves, the bivariate quantiles are estimated using the proposed method introduced in Section \ref{bqrquantilescurvesnew}. From the fitted pair copula (or vine copula), we simulate 10 000 data points from which the quantile curves  are estimated and the simulated points are shown as well. For all $\alpha$ levels, the  estimated empirical coverage probabilities $\hat{G}(\alpha)$ are evaluated so that we can estimate the adjustment $\hat{\beta}(\alpha)$ for the corresponding  quantile levels. 
	
	For the unconditional case, the estimated values for the adjustment to quantile curves (dashed lines) are $\hat{\beta}(0.25)= 0.14,$  $\hat{\beta}(0.75)= 0.59$ and $\hat{\beta}(0.05)= 0.02,$  $\hat{\beta}(0.95)= 0.86$. Using these values, we  construct the confidence regions $CI_{0.50}^{V_1,V_2 }$ and $CI_{0.90}^{V_1,V_2 }$, respectively. 
	The second row, first column shows the $CI^{V_1,  V_2 }_{0.50}$ (green  region, case 1.)) and $CI^{V_1 \perp V_2 }_{0.50}$ (gray  region, case 2.) ).  Last row, first column shows the $CI^{V_1,  V_2 }_{0.90}$  (red  region, case 1.)) and $CI^{V_1 \perp V_2 }_{0.90}$ (gray  region, case 2.)). The estimated empirical coverage probabilities and  the adjusted levels  for each case are given in the supplement in Table 4. The empirical coverage probability, based on the 10 000 samples, for the $CI_{0.50}^{V_1,V_2 }$ is 0.50, while the coverage probability below the level curve  at $\alpha = 0.25$ is 0.41, and below the level curve  at $\alpha = 0.75$ is 0.89. However, the empirical coverage probability for $CI_{0.50}^{V_1 \perp V_2 }$ is 0.65, thus we see the effect of falsely assuming independence between $V_1$ and $V_2$. The empirical coverage probability for $CI_{0.90}^{V_1,V_2 }$ is 0.90, while the coverage probability below the level curve  at $\alpha = 0.05$ is 0.10, below the level curve  at $\alpha = 0.95$ is 0.99.  However, the empirical coverage probability for $CI_{0.90}^{V_1 \perp V_2 }$ is 0.91. 
	
	For the conditional case, the estimated values for the adjustment to quantile curves (dashed lines) are given in the supplement in Tables 5 and 6. For the date 02.07.2017, the estimated values for the adjustment to quantile curves (dashed lines) are $\hat{\beta}(0.25)= 0.11,$  $\hat{\beta}(0.75)= 0.48$  and $\hat{\beta}(0.05)= 0.03,$  $\hat{\beta}(0.95)= 0.85$. The empirical coverage probability below the level curve  at $\alpha = 0.25$ is 0.52, below the level curve  at $\alpha = 0.75$ is 0.91 and below $\alpha = 0.05$ is 0.12, below $\alpha = 0.95$ is 0.99.
	For 21.08.2017, the estimated values for the adjustment to quantile curves (dashed lines) are $\beta(0.25)=0.16, $  $\beta(0.75)=0.66 $ and $\hat{\beta}(0.05)=0.025, $  $\hat{\beta}(0.95)=0.91. $ The empirical coverage probability below the level curve  at  $\alpha = 0.25$ is 0.37, below the $\alpha = 0.75$ is 0.83, below $\alpha = 0.05$ is 0.10 and below the level curve  at $\alpha = 0.95$ is 0.98. 
	The  second row shows $CI_{0.50}^{V_1,V_2| \mathbf{U}}$ (green, case 4.) ) and  $CI_{0.50}^{V_1 \perp V_2| \mathbf{U}}$ (gray, case 3.)), while the third row shows $CI_{0.90}^{V_1,V_2| \mathbf{U}}$ (red, case 4.)) and $CI_{0.90}^{V_1 \perp V_2| \mathbf{U}}$ (gray, case 3.)) for two different conditioning values for dates 02.07 and 21.08. While the confidence regions  $CI_{0.50}^{V_1,V_2| \mathbf{U}}$ have exact empirical coverage probabilities, for 02.07 the coverage probability of $CI_{0.50}^{V_1 \perp V_2| \mathbf{U}}$ is 0.17, and for 21.08 it is 0.25. Similarly, $CI_{0.90}^{V_1,V_2| \mathbf{U}}$ have exact empirical coverage probabilities, but for 02.07 the empirical coverage probability of $CI_{0.90}^{V_1 \perp V_2| \mathbf{U}}$ is 0.59, and for 21.08 it is 0.66. Thus, for the unconditional case, in case 2.) the empirical coverage probabilities are close to the expected value, but for the conditional case 3.) where we assume conditional independence the empirical coverage probabilities are much smaller than their expected value. Thus, in this case the empirical coverage probability is underestimated, leading to underestimation of the areas of interest, while in case 1.) and 4 .) the coverage probabilities are equal to the expected level.  
	In Figure \ref{figquantilesunconditionall}, all the  panels are given on the u-scale. However, using the transformations of the level curves between the u-scale and the x-scale, explained in Section \ref{BQR:bivariateuncondquant}, we also provide all the  level curves, quantile curves and the corresponding confidence regions on the transformed x-scale in the supplement in Figure 2. 
	
	There is obvious difference in the obtained shapes of confidence regions arising from bivariate quantiles (dependent responses) and the univariate quantiles based  regions (conditionally independent responses).  While the bivariate confidence regions are areas determined by two level curves, the regions obtained by the  univariate confidence intervals are bound to be rectangles. Also, for 21.08.2017  the univariate quantiles based confidence regions are partially contained in the  bivariate confidence regions obtained from the Y-vine regression and are of much smaller empirical coverage probabilities. So, there is many points that are excluded from the confidence region constructed from the univariate quantiles. For 02.07.2017, there is a very small overlap between  $CI^{V_1 \perp V_2| \mathbf{U} }_{0.50}$ and  the bivariate confidence regions obtained from the Y-vine regression $CI_{0.50}^{V_1,V_2 | \mathbf{U}}$, while  the $CI^{V_1 \perp V_2| \mathbf{U} }_{0.90}$  is a partially contained in the bivariate  $CI_{0.90}^{V_1,V_2 | \mathbf{U}}$. However, the empirical coverage probabilities are underestimated.  All in all, the univariate conditional quantiles based confidence regions  have not exact empirical coverage probabilities, are too small in area  and don't capture any joint conditional dependence between the responses. However, the  bivariate confidence regions we suggest have the expected empirical coverage probabilities and allow for  the dependence and the multidimensional nature of the problem. This example shows that even a small conditional dependence (estimated Kendall's $\hat{\tau}$ of $C_{V_1,V_2 | \mathbf{U}}= 0.09$) can make confidence regions based on conditional independence invalid (case 3.)).

	\section{Conclusions and outlook}\label{sc:conclussion}
	We studied the problem of bivariate (unconditional and conditional) quantiles using a  flexible class of models, vine copulas, allowing for asymmetric tail dependence. They are multivariate distributions constructed from bivariate blocks (pair copulas) using conditioning. We develop a  novel vine tree structure, the Y-vine tree structure, that is suitable for a regression problem containing  bivariate response variables. Also, a forward selection  of predictors procedure gives the best suitable fitted Y-vine. In addition, the Y-vine tree structure enables an easy way of obtaining the bivariate conditional density. We propose a numerical procedure for the determination of the level curves of a bivariate (conditional) distribution, and propose a simulation based adjustment of the level $\alpha$ of the level curve resulting in quantile curve with correct probability coverage. This way a  joint  analysis of the dependence structure of the responses given the predictors is  possible. This is a significant result especially when dealing with responses that are not (conditionally) independent. We develop a prediction method for bivariate responses given the predictor values using the Y-vine quantile regression. This enables us to not only jointly model, but also predict bivariate response conditional quantiles. Additionally, simulation from a Y-vine model for fixed predictor values is available.  We apply our proposed model on a real life data set containing a bivariate response, minimal and maximal daily temperature. We analyse the data with our new  approach for dependent responses and provide a joint vine copula model for the two responses. For this example, we  highlight the advantages of our joint bivariate modelling over independent and conditionally independent response modelling approaches with vine copulas.

	For future possible applications we think of adding a spatial and/or temporal component to our Y-vine based quantile regression. It would be interesting to see how the response dependence changes when the  spatial and/or temporal dependence component is also accounted for, but that is out of the scope of this paper. The standard lack of ability of  copula based models to include discrete variables is also an ongoing research topic. Some results from the univariate vine based quantile regression are available \citep{schallhorn2017d}, but it becomes even more complicated in our case, because of the multidimensionality of the problem and the numerical method for obtaining the bivariate quantile sets. Also, applications of different vine structures and variable selection methods, and subsequent comparisons of the performance, are left for further investigation and are expected to be heavily data specific problems.
	In addition, we can use the Y-vine tree structure for  testing of conditional independence between two variables given a set of conditioning variables. The Y-vines provide a  symmetric treatment of the two variables whose conditional independence is being tested. Using this way of testing for conditional independence we do not need joint normality nor rely on asymptotic normality results.
	A similar approach was proposed in \cite{bauer2016pair} using R-vines, 
	for non-Gaussian conditional independence testing  in continuous Bayesian networks. However, their approach needed, possibly high dimensional, integration for determining the required conditional distribution function and thus, is not applicable for large network problems. In contrast, we expect our approach to remain tractable in large networks.
	
	\section*{Acknowledgments}
	This work was supported by the Deutsche Forschungsgemeinschaft[DFG CZ 86/6-1]. We thank the anonymous referees and the associate editor for the various useful suggestions that helped improve the manuscript. Declarations of interest: none. 
	
	\newpage
	\section*{Appendix}
	\appendix

	\section{Appendix A: Algorithms for numerical estimation}\label{appendix:algs}
	\begin{algorithm}
		\SetAlgoLined
		\KwIn{\textbf{m} - granularity parameter (default = 1000), \\
			\hspace{1.35cm}\textbf{eq} - function based on which $C\left(u,v\right)$ is to be evaluated,\\
			\hspace{1.35cm}$err$ - accuracy of algorithm,\\
			\hspace{1.35cm}$\alpha$ - alpha level}
		\textbf{Initialization:} 
		\begin{equation*}
			\begin{aligned}
				&	M = \bigcup_{i=1}^m \left\lbrace \frac{i}{m}\right\rbrace, \\
				&L =	\left\lbrace line\left(\left(0,0\right), \left(q_1,q_2\right)\right)\right\rbrace, \;\; \left(q_1,q_2\right) \in \left\lbrace\left(w_i,1\right) \vert \forall w_i \in M \right\rbrace \cup \left\lbrace \left(1,w_i\right) \vert\forall w_i \in M  \right\rbrace ,\\
				& Points = \emptyset.
			\end{aligned}
		\end{equation*}
		\For{$l_s \in L$}{
			\eIf{
				$C\left(q_1,q_2\right) \geq \alpha$
			}
			{
				\begin{equation*}
					\begin{aligned}
						&point = \textrm{BinaryLineSearch}\left(l_s, \mathbf{eq}, err, \alpha\right) \\ 
						&Points = Points \cup point
					\end{aligned}
				\end{equation*}
			}
			{	
				\begin{equation*}
					\begin{aligned}
						&Points = Points 
					\end{aligned}
				\end{equation*}
			}
		}
		\Return $Points$
		\caption{PseudoInverse}
		\label{LineAlg}
	\end{algorithm}

	\begin{algorithm}
		\SetAlgoLined
		\KwIn{\textbf{l} - line defined by two coordinates, \\
			\hspace{1.35cm}\textbf{eq} - function based on which $C\left(u,v\right)$ is to be evaluated,\\
			\hspace{1.35cm}$err$ - accuracy of algorithm,\\
			\hspace{1.35cm}$\alpha$ - alpha level}
		\textbf{Initialization:} \\
		Introduce notation $\mathbf{l} = line\left(p_{start}=\left(0,0\right), p_{end}=\left(w_i,1\right)\right)$.\\
		\begin{equation*}
			\begin{aligned}
				&evl = \mathbf{eq}\left(\frac{p_{start} + p_{end}}{2}\right) \\
				&diff = \alpha - evl
			\end{aligned}
		\end{equation*}
		\While{$diff > err$}{
			\eIf{$diff > 0$}{$p_{start} = \frac{p_{start} + p_{end}}{2}$}{$p_{end} = \frac{p_{start} + p_{end}}{2}$}
			$evl = \mathbf{eq}\left( \frac{p_{start} + p_{end}}{2}\right)$}
		\Return $\frac{p_{start} + p_{end}}{2}$
		\caption{BinaryLineSearch}
		\label{BinAlg}
	\end{algorithm}

	\section{Appendix B: Proofs}
	\subsection{Proof of Proposition \ref{prop:cond}} \label{app:proposition3}
	\begin{proof}
		\begin{equation*}
			\small
			\begin{aligned}
				F_{Y_1,Y_2| \mathbf{X}} \left(y_1,y_2|\mathbf{x}  \right) & =     
				\int_{-\infty}^{y_1} \int_{-\infty}^{y_2}   f_{Y_1,Y_2|\mathbf{X}} \left(y_1',y_2'|\mathbf{x}  \right)   d y_2' \;  d y_1'\\
				& = \int_{-\infty}^{y_1} \int_{-\infty}^{y_2}       \frac{f_{Y_1,Y_2, \mathbf{X}} \left(y_1',y_2',\mathbf{x}  \right)  }{f_{\mathbf{X}} \left(  \mathbf{x}   \right)}	dy_2' \; dy_1'\\
				& = \frac{1}{f_{\mathbf{X}} \left(  \mathbf{x}   \right)} \int_{-\infty}^{y_1} \int_{-\infty}^{y_2}    \frac{\partial^{p+2}}{\partial y_1 \;\partial y_2 \; \partial x_1 \ldots \partial x_p} 	F_{Y_1,Y_2, \mathbf{X}} \left(y_1,y_2,\mathbf{x}  \right)  \Big|_{y_1=y_1',y_2=y_2'}  dy_2' \; dy_1' \\
				& = \frac{1}{f_{\mathbf{X}} \left(  \mathbf{x}   \right)} \cdot \frac{\partial^p}{\partial x_1 \ldots \partial x_p} F_{Y_1,Y_2, \mathbf{X}} \left(y_1,y_2,\mathbf{x}  \right)\\
				( by \; Sklar's \; theorem) \; \; &  = \frac{1}{f_{\mathbf{X}} \left(  \mathbf{x}   \right)} \cdot \frac{\partial^p}{\partial x_1 \ldots \partial x_p} C_{V_1,V_2,\mathbf{U}} \left(F_{Y_1}(y_1),F_{Y_2}(y_2),  F_{X_1}(x_1),\ldots, F_{X_p}(x_p)  \right)\\
				& = \frac{1}{f_{\mathbf{X}} \left(  \mathbf{x}   \right)} \cdot \frac{\partial^p}{\partial u_1 \ldots \partial u_p} C_{V_1,V_2,\mathbf{U}} \left(  v_1, v_2, u_1, \ldots , u_p \right) \Big|_{v_j=F_{Y_j}(y_j),u_i=F_{X_i}(x_i)  } \frac{\partial u_1 \ldots \partial u_p}{\partial x_1 \ldots \partial x_p} \\
				\left( \frac{\partial u_1 \ldots \partial u_p}{\partial x_1 \ldots \partial x_p}= \prod_{i=1}^{p} f_{X_i} \left( x_i   \right) \right) \; \; \;\; \;  & = \frac{\partial^p}{\partial u_1 \ldots \partial u_p} C_{V_1,V_2,\mathbf{U}} \left(  v_1, v_2, u_1, \ldots , u_p \right) \Big|_{v_j=F_{Y_j}(y_j),u_i=F_{X_i}(x_i)  } 			 \cdot \frac{\prod_{i=1}^{p} f_{X_i} \left( x_i   \right) }{f_{\mathbf{X}} \left(  \mathbf{x}   \right)} \\
				\left( \frac{\partial u_1 \ldots \partial u_p}{\partial x_1 \ldots \partial x_p}= \prod_{i=1}^{p} f_{X_i} \left( x_i   \right) \right) \; \; \;\; \;  & = \frac{\partial^p}{\partial u_1 \ldots \partial u_p} C_{V_1,V_2,\mathbf{U}} \left(  v_1, v_2, u_1, \ldots , u_p \right) \Big|_{v_j=F_{Y_j}(y_j),u_i=F_{X_i}(x_i)  } 		\cdot \frac{1}{c_{\mathbf{U}} \left(  \mathbf{u}   \right)} \\
				& = C_{V_1,V_2|\mathbf{U}} \left(F_{Y_1}(y_1), F_{Y_2}(y_2)| F_{X_1}(x_1),\ldots,  F_{X_p}(x_p) \right) ,
			\end{aligned}
		\end{equation*}	
			\noindent where $C_{V_1,V_2|\mathbf{U}} \left(F_{Y_1}(y_1), F_{Y_2}(y_2)| F_{X_1}(x_1),\ldots,  F_{X_p}(x_p) \right)$ or shortly  $ C_{V_1,V_2|\mathbf{U}} \left(v_1,v_2|\mathbf{u}  \right) $ is the conditional distribution of $V_1,V_2$ given $\mathbf{U}= \mathbf{u}$ and the joint copula distribution of $Y_1,Y_2, \mathbf{X}$ is denoted by $C_{V_1,V_2,\mathbf{U}}$. 
	\end{proof}

	\subsection{Proof of Theorem \ref{thm:density}} \label{app:proofcond}
	\begin{proof}
		By definition of a conditional density it follows that $f_{Y_1,Y_2\vert \mathbf{X}} = \frac{    f_{Y_1,Y_2, \mathbf{X}}}{    f_{\mathbf{X}}}.$
		The numerator $f_{Y_1,Y_2,\mathbf{X}}$  is expressed in Equation~\eqref{eq:jointdens}, and we  need to derive the denominator  $f_{ \mathbf{X}}$ in terms of copulas.
		Consider the part of the  Y-vine tree sequence  after removing the PITs of the responses $V_1$ and $V_2$, i.e., the tree sequence consisting of only the PITs of the predictors $(U_1,\ldots ,U_p)^T$. By  definition of the Y-vine tree structure, the predictors are arranged in a D-vine tree sequence with a specific order. Thus, the density of a D-vine with this given order (see more in \cite{czado2010pair})  can be expressed as
		\begin{equation}\label{eq:proofdensity1}
			\small
			\begin{split}
				f_{\mathbf{X}}  \left(\mathbf{x} \right) = &  \prod_{k=1}^{p}f_{X_{k}} (x_{k}) \cdot
				\prod_{k=1} ^{p-1}\prod_{i=1}^{p-k} c_{U_{i},U_{i+k};\mathbf{U}_{i+1:i+k-1}} \left(  F_{X_{i}|\mathbf{X}_{i+1:i+k-1}} (x_{i}|\mathbf{x}_{i+1:i+k-1}) , \right. \\
				& \hspace{5.3cm} \left.  F_{X_{i+k}|\mathbf{X}_{i+1: i+k-1}} (\mathbf{x}_{i+k}|x_{i+1:i+k-1})	
				\right) . 
			\end{split}
		\end{equation}
		
		\noindent	Canceling out all common terms in the expansions of the numerator and the denominator, given in  Equation~\eqref{eq:jointdens} and ~\eqref{eq:proofdensity1} respectively, we are left with the expression in  Equation~\eqref{eq:densitycond}.	
		All the required copulas in Equation~\eqref{eq:densitycond} are already derived in the Y-vine tree sequence, $c_{V_j,U_i; \mathbf{U}_{1:i-1}} \in \mathcal{B}\left(\mathcal{V}\right)$ for $j=1,2, \; i=1,\ldots,p$ and  $c_{V_1,V_2; \mathbf{U}} \in \mathcal{B}\left(\mathcal{V}\right)$ (these copulas can be seen as the  copulas on the furthest left side  of each tree in Figure \ref{figure:vine}).
		
	\end{proof}
	
	\subsection{Proof of Corollary \ref{corollary1}} \label{app:proofcorr}
	\begin{proof}
		Let's prove  part $\left. a.\right)$ for $j=1$. Due to symmetry the same proof  follows for $j=2$.
		By definition of a conditional density it follows that
		$ f_{Y_1\vert \mathbf{X}} = \frac{ f_{Y_1, \mathbf{X}} }{ f_{\mathbf{X}} }.$
		The denominator is expressed in  Equation~\eqref{eq:proofdensity1}, while the numerator needs to be expanded. Consider the random vector $(V_1,\mathbf{U})^T$  in the tree sequence of the Y-vine, i.e. remove the node of the PIT of the response $V_2$ from the first tree $T_1$ and all the nodes in the further trees that will  disappear by removing the variable $V_2$. By definition of the Y-vine,  the variables $(V_1, U_1,\ldots ,U_p)$ are arranged in a D-vine tree sequence with a specific order. Thus, the density of a D-vine with this given order (see more in \cite{czado2010pair})  is given as 
		\begin{equation}\label{eq:conddensy3}
			\small
			\begin{split}
				f_{ Y_1,\mathbf{X}} \left(y_1, y_2, \mathbf{x}\right) = & \prod_{k=1}^{p}f_{X_{k}} (x_{k}) \cdot f_{Y_{1}} (y_{1}) 	
				\cdot 
				\prod_{k=1} ^{p-1}    \left[  \prod_{i=1}^{p-k} c_{U_{i},U_{i+k};\mathbf{U}_{i+1:i+k-1}} \left(  F_{X_{i}|\mathbf{X}_{i+1:i+k-1}} (x_{i}|\mathbf{x}_{i+1:i+k-1}) , \right. \right. \\
				&  \hspace{5cm} \left.  F_{X_{i+k}|\mathbf{X}_{i+1:i+k-1}} (x_{i+k}|\mathbf{x}_{i+1:i+k-1})	
				\right) \Bigg] \\ 
				&\prod_{i=1}^p  \left[   c_{V_1,U_i; \mathbf{U}_{1:i-1}}  \left( F_{Y_1|\mathbf{X}_{1:i-1}}  \left( y_1|\mathbf{x}_{1:i-1}  \right), \right. \right.
				\left. \left.  F_{\mathbf{X}_i|X_{1:i-1} } \left( x_i|\mathbf{x}_{1:i-1}  \right)    \right) \right]	  .
			\end{split}
		\end{equation}
		Cancelling common terms of the numerator, Equation~\eqref{eq:conddensy3}, and the denominator, Equation~\eqref{eq:proofdensity1}, we are left with  Equation~\eqref{eq:densitycond5} for $j=1.$
		Now let's prove  part $\left. b.\right)$ for $(j,k) = (1,2).$  Due to symmetry the same proof  follows for $(j,k) = (2,1)$.
		Use that
		$	f_{Y_2\vert \mathbf{X}, Y_1} = \frac{ f_{Y_1,Y_2, \mathbf{X}} }{ f_{Y_1,\mathbf{X}} }$ holds.
		The numerator  is expressed in  Equation~\eqref{eq:jointdens}, and the denominator is expressed as in the part  $\left. a.\right)$ Equation~\eqref{eq:conddensy3}. Considering the associated ratio and cancelling all common terms, we are left with  Equation~\eqref{eq:densitycond4} for $(j,k) = (1,2).$
		Again, all the required copulas  are already derived in the Y-vine tree sequence, $c_{V_j,U_i; \mathbf{U}_{1:i-1}} \in \mathcal{B}\left(\mathcal{V}\right)$ for $j=1,2 \; i=1,\ldots,p$ and  $c_{V_1,V_2; \mathbf{U}} \in \mathcal{B}\left(\mathcal{V}\right)$, which means we don't require any additional calculations.
	\end{proof}
	
	\section{Appendix C: Theoretical  level curves of  bivariate copula distributions}\label{app:3dplot}
	\begin{figure}[H]
		\includegraphics[width = 0.85\textwidth, height=0.8\textheight]{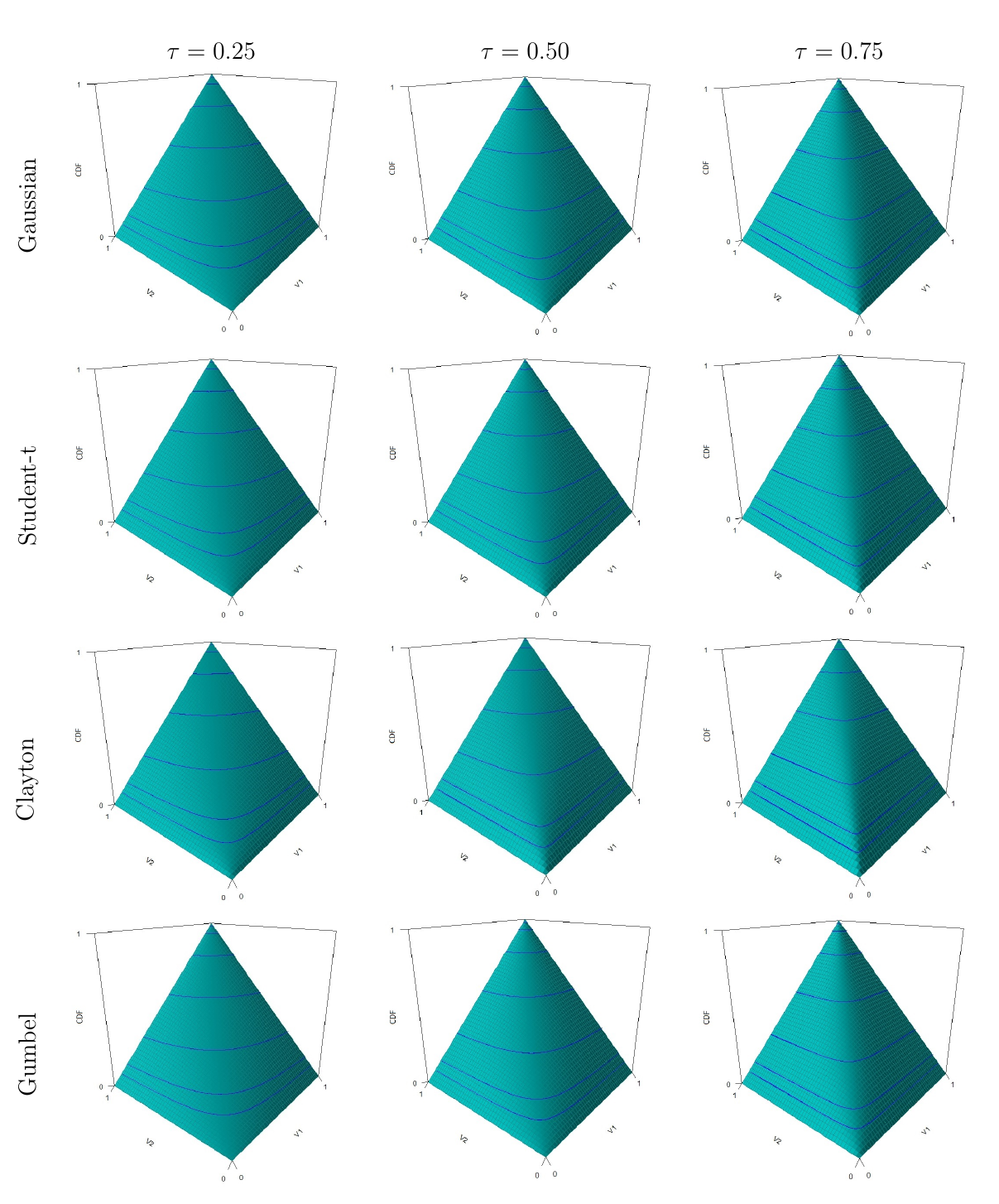}
		\centering
		\caption{A 3-dimensional plot of bivariate copula distribution with  theoretical level curves for $\alpha = 0.05,0.1,0.25,0.5,0.75,0.90,0.95$. Shown are Gaussian, Student-t ($df=5$), Clayton and Gumbel copulas (top to bottom) and $\tau = 0.25,0.5,0.75$ (left to right).}
		\label{fig:3dplot}
	\end{figure}

	\section{Appendix D: Pseudo-code for the bivariate vine based  regression algorithm} \label{app:code}
	\begin{algorithm}[H]
		\SetAlgoLined
		\KwIn{Data set $\mathbf{y}_n=\left(y_1^n,y_2^n\right)^T ,\; \mathbf{x}_n=\left(x_1^n,\ldots ,x_p^n\right)^T$, for $n=1,\ldots ,N$}
	\textbf{Initialization:} \\
	$acll_0=0$\\
	$NotChosenIndex=\left\lbrace 1, \ldots, p  \right\rbrace $\\
	$ChosenIndex=\emptyset$ \\
	\begin{enumerate}
		\item Estimate marginals   $F_{Y_j}, F_{X_i},\; j =1,2,\; i=1,\ldots,p,$  by a univariate kernel density estimator, implemented in \texttt{kde1d}.
		\item Obtain pseudo copula data  $u_i^n \coloneqq \hat{F}_{X_i}\left(x_i^n\right)$ for $i=1, \ldots, p$, $v_1^n \coloneqq \hat{F}_{Y_1}\left(y_1^n\right)$ and $v_2^n \coloneqq \hat{F}_{Y_2}\left(y_2^n\right)$.
	\end{enumerate} 
	\For{$j = 1,\ldots ,p$}
	{
		Calculate $acll_1^j$ as 
		$$acll_1^j = cll_0 + \ell(c_{V_1U_i})+ \ell(c_{V_2U_i}) + \ell(c_{V_1V_2;U_i}) $$
	}
	$r_1 \coloneqq \argmax_{j=1,\ldots, p} acll_1^j$ \\
	$NotChosenIndex = NotChosenIndex \setminus \left\lbrace r_1 \right\rbrace$\\
	$ChosenIndex = ChosenIndex \bigcup \left\lbrace r_1 \right\rbrace$\\
	$acll_1\coloneqq acll_1^{r_1}$\\
	\For{$k = 2,\ldots ,p$}
	{
		\For{$t \in NotChosenIndex$}
		{
			Calculate $acll_k^t$ as 
			\begin{equation*}
				acll_k^t =  acll_{k-1} + \ell(c_{V_1U_t;U_{r_1},\ldots, U_{r_{k-1}}})+ \ell(c_{V_2U_t;U_{r_1},\ldots ,U_{r_{k-1}}})
			\end{equation*}
		}
		$r_k \coloneqq \argmax_{t\in NotChosenIndex} acll_k^t$ \\
		$NotChosenIndex = NotChosenIndex \setminus \left\lbrace r_k \right\rbrace$\\
		$ChosenIndex = ChosenIndex \bigcup \left\lbrace r_k \right\rbrace$\\
		$acll_k\coloneqq acll_k^{r_k}$\\
	}
	\Return $ChosenIndex= \left\lbrace r_1,\ldots, r_p\right\rbrace$, i.e.  order of the predictors which uniquely determines the fitted bivariate  regression model. 
	\caption{Bivariate vine based  regression algorithm}
	\label{BagReg}
\end{algorithm}

\bibliographystyle{apalike}
\bibliography{References}
\end{document}


	{	\renewcommand*{\thefootnote}{\fnsymbol{footnote}}
		\title{\textbf{\sffamily   Bivariate vine copula based regression, bivariate level and quantile curves: supplementary material}}
		
		\date{\small \today}
		\newcounter{savecntr1}
		\newcounter{restorecntr1}
		\newcounter{savecntr2}
		\newcounter{restorecntr2}

		\author{Marija Tepegjozova\setcounter{savecntr1}{\value{footnote}}\thanks{Department of Mathematics, Technische Universit{\"a}t M{\"u}nchen, Boltzmannstra{\ss}e 3, 85748 Garching, Germany (email: \href{mailto:m.tepegjozova@tum.de}{m.tepegjozova@tum.de} (corresponding author))} $\;$ and
			Claudia Czado\setcounter{savecntr2}{\value{footnote}}\thanks{Department of Mathematics and Munich Data Science Institute, Technische Universit{\"a}t M{\"u}nchen, Boltzmannstra{\ss}e 3, 85748, Garching, Germany (email: \href{mailto:cczado@ma.tum.de}{cczado@ma.tum.de})}
		}
		\maketitle
	}

	\section{ Theoretical and estimated unconditional bivariate level curves}\label{appendix2}
	\subsection{Clayton copula} \label{app:clayton}
	The distribution function of a bivariate Clayton copula with parameter $\theta$ is  
	\begin{equation*}
		C_{V_1,V_2}\left(v_1,v_2;\theta\right) = \left(v_1^{-\theta} + v_2^{-\theta}-1\right)^{-1/\theta}.
	\end{equation*}
	By solving 
	$
	\left(v_1^{\theta} + v_2^{\theta}-1\right)^{-1/\theta} = \alpha
	$
	for $v_2$, we obtain the $\alpha$ bivariate level curve as
	\begin{equation*}
		Q^V_{\alpha} \coloneqq \left\lbrace \left(v_1,\left(\alpha^{-\theta} - v_1^{-\theta} + 1\right)^{-1/\theta}\right) \; \Bigg| \; \forall \; v_1\in \left[0,1\right] \right\rbrace.
	\end{equation*}
	\subsection{Gumbel copula} 
	\noindent The distribution function of a bivariate Gumbel copula with parameter $\theta$ is 
	\begin{equation*}
		C_{V_1,V_2}\left(v_1,v_2;\theta\right) = \exp\left\lbrace - \left[ \left(- \ln v_1\right)^{\theta} + \left(- \ln v_2\right)^{\theta}\right]^{1/\theta} \right\rbrace.
	\end{equation*}
	By solving 
	$	\exp\left\lbrace - \left[ \left(- \ln v_1\right)^{\theta} + \left(- \ln v_2\right)^{\theta}\right]^{1/\theta} \right\rbrace = \alpha
	$ for $v_2,$ we obtain the $\alpha$ bivariate level curve as
	\begin{equation*} 
		Q^V_{\alpha} \coloneqq \left\lbrace \left(v_1,\exp\left\lbrace - \left[ \left(- \ln \alpha\right)^{\theta} - \left(- \ln v_1\right)^{\theta}\right]^{1/\theta} \right\rbrace\right) \; \Bigg| \; \forall \; v_1\in \left[0,1\right] \right\rbrace.
	\end{equation*}
	\subsection{Gaussian copula} 
	In contrast to the  Archimedean copulas 
	as Clayton and Gumbel, for which there is a closed form solution of the level curve function, for the elliptical copulas, such as Gaussian and Student-t copula, there is no closed form solution. Thus, we use a numerical procedure to derive the theoretical level curves. \\
	The distribution function of the Gaussian pair copula is
	\begin{equation*}
		\begin{aligned}
			C_{V_1,V_2}\left(v_1,v_2;\theta\right) & = \Phi_2 \left( \Phi^{-1}\left(v_1\right),\Phi^{-1}\left(v_2\right)\right) \\
			& = \int_{-\infty}^{\Phi^{-1}\left(v_1\right)} \int_{-\infty}^{\Phi^{-1}\left(v_2\right)} \frac{1}{2\pi \sqrt{1 - \theta ^2}} \exp\left(-\frac{a^2-2\theta ab + b^2}{2\left(1- \theta^2\right) }\right) da\; db,
		\end{aligned}
	\end{equation*}
	where $\Phi$ and $\Phi_2$ are the univariate and bivariate standard normal distribution functions, respectively.
	As  already stated, the equation 
	\begin{equation*}
		\int_{-\infty}^{\Phi^{-1}\left(v_1\right)} \int_{-\infty}^{\Phi^{-1}\left(v_2\right)} \frac{1}{2\pi \sqrt{1 - \theta ^2}} \exp\left(-\frac{a^2-2\theta ab + b^2}{2\left(1- \theta^2\right) }\right) da\; db = \alpha
	\end{equation*}
	does not have a closed form solution. Thus, we evaluate $C_{V_1,V_2}$ using the integral of its h-function given as 
	\begin{equation*}
		h_{V_1\vert V_2}\left(\left(v_1,v_2;\theta\right)\right) = \Phi \left(\frac{\Phi^{-1}\left(v_1\right) - \theta\Phi^{-1}\left(v_2\right)}{\sqrt{1-\theta^2}} \right).
	\end{equation*}
	The distribution function is then evaluated at the point $\left(\tilde{v_1},\tilde{v_2}\right)$ as 
	\begin{equation}\label{gaussC}
		C_{V_1,V_2}\left(\tilde{v_1},\tilde{v_2};\theta\right) = \int_{0}^{\tilde{v_2}} h_{V_1\vert V_2}\left(\left(\tilde{v_1},v_2;\theta\right)\right) dv_2.
	\end{equation}
	Finally, the theoretical bivariate level curve is derived using the numerical evaluation defined in Section 3.4 of the manuscript and  Equation \eqref{gaussC}.
	\subsection{Student-t copula} \label{app:studentt}
	Similarly as with the Gaussian copula, the Equation 
	$	C_{V_1,V_2}\left(v_1,v_2;\boldsymbol{\theta} = \left(\theta, df\right)\right) = \alpha $
	does not have a closed form solution. Again, we evaluate $C_{V_1,V_2}$ using the integral of its h-function given as  
	\begin{equation*}
		h_{V_1\vert V_2}\left(\left(v_1,v_2;\boldsymbol{\theta} = \left(\theta, df\right)\right)\right) = t_{df+1} \left(\frac{t_{df}^{-1}\left(v_1\right) - \theta t_{df}^{-1}\left(v_2\right)}{\sqrt{\frac{\left(df +  t_{df}^{-1}\left(v_2\right)^2\right)\left(1-\theta^2\right)}{df +1}}}\right) ,
	\end{equation*}
	where $t_k$ is the distribution function of the Student-t distribution with $k$ degrees of freedom. The distribution function is then evaluated at the point $\left(\tilde{v_1},\tilde{v_2}\right)$ as 
	\begin{equation}	\label{StudentC}
		C_{V_1,V_2}\left(\tilde{v_1},\tilde{v_2};\theta\right) = \int_{0}^{\tilde{v_2}} h_{V_1\vert V_2}\left(\left(\tilde{v_1},v_2;\theta\right)\right) dv_2.
	\end{equation}
	The theoretical bivariate level curve is derived using the numerical evaluation defined in Section 3.4 and  Equation \eqref{StudentC}.
	
	\subsection{Estimated level curves}\label{app:estimatedcurves}
	Let $ \left\lbrace \left(v_1^i,v_2^i\right) \right\rbrace_{i=1}^n$ be a set of $n$ points randomly drawn from a bivariate copula distribution. Given an estimated parameter $\hat{\theta}$ (together with family) obtained from this set of points we propose to evaluate $\hat{C}_{V_1,V_2}$ at a point $\left(\tilde{v}_1, \tilde{v}_2\right)$ as 
	\begin{equation}\label{estimatedQuantUncond}
		\hat{C}_{V_1,V_2}\left(\tilde{v}_1,\tilde{v}_2\right) = \int_{0}^{\tilde{v}_1} \hat{C}_{V_2\vert V_1}\left(\tilde{v}_2\vert v_1'\right) dv_1'.
	\end{equation}
	The difference between the estimated and the theoretical level curves for copulas for which the numerical inverse procedure is used is that  in the theoretical case we use the theoretical h-function of a copula, while in the estimated case we use the estimated one. Basically, from the simulated data, we estimate a pair-copula, which has an h-function, and that estimated h-function is being used. The estimated bivariate level curves are  obtained using the numerical evaluation defined in Section 3.4 and  Equation \eqref{estimatedQuantUncond}.
	%
	
	\section{Y-vine copula is a  valid  regular vine copula }
	In this section we prove the following:
	\begin{proposition}\label{thm:vinetree} The $Y$-vine tree sequence from  Definition 4.1, satisfies the regular vine tree sequence conditions $(i)$-$(iii)$ from Section 2 and thus, represents a valid regular vine tree sequence.
	\end{proposition}
	\begin{proof} 
		We  prove that  a $Y$-vine tree sequence, $\left\lbrace T_1,\ldots ,T_{p+1}\right\rbrace$, satisfies conditions $(i)$-$(iii)$ from Section 2.
		The first condition $(i)$ is trivial and follows by definition of $T_1$. 
		\noindent The next condition requires that $N_k=E_{k-1}\; \forall\; k\geq 2.$ For $k=2$, $N_2= \left\lbrace V_1U_1, V_2U_1, U_1U_2, \ldots ,U_{p-1}U_p \right\rbrace = E_1$ follows directly from Definition 4.1.
		To prove the statement for $k>2$, we start with the edge set of tree $T_{k-1}$,  $E_{k-1}$ given as 
		\begin{equation*}
			\small
			\begin{split}
				E_{k-1} =& \bigcup_{j=1,2}\left\lbrace\left(V_jU_{k-2};\mathbf{U}_{1:k-3} ,\; \; U_1U_{k-1}; \mathbf{U}_{2:k-2} \right)\right\rbrace \\ 
				&\bigcup_{i=1}^{p-k+1} \left\lbrace\left( U_iU_{i+k-2}; \mathbf{U}_{i+1:i+k-3 }  ,\; \; U_{i+1}U_{i+k-1}; \mathbf{U}_{i+2:i+k-2} \right)\right\rbrace .
			\end{split}
		\end{equation*}	
		%
		%
		\noindent  Edge $\left(V_jU_{k-2}; \mathbf{U}_{1:k-3} ,\; \; U_1U_{k-1};\mathbf{U}_{2:k-2} \right)$  is associated with  node $V_jU_{k-1};\mathbf{U}_{1:k-2}$ in $T_k$ for $j=1,2$ and  edge $\left( U_iU_{i+k-2};  \mathbf{U}_{i+1:i+k-3} ,\; \; U_{i+1}U_{i+k-1}; \mathbf{U}_{i+2:i+k-2} \right)$ is associated with  node $U_{i}U_{i+k-1};  \mathbf{U}_{i+1:i+k-2} $ for $i=1,,\ldots,  p-k+1$ in $T_k$. 
		Therefore, by  Definition 4.1,  $N_k=E_{k-1}$ holds for all $k$ in the Y-vine tree sequence.
		The last condition,  the proximity condition, states that for $k\geq 2$ two nodes can be connected in $T_k$ only if the corresponding edges in the previous tree $T_{k-1}$ share a common node. Consider the part of the tree sequence that only contains the predictors $(X_1, \ldots, X_p)$. By definition of the Y-vine tree sequence, the predictors are arranged in a D-vine tree sequence, which is a known regular vine tree sequence subset, implying that for the nodes containing only the predictors the proximity condition is satisfied. So, we consider the remaining nodes that contain the response variables in the conditioned set and the node that connects them to the D-vine of the predictors. 
		For $T_2$,  nodes $V_1U_1$, $V_2U_1$ are both connected to $U_1U_2$. For $V_jU_1, \; j=1,2$ the corresponding edge in $T_1$ is $(V_jU_1)$ which shares the node $U_1$ with the corresponding edge of node  $U_1U_2$, edge  $(U_1U_2)$.
		For $k>2$ in $T_k$ the nodes $V_1U_{k-1};\mathbf{U}_{1:k-2} $ and $V_2U_{k-1}; \mathbf{U}_{1:k-2} $ are connected to $U_1U_k;\mathbf{U}_{2:k-1}  $. In $T_{k-1}$ the corresponding edge of node $V_jU_{k-1}; \mathbf{U}_{1:k-2} $  for $j=1,2$ is the edge $(V_jU_{k-2}; \mathbf{U}_{1:k-3} , \; \; U_1U_{k-2}; \mathbf{U}_{1:k-3})$ and for node $U_1U_k;  \mathbf{U}_{2:k-1}  $ the corresponding edge is $(U_1U_{k-2}; \mathbf{U}_{2:k-3} ,\; \; U_2U_k; \mathbf{U}_{3:k-1})$. They share a common node $U_1U_{k-2}; \mathbf{U}_{2:k-3} $ in  $T_{k-1}$, thus the proximity condition is satisfied.
		
	\end{proof}
	\section{Data analysis supplementary material}
	In Table~\ref{vardescription} given is a variable description, the unit of measurement and the range of possible values for the 2 predictors Next\_Tmax or T\_max,  Next\_Tmin or T\_min and the 13 possible continuous predictors we consider.
	
\begin{figure}[h]
							\includegraphics[width = \textwidth]{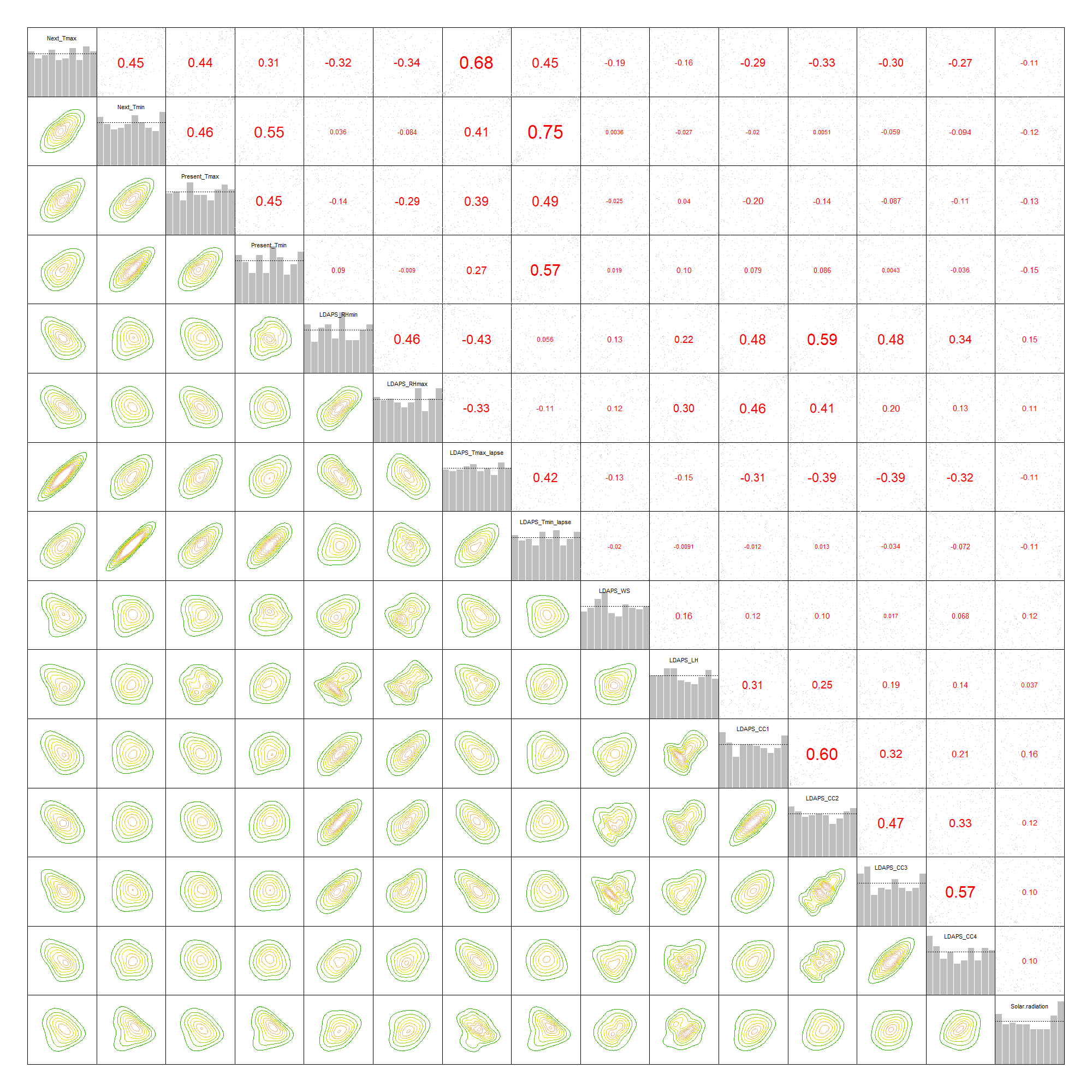}
							\centering
							\caption{Lower diagonal: normalized contour plots, upper diagonal: pairwise scatter plots with the associated empirical Kendall's $\hat{\tau}$ values and on the diagonal: histograms of the u-data.}
							\label{fig:pairsplot}
						\end{figure}

	In Figure~\ref{fig:pairsplot} shown are the empirical normalized contour plots for pairs of variables from the training set.
	On the lower diagonal, given are the normalized contour plots, where any deviance from elliptical shapes indicates a non-Gaussian dependence structure in the data. This is the case in almost all normalized contour plots and it supports our non-Gaussian approach with flexible vine copulas over any other modeling approach that assumes Gaussianity. On the upper diagonal, we see a scatter plot of the estimated u-data together with the corresponding empirical Kendall's $\hat{\tau}$. 
	\begin{landscape}
		\begin{table}[h]
			\centering
			\caption{Variable description, the unit of measurement and the range of possible values the considered variables can take.}
				\begin{tabular}{ l l c  } 
					Variable name& Description(unit) & Range \\		
					\hline
					Next\_Tmax & The next-day maximum air temperature $(^\circ C)$ & 17.4 to 38.9\\
					Next\_Tmin & The next-day minimum air temperature $(^\circ C)$ & 11.3 to 29.8\\
					Present\_Tmax & Maximum air temperature between 0 and 21 h on the present day $(^\circ C)$ & 20 to 37.6\\
					Present\_Tmin & Minimum air temperature between 0 and 21 h on the present day $(^\circ C)$& 11.3 to 29.9\\
					LDAPS\_RHmin& LDAPS model forecast of next-day minimum relative humidity (\%) &19.8 to 98.5\\
					LDAPS\_RHmax & LDAPS model forecast of next-day maximum relative humidity (\%) &58.9 to 100\\
					LDAPS\_Tmax\_lapse &LDAPS model forecast of next-day maximum air temperature applied lapse rate $(^\circ C)$& 17.6 to 38.5\\
					LDAPS\_Tmin\_lapse & LDAPS model forecast of next-day minimum air temperature applied lapse rate $(^\circ C)$ &  14.3 to 29.6\\
					LDAPS\_WS & LDAPS model forecast of next-day average wind speed (m/s)& 2.9 to 21.9\\
					LDAPS\_LH &LDAPS model forecast of next-day average latent heat flux $(W/m^2)$& -13.6 to 213.4\\
					LDAPS\_CC1 & LDAPS model forecast of next-day 1st 6-hour split average cloud cover (0-5 h) (\%)& 0 to 0.97\\
					LDAPS\_CC2 & LDAPS model forecast of next-day 2nd 6-hour split average cloud cover (6-11 h) (\%)& 0 to 0.97\\
					LDAPS\_CC3 & LDAPS model forecast of next-day 3rd 6-hour split average cloud cover (12-17 h) (\%)& 0 to 0.98\\
					LDAPS\_CC4 &LDAPS model forecast of next-day 4th 6-hour split average cloud cover (18-23 h) (\%)& 0 to 0.97\\
					Solar radiation & Daily incoming solar radiation $(wh/m^2)$& 4329.5 to 5992.9\\

				\end{tabular}
				\label{vardescription}
			\end{table}
		\end{landscape}
		\noindent We notice quite high dependence between both the responses, inbetween the predictors and between each other. Thus, to model the data properly one needs to account for this dependence, and our approach can do so. On the diagonal, given are histograms of the fitted u-data, showing that the estimated PITs are properly transformed on the u-scale.
		\subsection{Fitted Y-vine regression model}

						The variables  given below are  enumerated as follows, the response $T_{max}$ is enumerated with 1, $T_{max}=1$, the response $T_{min}$ is enumerated with 2, $T_{min}=2,$ then  $\texttt{LDAPS\_Tmin\_lapse}=7$ , $\texttt{LDAPS\_Tmax\_lapse}=6$, $\texttt{LDAPS\_CC1}=10$,  $\texttt{LDAPS\_WS}=8$, $\texttt{Present\_Tmin}=4$,
						$\texttt{LDAPS\_RHmax}=5$, $\texttt{	LDAPS\_CC3}=11$,  $\texttt{LDAPS\_LH}=9$ and  $\texttt{Present\_Tmax}=3$.
						Using that enumeration, in  Tables~\ref{families1} and \ref{families2} we show the  parametric   pair copulas that were fitted by our Y-vine regression model.
						In each tree we give  the pair copulas conditioned and conditioning sets, the estimated family, the rotation in degrees, the parameters, the degree of freedom (number of parameters) and the Kendall's $\hat{\tau}$ values. 
						\subsection{Fitted pair copulas}\label{app:bqrdatafamilies}
						\begin{table}[h] 
							\centering
							\caption{For the fitted $T_1$ to $T_5$ given are the conditioned and conditioning sets of the pair copulas, the estimated family, the rotation in degrees, the parameters, the degree of freedom and the Kendall's $\hat{\tau}$ values.}
							\begin{footnotesize}	
								\begin{tabular}{|rrllllrrr|}
									\hline
									tree & edge & conditioned & conditioning &  family & rotation & parameters & df & Kendall's $\hat{\tau}$\\
									\hline
									1 & 1 & 1, 7 &  & gaussian & 0 & 0.66  & 1 & 0.46\\
									1 & 2 & 2, 7 &  & gaussian & 0 & 0.91  & 1 & 0.73\\
									1 & 3 & 7, 6 &  &  gaussian & 0 & 0.64  & 1 & 0.44\\
									1 & 4 & 6, 10 &  &  gaussian & 0 & -0.45  & 1 & -0.30\\
									1 & 5 & 10, 8 &  & clayton & 180 & 0.27  & 1 & 0.12\\
									1 & 6 & 8, 4 &   & indep & 0 &  & 0 & 0.00\\
									1 & 7 & 4, 5 &  & indep & 0 &  & 0 & 0.00\\
									1 & 8 & 5, 11 &  & clayton & 180 & 0.48  & 1 & 0.19\\
									1 & 9 & 11, 9 &   & bb8 & 0 & 1.67 , 0.97  & 2 & 0.24\\
									1 & 10 & 9, 3 &   & student t & 0 & 0.04, 7.06  & 2 & 0.03\\
									\hline
									2 & 1 & 1, 6 & 7  & bb1 & 180 & 0.78 , 1.68  & 2 & 0.57\\
									2 & 2 & 2, 6 & 7  & indep & 0 &  & 0 & 0.00\\
									2 & 3 & 7, 10 & 6  & clayton & 0 & 0.64 & 1 & 0.24\\
									2 & 4 & 6, 8 & 10 & joe & 90 & 1.22  & 1 & -0.11\\
									2 & 5 & 10, 4 & 8 & clayton & 0 & 0.17  & 1 & 0.08\\
									2 & 6 & 8, 5 & 4& bb8 & 0 & 1.87, 0.70  & 2 & 0.13\\
									2 & 7 & 4, 11 & 5  & student t & 0 & 0.01, 5.21  & 2 & 0.01\\
									2 & 8 & 5, 9 & 11  & gaussian & 0 & 0.34  & 1 & 0.22\\
									2 & 9 & 11, 3 & 9  & gumbel & 270 & 1.12  & 1 & -0.10\\
									\hline
									3 & 1 & 1, 10 & 6, 7 & joe & 90 & 1.34  & 1 & -0.16\\
									3 & 2 & 2, 10 & 6, 7  & indep & 0 &  & 0 & 0.00\\
									3 & 3 & 7, 8 & 10, 6  & indep & 0 &  & 0 & 0.00\\
									3 & 4 & 6, 4 & 8, 10  & gaussian & 0 & 0.53  & 1 & 0.35\\
									3 & 5 & 10, 5 & 4, 8 & bb8 & 0 & 3.28, 0.85  & 2 & 0.42\\
									3 & 6 & 8, 11 & 5, 4  & joe & 0 & 1.17  & 1 & 0.09\\
									
									3 & 7 & 4, 9 & 11, 5  &student t & 0 & 0.15, 6.59  & 2 & 0.10\\
									3 & 8 & 5, 3 & 9, 11  & gumbel & 270 & 1.37  & 1 & -0.27\\
									\hline
									4 & 1 & 1, 8 & 10, 6, 7  & frank & 0 & -1.77  & 1 & -0.19\\
									4 & 2 & 2, 8 & 10, 6, 7  & student t & 0 & 0.10, 7.34  & 2 & 0.06\\
									4 & 3 & 7, 4 & 8, 10, 6 & gaussian & 0 & 0.65  & 1 & 0.45\\
									
									4 & 4 & 6, 5 & 4, 8, 10 & frank & 0 & -1.37  & 1 & -0.15\\
									4 & 5 & 10, 11 & 5, 4, 8  & frank & 0 & 2.21  & 1 & 0.23\\
									4 & 6 & 8, 9 & 11, 5, 4  & gumbel & 0 & 1.17  & 1 & 0.14\\
									4 & 7 & 4, 3 & 9, 11, 5  & gaussian & 0 & 0.68  & 1 & 0.47\\
									\hline
									5 & 1 & 1, 4 & 8, 10, 6, 7  &student t & 0 & 0.07, 11.79  & 2 & 0.05\\
									5 & 2 & 2, 4 & 8, 10, 6, 7  & bb8 & 0 & 1.44, 0.94  & 2 & 0.15\\
									5 & 3 & 7, 5 & 4, 8, 10, 6  & indep & 0 &  & 0 & 0.00\\
									5 & 4 & 6, 11 & 5, 4, 8, 10  & bb8 & 90 & 3.00, 0.82  & 2 & -0.36\\
									5 & 5 & 10, 9 & 11, 5, 4, 8  & frank & 0 & 1.28  & 1 & 0.14\\
									5 & 6 & 8, 3 & 9, 11, 5, 4  & indep & 0 &  & 0 & 0.00\\
									\hline
								\end{tabular}
								
							\end{footnotesize}
							\label{families1}
						\end{table}
						
						\begin{table}[h] 
							\centering
							\caption{For the fitted $T_6$ to $T_{10}$ given are the conditioned and conditioning sets of the pair copulas, the estimated family, the rotation in degrees, the parameters, the degree of freedom and the Kendall's $\hat{\tau}$ values.}
							\begin{footnotesize}	
								\begin{tabular}{|rrllllrrr|}
									\hline
									tree & edge & conditioned & conditioning &  family & rotation & parameters & df & Kendall's $\hat{\tau}$\\
									\hline
									
									6 & 1 & 1, 5 & 4, 8, 10, 6, 7  & frank & 0 & -0.91 & 1 & -0.10\\
									6 & 2 & 2, 5 & 4, 8, 10, 6, 7  & frank & 0 & 0.82  & 1 & 0.09\\
									6 & 3 & 7, 11 & 5, 4, 8, 10, 6  & clayton & 0 & 0.38  & 1 & 0.16\\
									6 & 4 & 6, 9 & 11, 5, 4, 8, 10  & joe & 0 & 1.09  & 1 & 0.05\\
									6 & 5 & 10, 3 & 9, 11, 5, 4, 8  & gumbel & 90 & 1.19  & 1 & -0.16\\
									\hline
									7 & 1 & 1, 11 & 5, 4, 8, 10, 6, 7 &  clayton & 270 & 0.18  & 1 & -0.08\\
									7 & 2 & 2, 11 & 5, 4, 8, 10, 6, 7 &  indep & 0 &  & 0 & 0.00\\
									7 & 3 & 7, 9 & 11, 5, 4, 8, 10, 6 &  clayton & 90 & 0.17  & 1 & -0.08\\
									7 & 4 & 6, 3 & 9, 11, 5, 4, 8, 10 &  bb7 & 180 & 1.18, 0.20  & 2 & 0.17\\
									\hline
									8 & 1 & 1, 9 & 11, 5, 4, 8, 10, 6, 7 &  indep & 0 &  & 0 & 0.00\\
									8 & 2 & 2, 9 & 11, 5, 4, 8, 10, 6, 7 & gaussian & 0 & -0.15  & 1 & -0.10\\
									8 & 3 & 7, 3 & 9, 11, 5, 4, 8, 10, 6  & bb1 & 0 & 0.24, 1.13  & 2 & 0.21\\
									\hline
									9 & 1 & 1, 3 & 9, 11, 5, 4, 8, 10, 6, 7 &  joe & 0 & 1.08  & 1 & 0.04\\
									9 & 2 & 2, 3 & 9, 11, 5, 4, 8, 10, 6, 7 &  gaussian & 0 & 0.16  & 1 & 0.10\\
									\hline
									10 & 1 & 1, 2 & 3, 9, 11, 5, 4, 8, 10, 6, 7 &  joe & 180 & 1.18  & 1 & 0.09\\
									\hline
								\end{tabular}
								
							\end{footnotesize}
							\label{families2}
						\end{table}
						
						\subsection{Bivariate confidence regions coverage probabilities}
						
						\begin{table}[H]
							\centering
							\begin{tabular}{|c|r|r|r|r|r|r|r|}
								\hline
								$\alpha$ & 0.05 & 0.10 &0.25 &0.50 & 0.75 & 0.90 & 0.95\\
								\hline
								$ \hat{G}(\alpha) $ &0.10&	0.20	& 0.41 &	0.67&	0.89&	0.97 &	0.99\\
								
								$	\hat{\beta}(\alpha) $ & 0.02 &	0.05 &	0.14 &	0.32 &	0.59 &	0.76 &	0.86 \\
								\hline
							\end{tabular}
							\caption{For all $\alpha$ levels,  estimated  coverage probabilities $\hat{G}(\alpha)$ and estimated adjustment $\hat{\beta}(\alpha)$ for the corresponding \textbf{unconditional}  quantile levels.}
							\label{table:coverageprobsunconditional}
						\end{table}
						
						\begin{table}[H]
							\centering
							\begin{tabular}{|c|r|r|r|r|r|r|r|}
								\hline
								$\alpha$ & 0.05 & 0.10 &0.25 &0.50 & 0.75 & 0.90 & 0.95\\
								\hline
								$ \hat{G}(\alpha) $ &0.12 &	0.24 &	0.52&	0.76&	0.91 &	0.97&	0.99\\
								
								$	\hat{\beta}(\alpha) $ & 0.03 &	0.05 &	0.11 &	0.23 &	0.48 &	0.73&	0.85 \\
								\hline
							\end{tabular}
							\caption{For all $\alpha$ levels,  estimated  coverage probabilities $\hat{G}(\alpha)$ and estimated adjustment $\hat{\beta}(\alpha)$ for the corresponding conditional  quantile levels for \textbf{02.07.2017}.}
							\label{table:coverageprobsconditional02}
						\end{table}
						
						\begin{table}[H]
							\centering
							\begin{tabular}{|c|r|r|r|r|r|r|r|}
								\hline
								$\alpha$ & 0.05 & 0.10 &0.25 &0.50 & 0.75 & 0.90 & 0.95\\
								\hline
								$ \hat{G}(\alpha) $ &	0.10&	0.17&	0.37&	0.64	&0.83&	0.95&	0.98\\
								$	\hat{\beta}(\alpha) $ & 0.025&	0.05&	0.16	&0.36 &	0.66 &	0.84 &	0.91 \\
								
								\hline
							\end{tabular}
							\caption{For all $\alpha$ levels,  estimated  coverage probabilities $\hat{G}(\alpha)$ and estimated adjustment $\hat{\beta}(\alpha)$ for the corresponding conditional  quantile levels for \textbf{21.08.2017}.}
							\label{table:coverageprobsconditional2108}
						\end{table}
						
						\begin{figure}[]
							\centering
							\begin{tikzpicture}
								
								
								\node[inner sep=0pt] (Col1) at (-20, 10)  {\includegraphics[width=0.29\textwidth]{plotTogether_Xscale.png}};
								
								\node (Col111) at (-20, 12.8)  {Unconditional};
								
								\node[inner sep=0pt] (Col1) at (-14, 10)  {\includegraphics[width=0.29\textwidth]{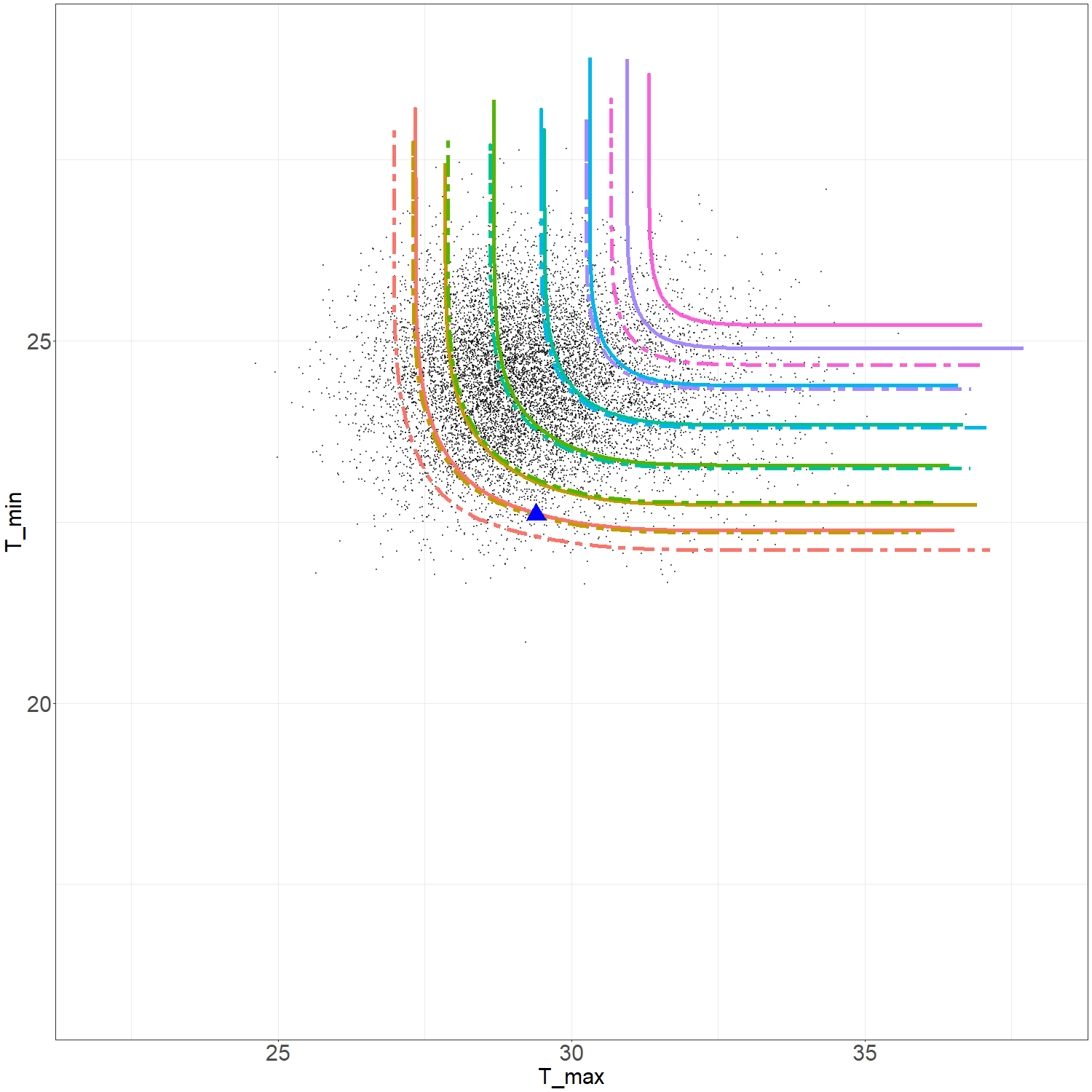}};
								
								\node (Col111) at (-14, 12.8)  {Conditional on 02.07};
								\node[inner sep=0pt] (Col1) at (-8, 10)  {\includegraphics[width=0.29\textwidth]{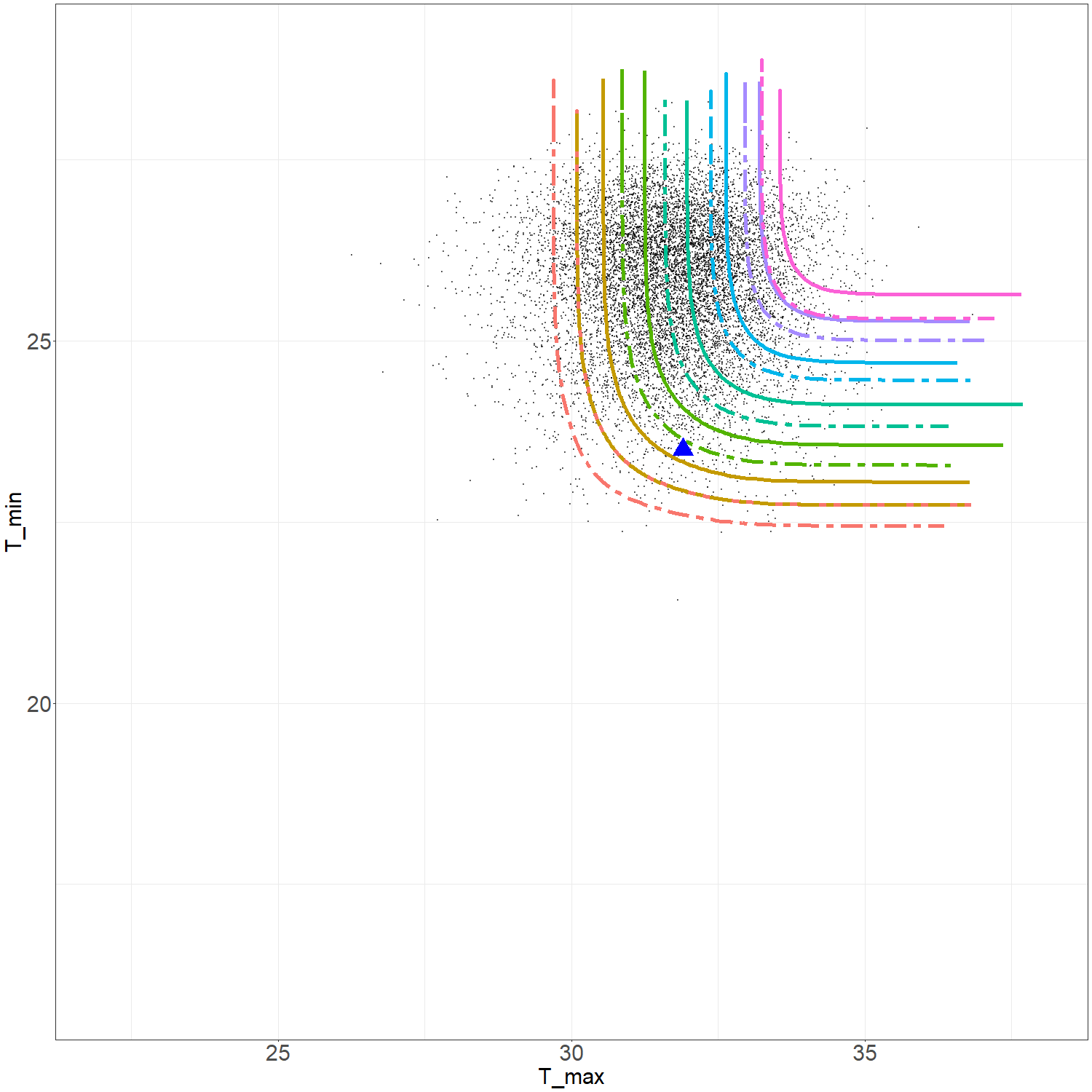}};
								\node (Col111) at (-8, 12.8)  {Conditional on 21.08};
								
								\node[inner sep=0pt] (Col1) at (-20, 4.5)  {\includegraphics[width=0.29\textwidth]{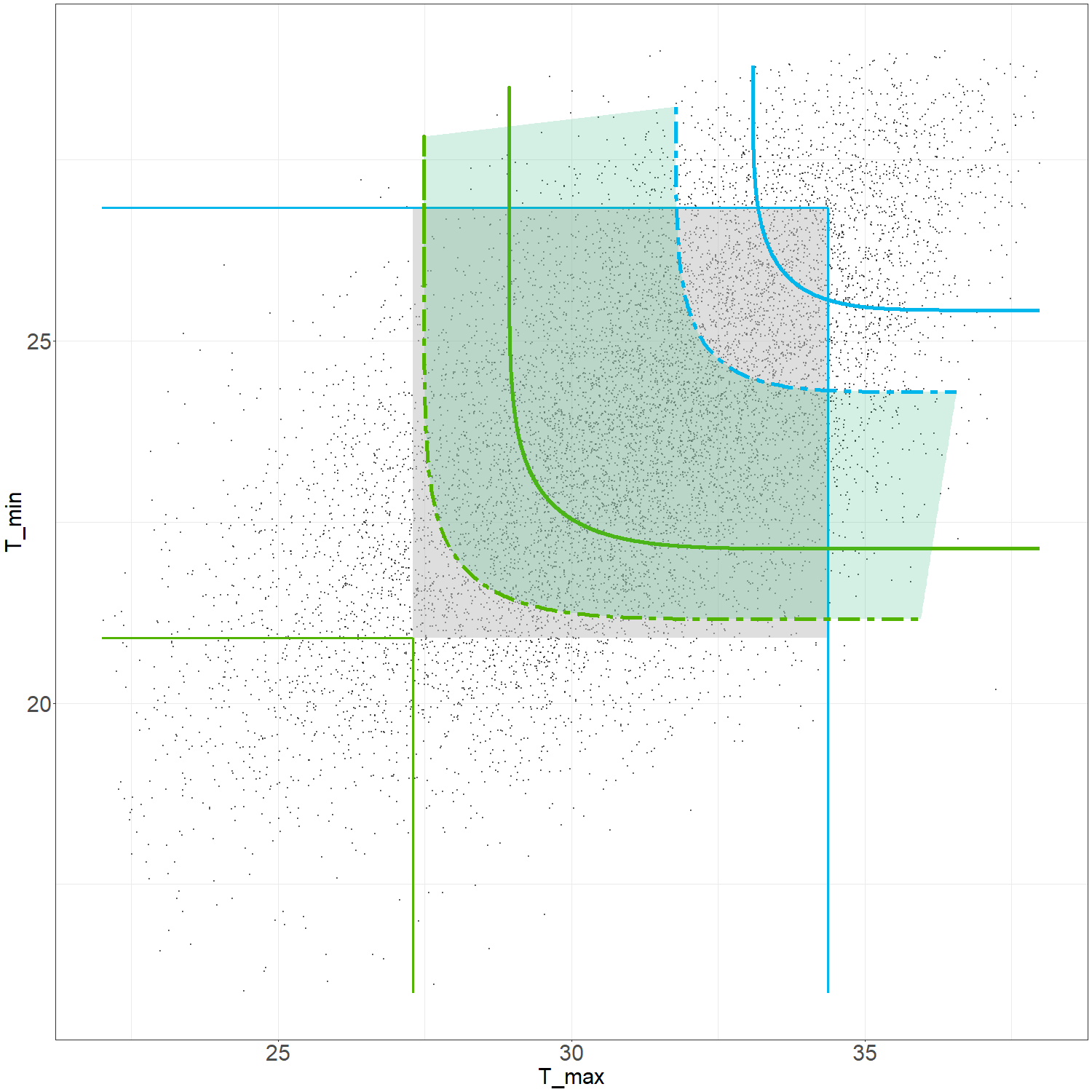}};

								\node[inner sep=0pt] (Col1) at (-14, 4.5)  {\includegraphics[width=0.29\textwidth]{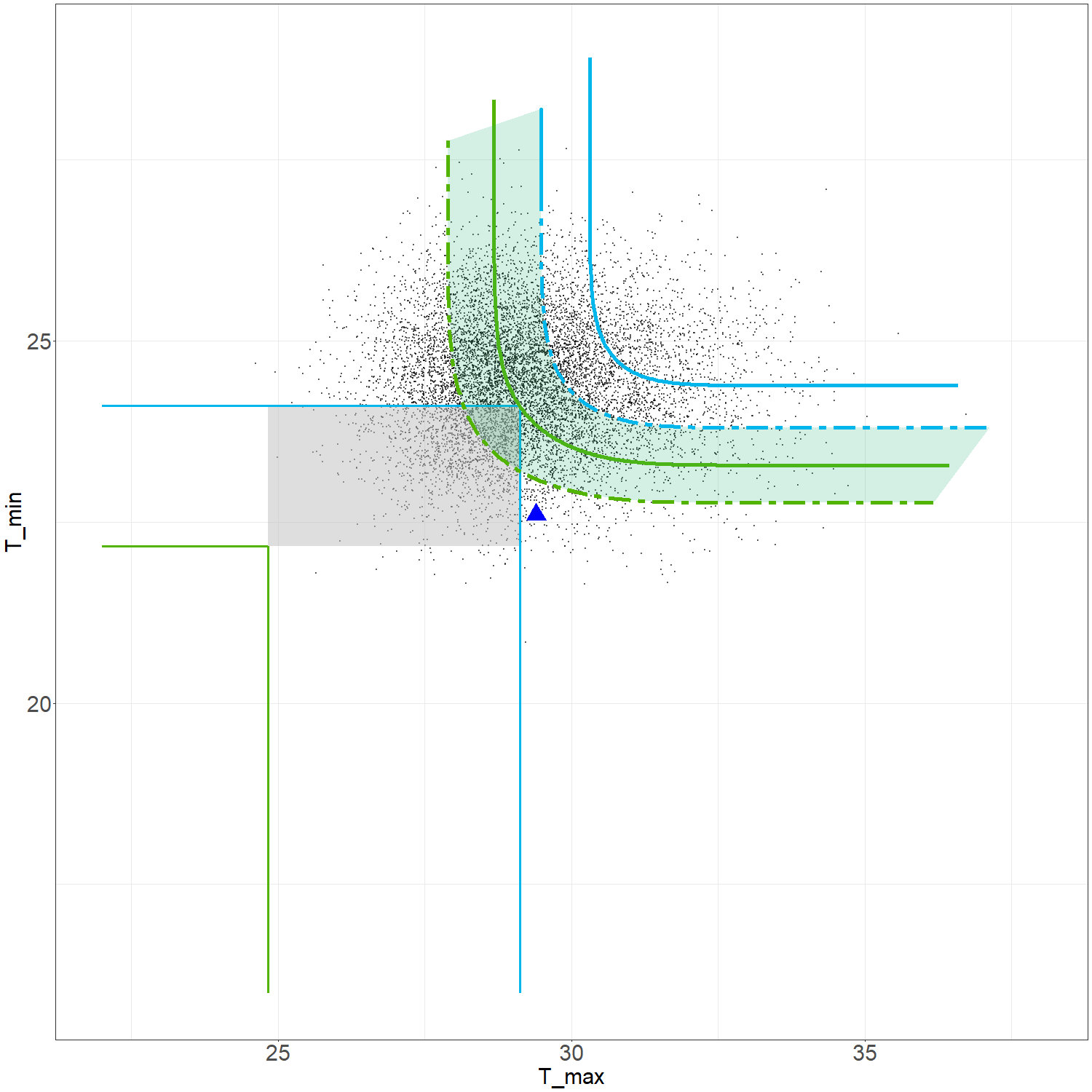}};

								\node[inner sep=0pt] (Col1) at (-8, 4.5)  {\includegraphics[width=0.29\textwidth]{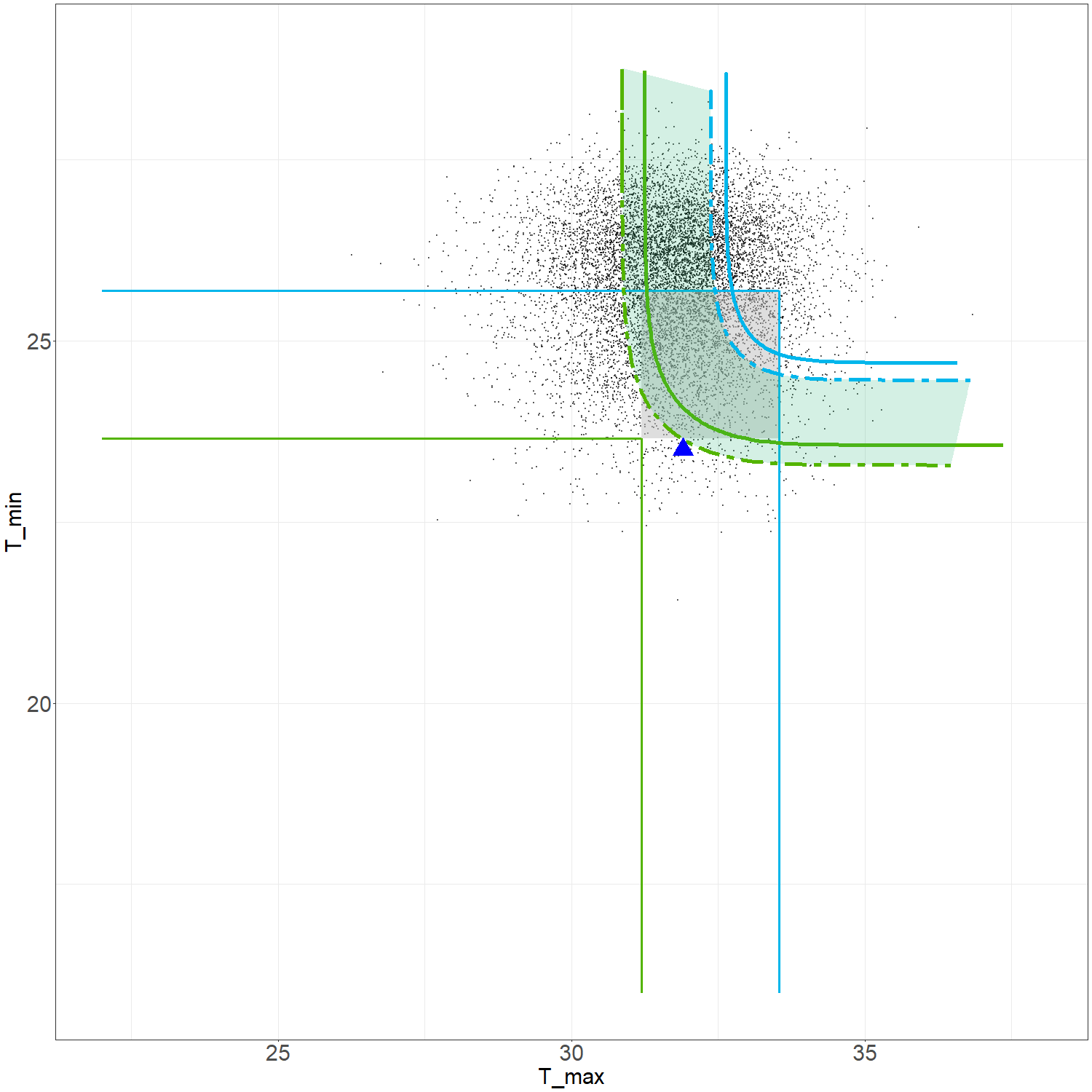}};
								
								\node[inner sep=0pt] (Col1) at (-20, -1)  {\includegraphics[width=0.29\textwidth]{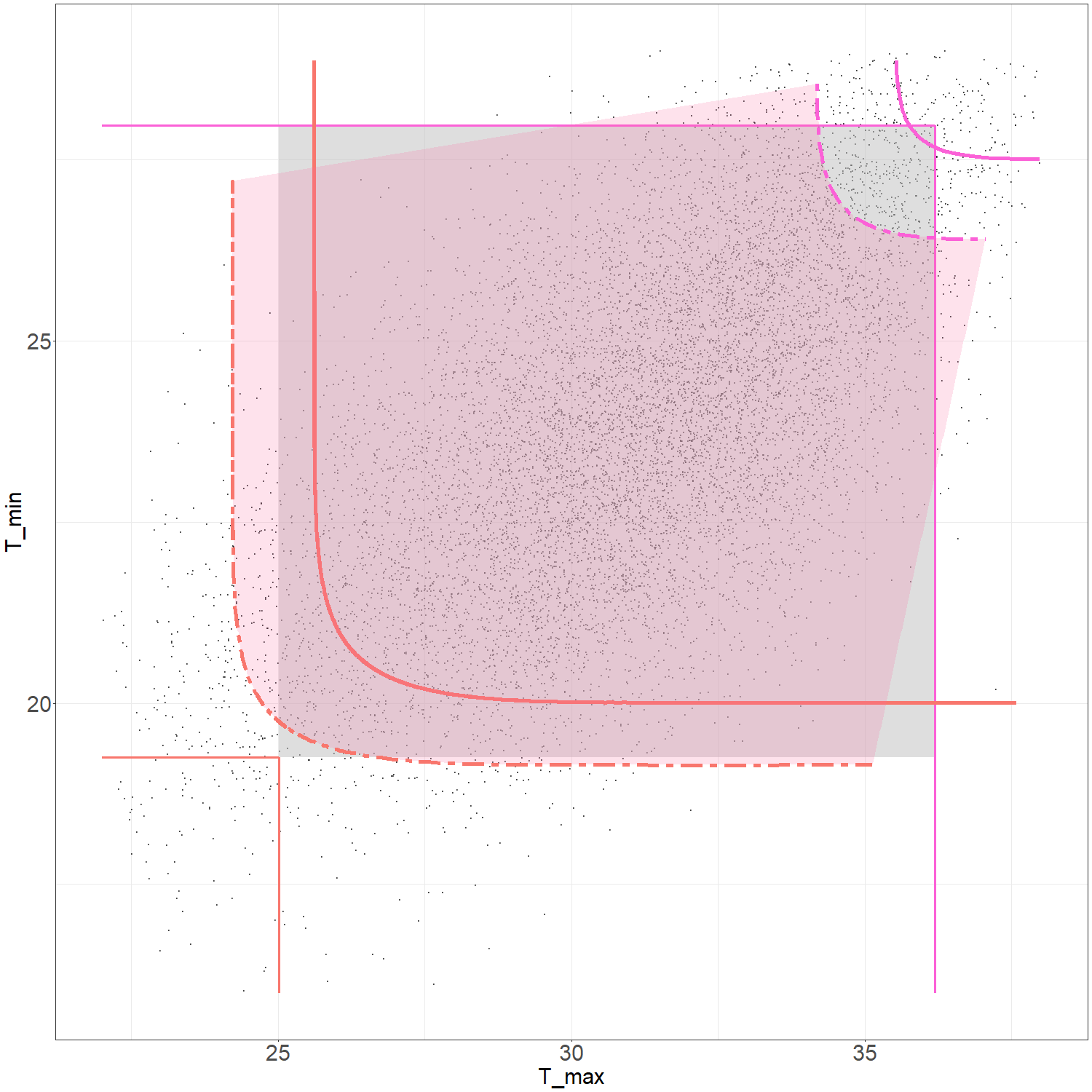}};

								\node[inner sep=0pt] (Col1) at (-14, -1)  {\includegraphics[width=0.29\textwidth]{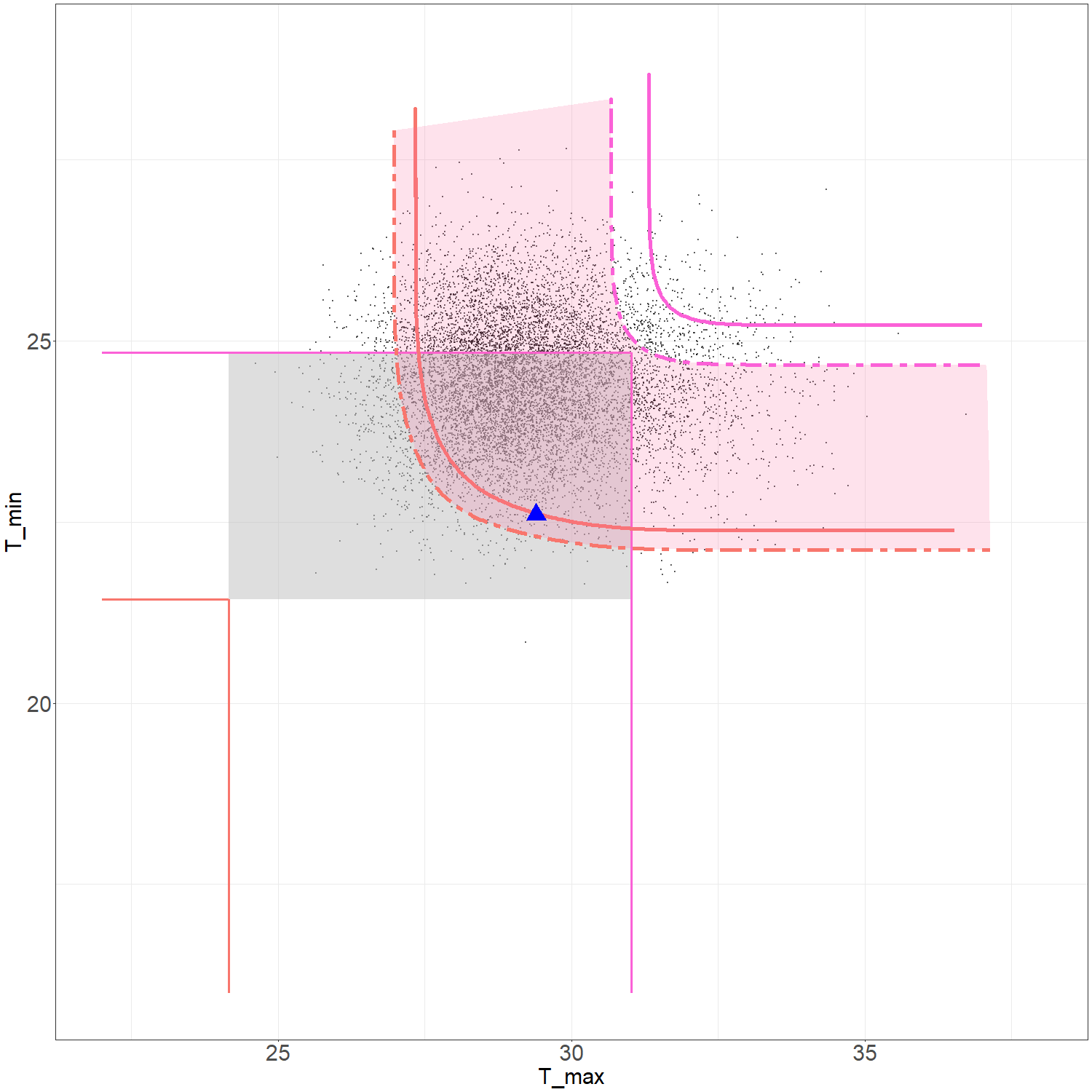}};

								\node[inner sep=0pt] (Col1) at (-8, -1)  {\includegraphics[width=0.29\textwidth]{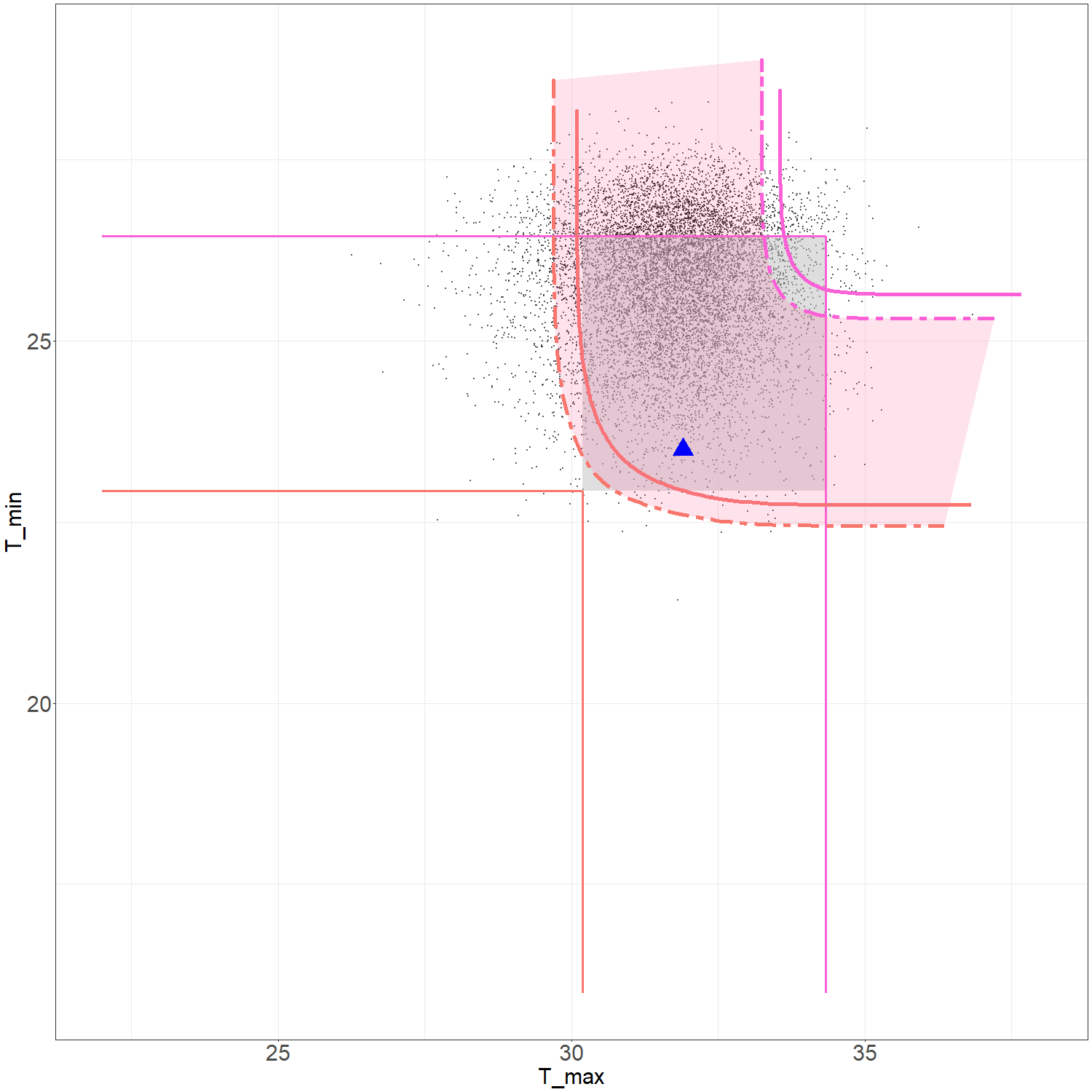}};
								
							\end{tikzpicture}
							\caption{First row: level curves (solid lines) and corresponding quantile curves (dashed lines) at $\alpha = 0.05,0.1,0.25,0.5, 0.75, 0.90,0.95$. Second row: first column $CI^{Y_1,  Y_2 }_{0.50}$ (green  region) and $CI^{Y_1 \perp Y_2 }_{0.50}$ (gray  region), second column  $CI^{Y_1,  Y_2 | \mathbf{X}}_{0.50}$ (green  region) and $CI^{Y_1 \perp Y_2 | \mathbf{X} }_{0.50}$ (gray  region). Third row: first column $CI^{Y_1,  Y_2 }_{0.90}$  (red  region) and $CI^{Y_1 \perp Y_2 }_{0.90}$ (gray  region), second column $CI^{Y_1,  Y_2 | \mathbf{X} }_{0.90}$ (red  region) and $CI^{Y_1 \perp Y_2 | \mathbf{X} }_{0.90}$ (gray region).}
							\label{figquantilesunconditionall}
						\end{figure}
